\RequirePackage{fix-cm} % Fix LaTeX2e bugs.
\documentclass[a4paper, twoside, reqno, dvips, 12pt]{amsart}
\usepackage{fixltx2e}   % Fix LaTeX2e bugs.

\usepackage[latin1]{inputenc}
\usepackage[T1]{fontenc}

\usepackage{esint}
\usepackage{amsbsy}
\usepackage{dsfont}
\usepackage{xspace}
\usepackage{amsgen}
\usepackage{amsthm}
\usepackage{amssymb}
\usepackage{amsmath}
\usepackage{upgreek}
\usepackage{amsfonts}
\usepackage{MnSymbol}
\usepackage{mathrsfs}
\usepackage{mathtools}
\usepackage{textgreek}
\usepackage[nice]{nicefrac}

\usepackage{relsize}
\usepackage{textcomp}
\usepackage[mathcal, mathscr]{euscript}

\usepackage{mathrsfs}
\DeclareMathAlphabet{\mathscrbf}{OMS}{mdugm}{b}{n}

\usepackage{a4wide}

\usepackage[dvipsnames, table]{xcolor}

\definecolor{bckg}{RGB}{20.8, 20.8, 20.8}
\definecolor{oneblue}{rgb}{0.0, 0.0, 0.85}
\definecolor{Lightblue}{RGB}{214, 214, 214}
\definecolor{bluepigment}{rgb}{0.2, 0.2, 0.6}
\definecolor{charcoal}{rgb}{0.21, 0.27, 0.31}
\definecolor{denimblue}{rgb}{0.08, 0.38, 0.74}
\definecolor{Lightgray}{rgb}{0.89, 0.89, 0.89}
\definecolor{darkgrey}{rgb}{0.273, 0.281, 0.30}
\definecolor{darkelectricblue}{rgb}{0.33, 0.41, 0.47}

\usepackage[sort&compress, comma, square, numbers]{natbib}

\usepackage{psfrag}
\usepackage{graphicx}
\usepackage{subfigure}
\usepackage{morefloats}
\usepackage{indentfirst}

\usepackage{acronym}
\usepackage{microtype}
\usepackage[labelsep=period,%
            labelfont={bf,sf,color=bluepigment},%
            justification=raggedright]{caption}

\usepackage[perpage, symbol]{footmisc}

\usepackage[usenames, dvipsnames, pdf]{pstricks}
\usepackage{epsfig}
\usepackage{pst-grad} % For gradients
\usepackage{pst-plot} % For axes

\usepackage[colorlinks,
           urlcolor=oneblue,
           linkcolor=denimblue,
           citecolor=NavyBlue,
           bookmarksopen=false,
           pdfpagemode=UseNone,
           pagebackref]{hyperref}

\usepackage[explicit]{titlesec}

\titleformat{\section}
  {\color{NavyBlue}\Large\sffamily\bfseries}
  {}
  {0em}
  {\colorbox{bckg!5}{\parbox{\dimexpr\linewidth-2\fboxsep\relax}{\centering\thesection. #1}}}
  [\vspace*{0.33em}]

\titleformat{name=\section,numberless}
  {\color{NavyBlue}\Large\sffamily\bfseries}
  {}
  {0.0em}
  {\colorbox{bckg!10}{\parbox{\dimexpr\linewidth-2\fboxsep\relax}{\centering#1}}}
  [\vspace*{0.33em}]

\titleformat{\subsection}
  {\color{NavyBlue}\large\sffamily\bfseries}
  {}
  {0.0em}
  {\colorbox{bckg!5}{\parbox{\dimexpr\linewidth-2\fboxsep\relax}{\centering\thesubsection. #1}}}
  [\vspace*{0.33em}]

\titleformat{name=\subsection,numberless}
  {\color{NavyBlue}\Large\sffamily\bfseries}
  {}
  {0em}
  {\colorbox{bckg!10}{\parbox{\dimexpr\linewidth-2\fboxsep\relax}{\centering#1}}}
  [\vspace*{0.33em}]

\titleformat{\subsubsection}
  {\color{bluepigment}\sffamily\normalsize\bfseries}
  {\thesubsubsection}
  {0.5em}
  {#1}
  [\vspace*{0.33em}]

\titleformat{\paragraph}[runin]
  {\color{bluepigment}\sffamily\small\bfseries}
  {}
  {0em}
  {#1}

\titlespacing{\section}{1.0em}{1.5em plus 2pt minus 2pt}%
{1.0em plus 2pt minus 2pt}[0em]
\titlespacing{\subsection}{1.0em}{1.5em plus 2pt minus 2pt}%
{1.0em}[0em]
\titlespacing{\subsubsection}{1.0em}{1.5em plus 2pt minus 2pt}%
{1.0em plus 2pt minus 2pt}[0em]

\usepackage{titletoc}

\contentsmargin{0.5em}
\newlength{\tocsep} 
\titlecontents{section}[\tocsep]
  {\addvspace{10pt}\bfseries\sffamily}
  {\contentslabel[\thecontentslabel]{\tocsep}}
  {}
  {\ \titlerule*[0.75pc]{.}\ \thecontentspage}
  []
\titlecontents{subsection}[\tocsep]
  {\addvspace{8pt}\sffamily}
  {\contentslabel[\thecontentslabel]{\tocsep}}
  {}
  {\ \titlerule*[0.5pc]{.}\ \thecontentspage}
  []
\titlecontents*{subsubsection}[\tocsep]
  {\addvspace{2pt}\footnotesize\sffamily}
  {}
  {}
  {\ \titlerule*[0.35pc]{.}\ \thecontentspage}
  [\\*]

\makeatletter
\def\@setauthors{%
  \begingroup
  \def\thanks{\protect\thanks@warning}%
  \trivlist
  \centering\footnotesize \@topsep30\p@\relax
  \advance\@topsep by -\baselineskip
  \item\relax
  \author@andify\authors
  \def\\{\protect\linebreak}%
  \textsc{\normalsize\textcolor{darkelectricblue}{\authors}}%
  \ifx\@empty\contribs
  \else
    ,\penalty-3 \space \@setcontribs
    \@closetoccontribs
  \fi
  \endtrivlist
  \endgroup
}
\def\@settitle{\begin{center}%
  \baselineskip14\p@\relax
    \bfseries
    \textsc{\Large\textcolor{charcoal}{\@title}}
  \end{center}%
}
\makeatother

\usepackage{enumitem}
\setlist[description]{%
  topsep=30pt,               % space before start / after end of list
  itemsep=5pt,               % space between items
  font={\bfseries\sffamily\color{NavyBlue}}, % if colour is needed
}

\usepackage{fancyhdr}
\usepackage{lastpage}

\newcommand*\Title{\textcolor{bluepigment}{Asymptotic nonlinear and dispersive pulsatile flow}}
\newcommand*\Authors{\textcolor{bluepigment}{D.~Mitsotakis, D.~Dutykh \& Q.~Li}}
\newcommand*{\plogo}{\textcolor{gray}{{\texttt{arXiv.org} / \textsc{hal}}}} % Generic publisher logo

\numberwithin{equation}{section}

\newtheorem{remark}{Remark}

\newcommand{\ud}{\mathrm{d}}
\newcommand{\ui}{\mathrm{i}}
\newcommand{\ue}{\mathrm{e}}
\renewcommand{\O}{\mathcal{O}}

\renewcommand{\eta}{\text{\texteta}}

\renewcommand{\leq}{\leqslant}
\renewcommand{\geq}{\geqslant}
\newcommand{\cf}{\emph{c.f.}\xspace}
\newcommand{\ie}{\emph{i.e.}\xspace}
\newcommand{\eg}{\emph{e.g.}\xspace}

\newcommand{\abs}[1]{\lvert\, #1\, \rvert}

\newcommand{\od}[2]{\frac{\mathrm{d} #1}{\mathrm{d}\/#2}}

\makeatletter
\renewcommand*\env@matrix[1][\arraystretch]{%
  \edef\arraystretch{#1}%
  \hskip -\arraycolsep
  \let\@ifnextchar\new@ifnextchar
  \array{*\c@MaxMatrixCols c}}
\makeatother

\usepackage{acronym}
\acrodef{BVP}{Boundary Value Problem}
\acrodef{NSWE}{Nonlinear Shallow Water Equations}

\begin{document}

\title[\Title]{Asymptotic nonlinear and dispersive pulsatile flow in elastic vessels with cylindrical symmetry}

\author[D.~Mitsotakis]{Dimitrios Mitsotakis$^*$}
\address{\textbf{D.~Mitsotakis:} Victoria University of Wellington, School of Mathematics and Statistics, PO Box 600, Wellington 6140, New Zealand}
\email{dimitrios.mitsotakis@vuw.ac.nz}
\urladdr{https://sites.google.com/site/dmitsot/}
\thanks{$^*$ Corresponding author}

\author[D.~Dutykh]{Denys Dutykh}
\address{\textbf{D.~Dutykh:} Univ. Grenoble Alpes, Univ. Savoie Mont Blanc, CNRS, LAMA, 73000 Chamb\'ery, France and LAMA, UMR 5127 CNRS, Universit\'e Savoie Mont Blanc, Campus Scientifique, 73376 Le Bourget-du-Lac Cedex, France}
\email{Denys.Dutykh@univ-smb.fr}
\urladdr{http://www.denys-dutykh.com/}

\author[Q.~Li]{Qian Li}
\address{\textbf{Q.~Li:} Victoria University of Wellington, School of Mathematics and Statistics, PO Box 600, Wellington 6140, New Zealand}
\email{liqian5@myvuw.ac.nz}

\keywords{Boussinesq systems; blood flow; elastic vessel}

%%% ------------------------------------------------------------------------ %%%

\begin{titlepage}
\thispagestyle{empty} % Remove page numbering on this page
\noindent
{\Large Dimitrios \textsc{Mitsotakis}}\\
{\it\textcolor{gray}{Victoria University of Wellington, New Zealand}}
\\[0.02\textheight]
{\Large Denys \textsc{Dutykh}}\\
{\it\textcolor{gray}{CNRS, Universit\'e Savoie Mont Blanc, France}}
\\[0.02\textheight]
{\Large Qian \textsc{Li}}\\
{\it\textcolor{gray}{Victoria University of Wellington, New Zealand}}
\\[0.16\textheight]

\vspace*{0.99cm}

\colorbox{Lightblue}{
  \parbox[t]{1.0\textwidth}{
    \centering\huge\sc
    \vspace*{0.75cm}
    
    \textcolor{bluepigment}{Asymptotic nonlinear and dispersive pulsatile flow in elastic vessels with cylindrical symmetry}
    
    \vspace*{0.75cm}
  }
}

\vfill % Whitespace between the title block and the publisher

\raggedleft     % Right-align all text
{\large \plogo} % Publisher and logo
\end{titlepage}

%%% ------------------------------------------------------------------------ %%%

\newpage
\thispagestyle{empty} % Remove page numbering on this page
\par\vspace*{\fill}   % Whitespace until the bottom
\begin{flushright} % Right-align all text
{\textcolor{denimblue}{\textsc{Last modified:}} \today}
\end{flushright}

%%% ------------------------------------------------------------------------ %%%

\newpage
\maketitle
\thispagestyle{empty}

%%% ------------------------------------------------------------------------ %%%

\begin{abstract}

The asymptotic derivation of a new family of one-dimensional, weakly nonlinear and weakly dispersive equations that model the flow of an ideal fluid in an elastic vessel is presented. Dissipative effects due to the viscous nature of the fluid are also taken into account. The new models validate by asymptotic reasoning other non-dispersive systems of equations that are commonly used, and improve other nonlinear and dispersive mathematical models derived to describe the blood flow in elastic vessels. The new systems are studied analytically in terms of their basic characteristic properties such as the linear dispersion characteristics, symmetries, conservation laws and solitary waves. Unidirectional model equations are also derived and analysed in the case of vessels of constant radius. The capacity of the models to be used in practical problems is being demonstrated by employing a particular system with favourable properties to study the blood flow in a large artery. Two different cases are considered: A vessel with constant radius and a tapered vessel. Significant changes in the flow can be observed in the case of the tapered vessel.

\bigskip
\noindent \textbf{\keywordsname:} Boussinesq systems; blood flow; elastic vessel \\

\smallskip
\noindent \textbf{MSC:} \subjclass[2010]{ 35Q35 (primary), 74J30, 92C35 (secondary)}
\smallskip \\
\noindent \textbf{PACS:} \subjclass[2010]{ 87.85.gf (primary), 87.19.U- (secondary)}

\end{abstract}

%%% ------------------------------------------------------------------------ %%%

\newpage
\tableofcontents
\thispagestyle{empty}

%%% ------------------------------------------------------------------------ %%%

\newpage
\section{Introduction}

The study of the axisymmetric flow of an inviscid fluid in elastic vessels is important on several accounts but especially because of its applications to the blood flow in arteries. The mathematical modelling of arterial systems is based on the equations of continuum mechanics for the flow of an incompressible fluid in vessels known as the \textsc{Navier--Stokes} equations, \cite{Fung1997a}. The incompressible \textsc{Navier--Stokes} formulation of the blood flow has the advantage of taking into account the dissipative effects of the flow due to viscosity. On the other hand, the flow exhibits a rather complex structure due to the mechanical interaction between the fluid and the vessel walls. Another very important factor that influences the blood flow is the viscoelastic nature of vessel walls. For example, large arteries deform under blood pressure and they are capable of storing elastic energy during the systolic phase and release it during the diastolic phase. Modelling the elastic properties of the vessels appears to have significant difficulties of mathematical and numerical nature, \cite{Fung1997a, Chandran2012, Quarteroni2004, VandeVosse2011, Nichols2011, Smith2002}.

Several attempts to simplify the study of the blood flow have been made, especially in the case of large vessels with elastic wall that are capable to deform under pressure. In many recent studies the viscosity of the flow has been neglected since otherwise the mathematical modelling becomes very complicated. To this end, the focus is mainly on the inviscid, incompressible and radially symmetric fluid flow equations known to as the Euler equations. These equations written in cylindrical coordinates take the form:
\begin{align}
& u_{\,t}\ +\ u\,u_{\,x}\ +\ v\,u_{\,r}\ +\ \frac{1}{\rho}\;p_{\,x}\ =\ 0\,, \label{eq:euler1}\\
& v_{\,t}\ +\ u\,v_{\,x}\ +\ v\,v_{\,r}\ +\ \frac{1}{\rho}\;p_{\,r}\ =\ 0\,, \label{eq:euler2}\\
& u_{\,x}\ +\ v_{\,r}\ +\ \frac{1}{r}\;v\ =\ 0\,, \label{eq:euler3}
\end{align}
where $u\ =\ u\,(x,\,r,\,t)\,$, $v\ =\ v\,(x,\,r,\,t)$ are the horizontal and radial velocity of fluid respectively, $p\ =\ p\,(x,\,r,\,t)$ is the pressure of the fluid, while $\rho$ is the constant density of the fluid.

\begin{figure}
  \centering
  \includegraphics[width=0.8\columnwidth]{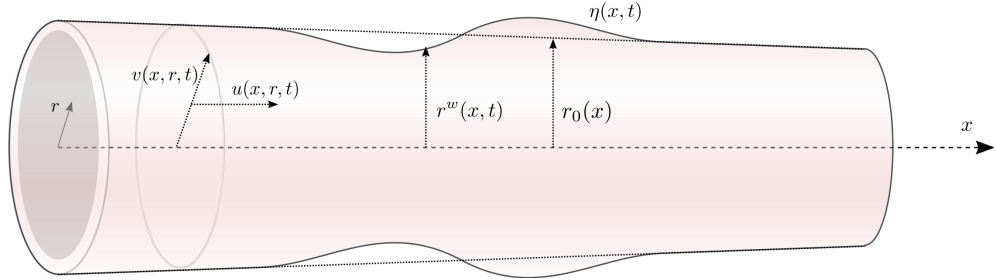}
  \caption{\small\em Sketch of the physical domain for a single vessel segment with elastic and impenetrable wall.}
  \label{fig:vessel}
\end{figure}

A sketch of the physical domain of this problem is presented in Fig.~\ref{fig:vessel}, where the distance of vessel's wall from the centre of the vessel in a cross section is denoted by $r^{\,w}\,(x,\,t)$ and depends on $x$ and $t$ while the radius of the vessel at rest is the function $r_{\,0}\,(x)\,$. In general the deformation of the wall will be a function of $x$ and $t\,$. If we denote the radial displacement of the wall will by $\eta\,(x,\,t)$ then the vessel wall radius can be written as $r^{\,w}\,(x,\,t)\ =\ r_{\,0}\,(x)\ +\ \eta\,(x,\,t)\,$.

The governing equations \eqref{eq:euler1} -- \eqref{eq:euler3} combined with initial and boundary conditions form a closed system. A compatibility condition is also applied at the centre of the vessel (due to cylindrical symmetry). Specifically, we assume that
\begin{equation}\label{eq:bc1}
  v\,(x,\,r,\,t)\ =\ 0\,, \mbox{ for } r\ =\ 0\,.
\end{equation}
On the vessel wall the impermeability condition can be written in the form:
\begin{equation}\label{eq:bc2}
  v\,(x,\,r,\,t)\ =\ \eta_{\,t}\,(x,\,t)\ +\ (r_{\,0}\,(x)\ +\ \eta\,(x,\,t))_{\,x}\,u\,(x,\,r,\,t)\,, \mbox{ for } r\ =\ r^{\,w}\,(x,\,t)\,,
\end{equation}
and expresses that the fluid velocity equals the wall speed $v\ =\ r^{\,w}_{\,t}\,$. The second boundary condition is actually \textsc{Newton}'s second law on the vessel wall written in the form:
\begin{equation}\label{eq:bc3}
  \rho^{\,w}\,h\,\eta_{\,t\,t}\,(x,\,t)\ =\ p^{\,w}\,(x,\,t)\ -\ \frac{E_{\,\sigma}\,h}{r_{\,0}^{\,2}\,(x)}\;\eta\,(x,\,t)\,,
\end{equation}
where $\rho^{\,w}$ is the wall density, $p^{\,w}$ is the transmural pressure, $h$ is the thickness of the vessel wall, $E_{\,\sigma}\ =\ \frac{E}{1\ -\ \sigma^{\,2}}$ where $E$ is the \textsc{Young} modulus of elasticity with $\sigma$ denoting the \textsc{Poisson} ratio of the elastic wall. In this study we assume that $E$ is a constant and in general we will replace in the notation $E_{\,\sigma}$ by $E\,$. It is noted that because the flow is pressure-driven the effect of gravity is neglected. For more information about the derivation of the \textsc{Euler} equations and the boundary conditions we refer to \cite{Zamir2000, Chandran2012}. It is noted that assuming a laminar flow and small viscosity the \textsc{Navier--Stokes} equations can be reduced to a modified system which is very similar to the \textsc{Euler} equations, \cite{Smith2002}, and therefore an analysis on the \textsc{Euler} equations can easily be generalised to the specific simplified viscous case.

Due to the complexity of the \textsc{Euler} equations several one-dimensional models have been introduced, \cite{Hughes1973, Zagzoule1986, Tait1981, Stergiopulos1992, Olufsen2000, Alastruey2011, Reymond2011, Wang2015}. The models include unidirectional, \cf \eg  \cite{Erbay1992, Yomosa1987, Muller2013, Demiray2007, Cascaval2003}, and bidirectional models, \cf \eg \cite{Bessems2007, Mynard2008, Cascaval2012}. Although these models usually are neither asymptotic models nor dispersive, systematic comparisons between one and three dimensional idealised arterial blood flow models showed a very good agreement, \cf \eg \cite{Xiao2014}. Moreover, one-dimensional models can also be used to compute inflow boundary conditions to three-dimensional models. However, one-dimensional models cannot handle curved vessels unless the central axis is a graph of a function. For this reason three-dimensional models cannot be totally replaced by the simple one-dimensional models, \cf \eg \cite{Xiao2014, Perktold1995, Taylor1998, Figueroa2006, Berntsson2016}. It is noted that the literature is not limited in the above references but is very extensive and we apologise if we do not include the complete literature in the field.

In this paper we derive some new asymptotic one-dimensional model equations of \textsc{Boussinesq} type (weakly non-linear and weakly dispersive) that approximate the system \eqref{eq:euler1} -- \eqref{eq:euler3} with boundary conditions \eqref{eq:bc1} -- \eqref{eq:bc3}. The new systems describe inviscid and irrotational fluid flow in elastic vessels of variable diameter and can be used as an alternative to the \textsc{Euler} equations \eqref{eq:euler1} -- \eqref{eq:euler3}. We also derive dissipative \textsc{Boussinesq} equations from the \textsc{Navier--Stokes} equations using standard arguments on the velocity profile, \cite{Smith2002}, and extending the asymptotic reasoning of the inviscid case to the viscous case. The new models are generic and can be used to study the blood flow in large arteries while discarding the dispersive terms then classical non-dispersive models can be recovered. For this reason we validate the new mathematical models by studying the blood flow in ideal arterial segments of the human body using a numerical method based on simple finite-difference methods. The new model equations are of significant importance since they validate using asymptotic reasoning standard 1D models commonly used to simulate blood flow in arteries, they extend these models to more accurate weakly nonlinear and weakly dispersive models, they can be integrated numerically efficiently, and also used to simulate fluid flows in elastic tubes.

The paper is organised as follows: In Section~\ref{sec:review} we review some recent 1D model equations describing nonlinear and dispersive fluid flow in elastic vessels. In Sections~\ref{sec:deriv1} and \ref{sec:deriv2} we derive the new asymptotic models ignoring the viscosity of the flow. In Section  \ref{sec:dissipative} we extend the new model equations by incorporating dissipative effects due to viscosity. In Sections~\ref{sec:properties} and \ref{sec:dispersion} we limit the analysis to vessels with constant initial radius. In Section~\ref{sec:properties} we derive unidirectional models and discuss some important characteristic properties of the new models such as symmetries and conservation laws. In Section~\ref{sec:dispersion} we study the linear dispersion relation of the derived systems and we discuss the existence and computation of traveling wave solutions. We also present model equations with improved dispersion characteristics. In Section~\ref{sec:application} we employ a particular system which is characterised by its simplicity and its accuracy and we validate the new models by numerical means in studies of blood flow. We close this paper with some conclusions.

%%% ------------------------------------------------------------------------ %%%

\section{Review of 1D models for pulsatile flow in elastic vessels}
\label{sec:review}

Recent studies of fluid flows in vessels focus on the derivation of simplified models that take into account the elasticity of the vessels of constant radius $r_{\,0}\ >\ 0$ (or constant cross-sectional area $A$). These models usually ignore the viscous nature of the fluid, and are based on the assumption that the flow is symmetric around the axis of the vessel. All these models are systems of partial differential equations in one space dimension, \cf \cite{Formaggia2003, Sherwin2003, Cascaval2012}, and are based on the mass and momentum conservation of fluid flow in an elastic vessel, \cite{Zagzoule1986}:
\begin{align}
  & \eta_{\,t}\ +\ \eta_{\,x}\,u\ +\ \frac{1}{2}\;\eta\,u_{\,x}\ +\ \frac{1}{2}\;r_{\,0}\,u_{\,x}\ =\ 0\,, \label{eq:j1}\\
  & u_{\,t}\ +\ u\,u_{\,x}\ +\ \frac{1}{\rho}\;p_{\,x}\ =\ 0\,, \label{eq:j2}
\end{align}
where $\eta\ =\ \eta\,(x,\,t)$ denotes the radial displacement of the vessel wall from its rest position, $u\ =\ u\,(x,\,t)$ and $p\ =\ p\,(x,\,t)$ is the cross-section averaged velocity and pressure of a fluid of constant density $\rho$ at point $x$ at time $t\,$. The choice of the pressure $p$ appeared to be crucial in the mathematical classification of the system \eqref{eq:j1} -- \eqref{eq:j2}. For example, using the simple Laplace law to a pressure-area relation leads to the formula
\begin{equation}\label{eq:p1}
  p\ =\ \frac{E\,h}{r_{\,0}^{\,2}}\;\eta\,,
\end{equation}
where $E$ is the \textsc{Young} modulus, $h$ the thickness of the vessel and $r_{\,0}$ the constant radius of the relaxed vessel leads to a hyperbolic system of conservation laws, \cite{Sherwin2003}:
\begin{align}
  & \eta_{\,t}\ +\ \eta_{\,x}\,u\ +\ \frac{1}{2}\;\eta\,u_{\,x}\ +\ \frac{1}{2}\;r_{\,0}\,u_{\,x}\ =\ 0\,, \label{eq:j3}\\
  & u_{\,t}\ +\ u\,u_{\,x}\ +\ \frac{E\,h}{\rho\,r_{\,0}^{\,2}}\;\eta_{\,x}\ =\ 0\,. \label{eq:j4}
\end{align}
On the other hand, using \eqref{eq:bc3} to the pressure/displacement relationship, \cite{Cascaval2012}:
\begin{equation}\label{eq:p2}
  \rho^{\,w}\,h\,\eta_{\,t\,t}\ =\ p\ -\ \frac{E\,h}{r_{\,0}^{\,2}}\;\eta\,,
\end{equation}
and substituting $p$ into \eqref{eq:j2} one obtains the nonlinear and dispersive system, \cite{Cascaval2012}:
\begin{align}
  & \eta_{\,t}\ +\ \eta_{\,x}\,u\ +\ \frac{1}{2}\;\eta\,u_{\,x}\ +\ \frac{1}{2}\;r_{\,0}\,u_{\,x}\ =\ 0\,, \label{eq:j5}\\
  & u_{\,t}\ +\ u\,u_{\,x}\ +\ \frac{E\,h}{\rho\,r_{\,0}^{\,2}}\;\eta_{\,x}\ +\ \frac{\rho^{\,w}\,h}{\rho}\;\eta_{\,x\,t\,t}\ =\ 0\,. \label{eq:j6}
\end{align}

The last \textsc{Boussinesq}-type system is unfortunately not very useful for practical problems due to its poor stability properties, \cite{BCS}. For this reason further investigation can lead to other {\em regularised} \textsc{Boussinesq} systems. For example, consider the non-dimensional variables:
\begin{equation}\label{eq:ndv1}
  \eta^{\,\ast}\ =\ \frac{\eta}{a}\,,\quad x^{\,\ast}\ =\ \frac{x}{\lambda}\,,\quad u^{\,\ast}\ =\ \frac{u}{c_{\,0}}\,,\quad t^{\,\ast}\ =\ \frac{t}{T}\,,
\end{equation}
where $a$ is a typical excitation of a pulse, $\lambda$ a typical wavelenght of a pulse, $T\ =\ 2\,a\,\lambda/r_{\,0}\,c_{\,0}$ the characteristic time scale, while
\begin{equation}\label{eq:ndv2}
  c_{\,0}\ =\ \frac{a}{r_{\,0}}\;\sqrt{\frac{2\,E\,h}{\rho\,r_{\,0}}}\,, 
\end{equation}
is a modified \textsc{Moens--Korteweg} characteristic speed, \cite{Korteweg1878, Moens1877}. Then, non-dimensionalisation and scaling of \eqref{eq:j5} -- \eqref{eq:j6} using \eqref{eq:ndv1} -- \eqref{eq:ndv2} leads to the non-dimensional system:
\begin{align}
  \eta^{\,\ast}_{\,t^{\,\ast}}\ +\ 2\,\varepsilon\,\eta^{\,\ast}_{\,x^{\,\ast}}\,u^{\,\ast}\ +\ \varepsilon\,\eta^{\,\ast}\,{u^{\,\ast}}_{\,x^{\,\ast}}\ +\ {u^{\,\ast}}_{\,x^{\,\ast}}\ =\ 0\,,\label{neweq1}\\
  u^{\,\ast}_{\,t^{\,\ast}}\ +\ \eta^{\,\ast}_{\,x^{\,\ast}}\ +\ 2\,\varepsilon\,u^{\,\ast}\,u^{\,\ast}_{\,x^{\,\ast}}\ +\ \delta^{\,2}\,\alpha\,\eta^{\,\ast}_{\,x^{\,\ast}\,t^{\,\ast}\,t^{\,\ast}}\ =\ 0\,, \label{neweq2}
\end{align}
where here $\varepsilon\ =\ a/r_{\,0}\,$, $\delta^{\,2}\ =\ r_{\,0}^{\,2}/\lambda^{\,2}\,$, $\varepsilon,\,\delta^{\,2}\ \ll\ 1$ and $\alpha\ =\ \rho^{\,w}\,h\,/\,2\,\rho\,r_{\,0}\,$. From \eqref{neweq1} -- \eqref{neweq2} we observe that $\eta^{\,\ast}_{\,t^{\,\ast}}\ =\ -\,u^\ast_{\,x^{\,\ast}}\ +\ \O\,(\varepsilon,\,\delta^{\,2})$ and thus $\eta^{\,\ast}_{\,x^{\,\ast}\,t^{\,\ast}\,t^{\,\ast}}\ =\ -\,u^{\,\ast}_{\,x^{\,\ast}\,x^{\,\ast}\,t^{\,\ast}}\ +\ \O\,(\varepsilon,\,\delta^{\,2})\,$. After substituting the last relationship into \eqref{neweq2} and omitting the high-order terms of $\O\,(\varepsilon\,\delta^{\,2},\,\delta^{\,4})$ we obtain the regularised system of \textsc{Boussinesq} equations
\begin{align}
  \eta^{\,\ast}_{\,t^{\,\ast}}\ +\ 2\,\varepsilon\,\eta^{\,\ast}_{\,x^{\,\ast}}\,u^{\,\ast}\ +\ \varepsilon\,\eta^{\,\ast}\,u^{\,\ast}_{\,x^{\,\ast}}\ +\ u^{\,\ast}_{\,x^{\,\ast}}\ =\ 0\,,\label{casc1} \\
  u^{\,\ast}_{\,t^{\,\ast}}\ +\ \eta^{\,\ast}_{\,x^{\,\ast}}\ +\ 2\,\varepsilon\,u^{\,\ast}\,u^{\,\ast}_{\,x^{\,\ast}}\ -\ \delta^{\,2}\,\alpha\,u^{\,\ast}_{\,x^{\,\ast}\,x^{\,\ast}\,t^{\,\ast}}\ =\ 0\,,\label{casc2}
\end{align}
which can be written in dimensional variables in the form:
\begin{align}
  \eta_{\,t}\ +\ \eta_{\,x}\,u\ +\ \frac{1}{2}\;(\eta\ +\ r_{\,0})\,u_{\,x}\ =\ 0\,,\label{vcasc1} \\
  u_{\,t}\ +\ u\,u_{\,x}\ +\ \frac{E\,h}{\rho\,r_{\,0}^{\,2}}\;\eta_{\,x}\ -\ \frac{\rho^{\,w}\,h}{2\,\rho\,E}\;u_{\,x\,x\,t}\ =\ 0\,. \label{vcasc2}
\end{align}

One can also derive one-way propagation model equations of KdV and BBM type by assuming that
\begin{equation}\label{bbmassumpt}
  u^{\,\ast}\ =\ \eta^{\,\ast}\ +\ \varepsilon\,F\ +\ \delta^{\,2}\,G\,,
\end{equation}
where $F\ =\ F\,(x,\,t)\,$, $G\ =\ G\,(x,\,t)$ are unknown functions. Substitution of the last equation into the system \eqref{casc1} -- \eqref{casc2} and equating coefficients of same order, \cite{Whitham1999}, leads to the KdV-type equation:
\begin{equation*}
  \eta^{\,\ast}_{\,t^{\,\ast}}\ +\ \frac{5}{2}\;\varepsilon\,\eta^{\,\ast}\,\eta^{\,\ast}_{\,x^{\,\ast}}\ +\ \eta^{\,\ast}_{\,x^{\,\ast}}\ +\ \delta^{\,2}\,\frac{\alpha}{2}\,\eta^{\,\ast}_{\,x^{\,\ast}\,x^{\,\ast}\,x^{\,\ast}}\ =\ 0\,,
\end{equation*}
and the BBM-type equation
\begin{equation*}
  \eta^{\,\ast}_{\,t^{\,\ast}}\ +\ \frac{5}{2}\;\varepsilon\,\eta^{\,\ast}\,\eta^{\,\ast}_{\,x^{\,\ast}}\ +\ \eta^{\,\ast}_{\,x^{\,\ast}}\ -\ \delta^{\,2}\,\frac{\alpha}{2}\;\eta^{\,\ast}_{\,x^{\,\ast}\,x^{\,\ast}\,t^{\,\ast}}\ =\ 0\,,
\end{equation*}
or in dimensional variables
\begin{equation*}
  \eta_{\,t}\ +\ \tilde{c}\,\eta_{\,x}\ +\ \frac{5}{2\,r_{\,0}}\;\tilde{c}\,\eta\,\eta_{\,x}\ +\ \frac{\rho^{\,w}\,h\,r_{\,0}}{4\,\rho}\,\tilde{c}\,\eta_{\,x\,x\,x}\ =\ 0\,,
\end{equation*}
and
\begin{equation}
  \eta_{\,t}\ +\ \tilde{c}\,\eta_{\,x}\ +\ \frac{5}{2\,r_{\,0}}\;\tilde{c}\,\eta\,\eta_{\,x}\ -\ \frac{\rho^{\,w}\,h\,r_{\,0}}{4\,\rho}\;\eta_{\,x\,x\,t}\ =\ 0\,,
\end{equation}
where $\tilde{c}\ =\ \sqrt{\dfrac{E\,h}{2\,\rho\,r_{\,0}}}$ is the \textsc{Moens--Korteweg} characteristic speed, \cite{Korteweg1878, Moens1877}.

Although equations \eqref{eq:j1} -- \eqref{eq:j2} (and the derived systems \eqref{eq:j3} -- \eqref{eq:j4} and \eqref{eq:j5} -- \eqref{eq:j6} appeared to be useful and quite accurate in the approximation of the blood flow in large arteries, \cite{Cascaval2012, Formaggia2003, Sherwin2003, Quarteroni2004}, their derivation is based on the assumption:
\begin{equation*}
  \int_{\,S} \hat{u}^{\,2}\; \ud \sigma\ =\ A\,u^{\,2}\,,
\end{equation*}
where in this notation $S$ is a cross-section of the vessel and $A$ its area, $\hat{u}\ =\ \hat{u}\,(x,\,r,\,t)$ is the horizontal velocity of the fluid measured at distance $r$ from the centre of the vessel. Moreover, the equations for the cross-sectional averaged pressure \eqref{eq:p1} and \eqref{eq:p2} are not asymptotic and therefore the derivations following \eqref{eq:j1} -- \eqref{eq:j2} is not clear that are valid by asymptotically means.

An asymptotic unidirectional model equation was derived in \cite{Cascaval2003}. This equation is a variable coefficient KdV-type equation and describes the unidirectional propagation of pulse waves in an elastic vessel of variable radius. Other model equations for the flow in cylindrical tubes of constant cross section in equilibrium has been derived in \cite{Ravindran1979}. These equations are asymptotic approximations of \textsc{Euler}'s equations but again written in their simplest form are difficult to use in practice since each equation contains temporal derivatives of both the excitation of tube's wall and velocity of the fluid. Although asymptotic unidirectional models exist in the literature, bidirectional models are of significant importance since they have the capacity to model wave reflections from the walls and boundaries of the fluid domain. Next section focuses on the asymptotic derivation of weakly nonlinear and weakly dispersive model equations that describe the two-way propagation of waves in elastic vessels of variable radius that are easy to use in practice.

\begin{remark}
Although the viscous nature of the fluid has been neglected in all these models, it can be recovered by adding an extra dissipative term. This has been considered in several works \cf \eg \cite{Sherwin2003, Formaggia2003, Quarteroni2004} and also in Section~\ref{sec:dissipative} bellow.
\end{remark}

%%% ------------------------------------------------------------------------ %%%

\section{New model equations}
\label{sec:deriv1}

In order to derive an asymptotic model we first write the \textsc{Euler} equations in a non-dimensional and scaled form using the following change of variables:
\begin{multline}\label{eq:ndv}
  \eta^{\,\ast}\ =\ \frac{\eta}{a}\,,\quad x^{\,\ast}\ =\ \frac{x}{\lambda}\,,\quad r^{\,\ast}\ =\ \frac{r}{R}\,,\quad t^{\,\ast}\ =\ \frac{t}{T}\,,\\ 
  u^{\,\ast}\ =\ \frac{1}{\varepsilon\,\tilde{c}}\;u\,,\quad v^{\,\ast}\ =\ \frac{1}{\varepsilon\,\delta\,\tilde{c}}\;v\,,\quad p^{\,\ast}\ =\ \frac{1}{\varepsilon\,\rho\,\tilde{c}^{\,2}}\;p\,,
\end{multline}
where $a$ is a typical excitation of the vessel wall, $\lambda$ a typical wavelenght of a pulse, $R$ is a vessel's typical radius, $T\ =\ \lambda\,/\,\tilde{c}$ the characteristic time scale, while $\tilde{c}\ =\ \sqrt{E\,h\,/\,2\,\rho\,R}$ is the \textsc{Moens--Korteweg} characteristic speed. It is noted that the external pressure is considered zero and is neglected. The parameters $\varepsilon$ and $\delta$ characterise the nonlinearity and the dispersion of the system:
\begin{equation*}
  \varepsilon\ =\ \frac{a}{R}\,, \qquad \delta\ =\ \frac{R}{\lambda}\,.
\end{equation*}
Usually, $\varepsilon$ and $\delta$ are very small. Specifically, we assume that $\varepsilon\ \ll\ 1\,$, $\delta^{\,2}\ \ll\ 1\,$, while the \textsc{Stokes}--\textsc{Ursell} number is of order 1\,: $\varepsilon\,/\,\delta^{\,2}\ =\ \O\,(1)\,$.

Omitting the $\ast$ from the notation below, the non-dimensional form of the \textsc{Euler} equations takes the form:
\begin{align}
  & u_{\,t}\ +\ \varepsilon\,u\,u_{\,x}\ +\ \varepsilon\,v\,u_{\,r}\ +\ p_{\,x}\ =\ 0\,, \label{eq:mom1} \\
  & \delta^{\,2}\,(v_{\,t}\ +\ \varepsilon\,u\,v_{\,x}\ +\ \varepsilon\,v\,v_{\,r})\ +\ p_{\,r}\ =\ 0\,, \qquad \mbox{ for } 0\ \leq\ r\ \leq\ r^{\,w}\doteq\ r_{\,0}\ +\ \varepsilon\,\eta \label{eq:mom2}\\
  & r\,u_{\,x}\ +\ (r\,v)_{\,r}\ =\ 0\,, \label{eq:mas} \\
  & \delta^{\,2}\,v_{\,x}\ =\ u_{\,r}\,, \label{eq:iro}
\end{align}
while the boundary and compatibility conditions are written as:
\begin{align}
  & v\,(x,\,r^{\,w},\,t)\ =\ \eta_{\,t}\,(x,\,t)\ +\ r^{\,w}_{\,x}\,u\,(x,\,r^{\,w},\,t)\,, \label{eq:dbc1} \\
  & p^{\,w}\,(x,\,t)\ \doteq\ p\,(x,\,r^{\,w},\,t)\ =\ \alpha\,\delta^{\,2}\,\eta_{\,t\,t}\,(x,\,t)\ +\ \beta\,(x)\,\eta\,(x,\,t)\,, \label{eq:dbc2} \\
  & v\,(x,\,0,\,t)\ =\ 0\,, \label{eq:dbc3}
\end{align}
where denoting
\begin{equation}\label{eq:parsv1}
  \bar{\alpha}\ =\ \frac{\rho^{\,w}\,h}{\rho}\ \mbox{ and }\ \bar{\beta}\,(x)\ =\ \frac{E\,h}{\rho\,r_{\,0}^{\,2}\,(x)}\,,
\end{equation}
then
\begin{equation}\label{eq:parsv2}
  \alpha\ =\ \frac{\bar{\alpha}}{R}\ \mbox{ and }\ \beta\,(x)\ =\ \frac{2\,R^{\,2}\,\rho}{E\,h}\;\bar{\beta}\,(x)\,.
\end{equation}
Equation \eqref{eq:iro} represents the irrotationality of the flow and it is equivalent with the assumption that the flow is potential, \cite{Landau1987}.

In the long wave approximation the horizontal velocity is a small perturbation of the velocity of the fluid at the vessel wall. This can be seen by integrating \eqref{eq:iro} from $r$ to $r^{\,w}$ and solving for $u\,$:
\begin{equation}\label{E7}
  u\,(x,\,r,\,t)\ =\ u\,(x,\,r^{\,w},\,t)\ -\ \delta^{\,2}\,\int_{\,r}^{\,r^{\,w}}\,v_{\,x}\,(x,\,s,\,t)\;\ud s\,.
\end{equation}
The last equation gives a low-order asymptotic relation for the horizontal velocity $u\,(x,\,r,\,t)\ =\ u\,(x,\,r^{\,w},\,t)\ +\ \O\,(\delta^{\,2})$ and shows that the velocity is almost uniform along the radius $r\,$. Moreover, differentiation of the last equation yields
\begin{equation}\label{E6}
  u_{\,r}\,(x,\,r,\,t)\ =\ \O\,(\delta^{\,2})\,.
\end{equation}
In the sequel we will denote $u^{\,w}\,(x,\,t)\ \doteq\ u\,(x,\,r^{\,w},\,t)\,$. Using this notation we observe that $u_{\,x}\,(x,\,r,\,t)\ =\ u_{\,x}\,(x,\,r^{\,w},\,t)\ +\ r^{\,w}_{\,x}\,u_{\,r}\,(x,\,r^{\,w},\,t)\ +\ \O\,(\delta^{\,2})$ and $u_{\,t}\,(x,\,r,\,t)\ =\ u_{\,t}\,(x,\,r^{\,w},\,t)\ +\ r^{\,w}_{\,t}\,u_{\,r}\,(x,\,r^{\,w},\,t)\ +\ \O\,(\delta^{\,2})\,$. But since $u_{\,r}\ =\ \O\,(\delta^{\,2})$ then we have that
\begin{align}
 u_{\,x}\,(x,\,r,\,t)\ &=\ u_{\,x}^{\,w}\,(x,\,t)\ +\ \O\,(\delta^{\,2})\,, \label{eq:deriv1} \\
 u_{\,t}\,(x,\,r,\,t)\ &=\ u_{\,t}^{\,w}\,(x,\,t)\ +\ \O\,(\delta^{\,2})\,. \label{eq:deriv2}
\end{align}
Since the radial accelerations are negligible we can simplify the momentum conservation Equation~\eqref{eq:mom1} into
\begin{equation}\label{eq:mom1b}
  u_{\,t}\,(x,\,r,\,t)\ +\ \varepsilon\,u\,(x,\,r,\,t)\,u_{\,x}\,(x,\,r,\,t)\ +\ p_{\,x}\,(x,\,r,\,t)\ =\ \O\,(\varepsilon\,\delta^{\,2})\,,
\end{equation}

We continue with the derivation of analogous asymptotic approximations for the radial velocity $v\,$. We consider the function
\begin{equation*}
  Q\,(x,\,r,\,t)\ =\ \frac{1}{r}\;\int_{\,0}^{\,r}\,s\,u\,(x,\,s,\,t)\;\ud s\,.
\end{equation*}
Using \eqref{E7} we obtain the relation
\begin{equation*}
  Q\,(x,\,r,\,t)\ =\ \frac{1}{r}\;\int_{\,0}^{\,r}\,s\,u^{\,w}\,(x,\,t)\;\ud s\ +\ \O\,(\delta^{\,2})\ =\ \frac{r}{2}\;u^{\,w}\,(x,\,t)\ +\ \O\,(\delta^{\,2})\,.
\end{equation*}

Integrating the mass conservation equation \eqref{eq:mas} we have
\begin{equation*}
 \int_{\,0}^{\,r}\,(s\,v)_{\,s}\;\ud s\ =\ -\,\int_{\,0}^{\,r}\,s\,u_{\,x}\;\ud s\,,  
\end{equation*}
from which, after integration by parts we obtain:
\begin{equation*}
  v\,(x,\,r,\,t)\ =\ -\,\frac{1}{r}\;\int_{\,0}^{\,r}\,s\,u_{\,x}\;\ud s\ =\ -\,Q_{\,x}\,(x,\,r,\,t)\,, 
\end{equation*}
or using asymptotic reasoning
\begin{equation}\label{E9}
  v\,(x,\,r,\,t)\ =\ -\,\frac{r}{2}\;u^{\,w}_{\,x}\,(x,\,t)\ +\ \O\,(\delta^{\,2})\,.
\end{equation}

The boundary condition \eqref{eq:dbc1} using \eqref{E9} is reduced to the equation
\begin{equation*}
  -\,\frac{r^{\,w}}{2}\;u^{\,w}_{\,x}\,(x,\,t)\ =\ \eta_{\,t}\,(x,\,t)\ +\ r^{\,w}_{\,x}\,u^{\,w}\,(x,\,t)\ +\ \O\,(\delta^{\,2})\,,
\end{equation*}
which gives an equation for $\eta$
\begin{equation}\label{eq:lowmas}
  \eta_{\,t}\,(x,\,t)\ +\ \frac{1}{2}\;r^{\,w}\,u^{\,w}_{\,x}\,(x,\,t)\ +\ r^{\,w}_{\,x}\,u^{\,w}\,(x,\,t)\ =\ \O\,(\delta^{\,2})\,.
\end{equation}
It is noted that this equation is a low order approximation of the mass conservation equation (similar to that of \cite{Cascaval2012}, see also Eq.~\eqref{casc1}).

In order to eliminate the pressure from \eqref{eq:mom1b} we need to derive high-order approximations of the non-hydrostatic pressure as follows. Equation~\eqref{E9} implies:
\begin{equation}\label{E10}
  v_{\,t}\,(x,\,r,\,t)\ =\ -\,\frac{r}{2}\;u^{\,w}_{\,x\,t}\,(x,\,t)\ +\ \O\,(\delta^{\,2})\,,
\end{equation}
while momentum equation \eqref{eq:mom2} implies that
\begin{equation}\label{Estar}
  \delta^{\,2}\,v_{\,t}\ +\ p_{\,r}\ =\ \O\,(\varepsilon\,\delta^{\,2})\,.
\end{equation}
Substituting \eqref{E10} into \eqref{Estar} we have the expression for the pressure:
\begin{equation*}
  p_{\,r}\,(x,\,r,\,t)\ =\ \delta^{\,2}\;\frac{r}{2}\;u^{\,w}_{\,x\,t}\,(x,\,t)\ +\ \O\,(\varepsilon\,\delta^{\,2},\,\delta^{\,4})\,,
\end{equation*}
which after integration leads to the formula:
\begin{equation}\label{E11}
  p\,(x,\,r,\,t)\ =\ p^{\,w}\,(x,\,t)\ -\ \delta^{\,2}\,u^{\,w}_{\,x\,t}\,(x,\,t)\,\frac{(r^{\,w})^{\,2}\ -\ r^{\,2}}{4}\ +\ \O\,(\varepsilon\,\delta^{\,2},\,\delta^{\,4})\,.
\end{equation}
Differentiation gives
\begin{equation}\label{E13}
  p_{\,x}\,(x,\,r,\,t)\ =\ p^{\,w}_{\,x}\,(x,\,t)\ -\ \delta^{\,2}\,u^{\,w}_{\,x\,t}\,(x,\,t)\;\frac{(r^{\,w})^{\,2}\ -\ r^{\,2}}{4}\ -\ \delta^{\,2}\;\frac{r^{\,w}\,r^{\,w}_{\,x}}{2}\;u^{\,w}_{\,x\,x\,t}\,(x,\,t)\ +\ \O\,(\varepsilon\,\delta^{\,2},\,\delta^{\,4})\,.
\end{equation}

Finally, we derive high-order asymptotic expansions for the velocities and eventually for the mass and momentum conservation equations. Starting from the horizontal velocity $u\,(x,\,r,\,t)$ we combine \eqref{E7} and \eqref{E9} to obtain
\begin{equation*}
  u\,(x,\,r,\,t)\ =\ u^{\,w}\,(x,\,t)\ +\ \delta^{\,2}\,\int_{\,r}^{\,r^{\,w}}\,\frac{s}{2}\;u^{\,w}_{\,x\,x}\,(x,\,t)\;\ud s\ +\ \O\,(\delta^{\,4})\,, 
\end{equation*}
which leads to the high-order approximation of the horizontal velocity:
\begin{equation}\label{E12}
  u\,(x,\,r,\,t)\ =\ u^{\,w}\,(x,\,t)\ +\ \delta^{\,2}\,u^{\,w}_{\,x\,x}\,(x,\,t)\;\frac{(r^{\,w})^{\,2}\ -\ r^{\,2}}{4}\ +\ \O\,(\delta^{\,4})\,,
\end{equation}
and of the time derivative of the horizontal velocity (since $r^{\,w}\ =\ r_{\,0}\ +\ \varepsilon\,\eta$):
\begin{equation}\label{E12t}
  u_{\,t}\,(x,\,r,\,t)\ =\ u^{\,w}_{\,t}\,(x,\,t)\ +\ \delta^{\,2}\,u^{\,w}_{\,x\,x\,t}\,(x,\,t)\;\frac{(r^{\,w})^{\,2}\ -\ r^{\,2}}{4}\ +\ \O\,(\delta^{\,4},\,\epsilon\,\delta^{\,2})\,, 
\end{equation}
In order to compute high-order approximation to the radial velocity we integrate \eqref{eq:mas} from $0$ to $r^{\,w}$ we get
\begin{equation*}
  r^{\,w}\,v\,(x,\,r^{\,w},\,t)\ =\ -\,\int_{\,0}^{\,r^{\,w}}\,s\,u_{\,x}\,(x,\,s,\,t)\;\ud s\,.
\end{equation*}

The last equation with the help of the boundary condition \eqref{eq:dbc1} leads to
\begin{align*}
  r^{\,w}\,(\,\eta_{\,t}\ +\ r_{\,x}^{\,w}\,u^{\,w}\,)\ &=\ -\,\int_{\,0}^{\,r^{\,w}}\,s\,u_{\,x}\;\ud s\,,\\
  \mbox{(from \eqref{E12})}\ &=\ -\,\int_{\,0}^{\,r^{\,w}}\,s\,\left[\,u^{\,w}_{\,x}\ +\ \delta^{\,2}\,\left(u^{\,w}_{\,x\,x}\;\frac{(r^{\,w})^{\,2}\ -\ s^{\,2}}{4}\right)_{\,x}\,\right]\;\ud s\ +\ \O\,(\delta^{\,4})\,, \\
  &=\ -\,\frac{1}{2}\;(r^{\,w})^{\,2}\,u^{\,w}_{\,x}\ -\ \frac{\delta^{\,2}}{16}\;(r^{\,w})^{\,4}\,u^{\,w}_{\,x\,x\,x}\ -\ \frac{\delta^{\,2}}{4}\;(r^{\,w})^{\,3}\,r^{\,w}_{\,x}\,u^{\,w}_{\,x\,x}\ +\ \O\,(\delta^{\,4})\,.
\end{align*}
Solving the last relationship for $\eta_{\,t}$ we obtain the approximate mass conservation equation:
\begin{equation}\label{eq:mass}
  \eta_{\,t}\ +\ \frac{1}{2}\;r^{\,w}\,u^{\,w}_{\,x}\ +\ r^{\,w}_{\,x}\,u^{\,w}\ +\ \frac{\delta^{\,2}}{4}\;(r^{\,w})^{\,2}\,r^{\,w}_{\,x}\,u^{\,w}_{\,x\,x}\ +\ \frac{\delta^{\,2}}{16}\;(r^{\,w})^{\,3}\,u^{\,w}_{\,x\,x\,x}\ =\ \O\,(\delta^{\,4})\,. 
\end{equation}

Furthermore, substituting \eqref{E13} into \eqref{eq:mom1b}, and using \eqref{E12t}, \eqref{eq:deriv1}, \eqref{eq:deriv2} and the boundary condition \eqref{eq:dbc2}, yields the approximate momentum conservation equation,
\begin{equation}\label{eq:mom1c}
  u^{\,w}_{\,t}\ +\ \varepsilon\,u^{\,w}\,u^{\,w}_{\,x}\ +\ [\,\beta\,(x)\,\eta\,]_{\,x}\ -\ \delta^{\,2}\;\frac{r^{\,w}\,r^{\,w}_{\,x}}{2}\;u^{\,w}_{\,x\,t}\ +\ \alpha\,\delta^{\,2}\,\eta_{\,x\,t\,t}\ =\ \O(\varepsilon\delta^{\,2})\,.
\end{equation}

We can further use the low order approximation of the mass conservation equation \eqref{eq:lowmas} to improve the term $\eta_{\,x\,t\,t}$ in \eqref{eq:mom1c}. Specifically, differentiating \eqref{eq:lowmas} we obtain $\eta_{\,x\,t\,t}\ =\ -\,\Bigl(\frac{1}{2}\;r^{\,w}\,u^{\,w}_{\,x}\ +\ r^{\,w}_{\,x}\,u^{\,w}\Bigr)_{\,x\,t}\ +\ \O\,(\delta^{\,2})\,$. Substituting the last equation into \eqref{eq:mom1c} we obtain the BBM-type equation:
\begin{equation}\label{eq:momentum}
  u^{\,w}_{\,t}\ +\ [\,\beta\,(x)\,\eta\,]_{\,x}\ +\ \varepsilon\,u^{\,w}\,u^{\,w}_{\,x}\ -\ \alpha\,\delta^{\,2}\,(r^{\,w}_{\,x}\,u^{\,w})_{\,x\,t}\ -\ \frac{\delta^{\,2}\,r^{\,w}\,r^{\,w}_{\,x}}{2}\;u^{\,w}_{\,x\,t}\ -\ \frac{\alpha\,\delta^{\,2}}{2}\;\left(r^{\,w}\,u^{\,w}_{\,x}\right)_{\,x\,t}\ =\ \O\,(\epsilon\,\delta^{\,2},\,\delta^{\,4})\,.
\end{equation}

Finally, we recollect the high-order approximations of mass and momentum equations. Since $r^{\,w}\,(x,\,t)\ =\ r_{\,0}\,(x)\ +\ \varepsilon\,\eta\,(x,\,t)$ we can further simplify the terms multiplied by $\delta^{\,2}$ or $\varepsilon$ in \eqref{eq:mass} and \eqref{eq:momentum} and obtain:
\begin{align}
  & \eta_{\,t}\ +\ \frac{1}{2}\;(r_{\,0}\ +\ \varepsilon\,\eta)\,u^{\,w}_{\,x}\ +\ ({r_{\,0}}_{\,x}\ +\ \varepsilon\,\eta_{\,x})\,u^{\,w}\ +\ \frac{\delta^{\,2}}{4}\;{r_{\,0}}^{\,2}\,r_{\,0\,x}\,u^{\,w}_{\,x\,x}\ +\ \frac{\delta^{\,2}}{16}\;{r_{\,0}}^{\,3}\,u^{\,w}_{\,x\,x\,x}\ =\ \O\,(\varepsilon\,\delta^{\,2},\,\delta^{\,4})\,, \label{mass} \\
  & u^{\,w}_{\,t}\ +\ [\,\beta\,(x)\,\eta\,]_{\,x}\ +\ \varepsilon\,u^{\,w}\,u^{\,w}_{\,x}\ -\ \alpha\,\delta^{\,2}\,({r_{\,0}}_{\,x}\,u^{\,w}_{\,t})_{\,x}\ -\ \frac{\delta^{\,2}\,r_{\,0}\,{r_{\,0}}_{\,x}}{2}\;u^{\,w}_{\,x\,t}\ -\ \frac{\alpha\,\delta^{\,2}}{2}\;\left(r_{\,0}\,u^{\,w}_{\,x\,t}\right)_{\,x}\ =\ \O\,(\varepsilon\,\delta^{\,2},\,\delta^{\,4})\,. \label{momentum}
\end{align}
Discarding the high-order terms system \eqref{mass} -- \eqref{momentum} can be written in dimensional form as  
\begin{align}
  & \eta_{\,t}\ +\ \frac{1}{2}\;(r_{\,0}\ +\ \eta)\,u^{\,w}_{\,x}\ +\ ({r_{\,0}}_{\,x}\ +\ \eta_{\,x})\,u^{\,w}\ +\ \frac{{r_{\,0}}^{\,2}\;{r_{\,0}}_{\,x}}{4} u^{\,w}_{\,x\,x}\ +\ \frac{{r_{\,0}}^{\,3}}{16}\;u^{\,w}_{\,x\,x\,x}\ =\ 0\,, \label{dmass} \\
  & u^{\,w}_{\,t}\ +\ \left[\,\frac{E\,h}{r_{\,0}^{\,2}\,\rho}\;\eta\,\right]_{\,x}\ +\ u^{\,w}\,u^{\,w}_{\,x}\ -\ \frac{\rho^{\,w}\,h}{\rho}\;({r_{\,0}}_{\,x}\,u^{\,w}_{\,t})_{\,x}\ -\ \frac{r_{\,0}\,{r_{\,0}}_{\,x}}{2}\;u^{\,w}_{\,x\,t}\ -\ \frac{\rho^{\,w}\,h}{2\,\rho}\left(r_{\,0}\,u^{\,w}_{\,x\,t}\right)_{\,x}\ =\ 0\,. \label{dmomentum}
\end{align}
We mention that the last system differs from the non-asymptotic model derived in \cite{Cascaval2012} and it has similarities with analogous asymptotic models derived in the context of water waves, \cite{Nwogu1993}.

\begin{remark}
The viscoelastic nature of a vessel can be taken into account by considering in the left hand side of the boundary condition \eqref{eq:bc3} an additional term proportional to the displacement velocity $\gamma\,\eta_{\,t}$ with $\gamma$ a positive damping parameter. This is the case of a simple \textsc{Voigt/Kelvin} model, \cf \eg \cite{Fung1993}. The addition of this term will result in additional dissipative terms in the momentum equation that we will study in detail in the future.
\end{remark}

%%% ------------------------------------------------------------------------ %%%

\section{Further developments}
\label{sec:deriv2}

In the previous section we approximated the velocity $u\,(x,\,r,\,t)$ by the velocity $u\,(x,\,r^{\,w},\,t)$ eliminating in this way the radial component of the velocity from the equations and reducing the dimensionality of the system to one. Using similar arguments to \cite{BS, BCS} and approximating the velocity at any radius $r$ we can derive a whole new class of systems.

Using the high-order approximation \eqref{E12} to the horizontal velocity we observe that
\begin{equation*}
  u^{\,w}\,(x,\,t)\ =\ u\,(x,\,r,\,t)\ -\ \delta^{\,2}\,u_{\,x\,x}\,(x,\,r,\,t)\;\frac{(r^{\,w})^{\,2}\ -\ r^{\,2}}{4}\ +\ \O\,(\delta^{\,4})\,.
\end{equation*}
Taking $r\ =\ \theta\,r^{\,w}$ with $0\ \leq\ \theta\ \leq\ 1$ and denoting $u\,(x,\,\theta\,r^{\,w},\,t)$ by $u^{\,\theta}\,(x,\,t)$ we have,
\begin{equation*}
  u^{\,w}\ =\ u^{\,\theta}\ -\ \delta^{\,2}\,u^{\,\theta}_{\,x\,x}\;\frac{(1\ -\ \theta^{\,2})(r^{\,w})^{\,2}}{4}\ +\ \O\,(\delta^{\,4})\,.
\end{equation*}
Using the previous expression to low-order terms and taking into account $u^{\,w}\ =\ u^{\,\theta}\ +\ \O\,(\delta^{\,2})$ for the high-order terms in Equations~\eqref{mass} -- \eqref{momentum} we have:
\begin{multline}\label{masst}
  \eta_{\,t}\ +\ \frac{1}{2}\;(r_{\,0}\ +\ \varepsilon\,\eta)\,u^{\,\theta}_{\,x}\ +\ ({r_{\,0}}_{\,x}\ +\ \varepsilon\,\eta_{\,x})\,u^{\,\theta}\\ 
  -\ \delta^{\,2}\;\frac{{r_{\,0}}^{\,2}\,{r_{\,0}}_{\,x}\,(1\ -\ 2\,\theta^{\,2})}{4}\;u^{\,\theta}_{\,x\,x}\ +\ \delta^{\,2}\;\frac{{r_{\,0}}^{\,3}\,(2\,\theta^{\,2}\ -\ 1)}{16}\;u^{\,\theta}_{\,x\,x\,x}\ =\ \O(\varepsilon\,\delta^{\,2},\,\delta^{\,4})\,,
\end{multline}
\begin{multline}\label{momentumt}
  (1\ -\ \alpha\,\delta^{\,2}\,{r_{\,0}}_{\,x\,x})\,u^{\,\theta}_{\,t}\ +\ [\,\beta\,\eta\,]_{\,x}\ +\ \varepsilon\,u^{\,\theta}\,u^{\,\theta}_{\,x}\\ 
  -\ \delta^{\,2}\,{r_{\,0}}_{\,x}\;\frac{3\,\alpha\ +\ r_{\,0}}{2}\;u^{\,\theta}_{\,x\,t}\ -\ \delta^{\,2}\;\frac{[\,2\,\alpha\ +\ (1\ -\ \theta^{\,2})\,r_{\,0}\,]\,r_{\,0}}{4}\;u^{\,\theta}_{\,x\,x\,t}\ =\ \O\,(\varepsilon\,\delta^{\,2},\,\delta^{\,4})\,.
\end{multline}
We can further use the low-order terms:
\begin{align}
  & \eta_{\,t}\ =\ -\,\frac{1}{2}\;r_{\,0}\,u^{\,\theta}_{\,x}\ -\ {r_{\,0}}_{\,x}\,u^{\,\theta}\ +\ \O\,(\varepsilon,\,\delta^{\,2})\,,\label{bbm1}\\
  & u^{\,\theta}_{\,t}\ =\ -\,[\,\beta\,\eta\,]_{\,x}\ +\ \O\,(\varepsilon,\,\delta^{\,2})\,. \label{bbm2}
\end{align}
The BBM-type relation \eqref{bbm1} implies that
\begin{equation}\label{bbm1a}
  r_{\,0}\,u^{\,\theta}_{\,x\,x\,x}\ =\ -2\,\eta_{\,x\,x\,t}\ -\ 5\,{r_{\,0}}_{\,x\,x}\,u^{\,\theta}_{\,x}\ -\ 4\,{r_{\,0}}_{\,x}\,u^{\,\theta}_{\,x\,x}\ -\ 2\,{r_{\,0}}_{\,x\,x\,x}\,u^{\,\theta}\ +\ \O(\varepsilon,\,\delta^{\,2})\,,
\end{equation}
while the BBM-type relation \eqref{bbm2} implies that
\begin{equation}\label{bbm2a}
  u^{\,\theta}_{\,x\,x\,t}\ =\ -\,[\,\beta\,\eta\,]_{\,x\,x\,x}\ +\ \O(\varepsilon,\,\delta^{\,2})\,.
\end{equation}
Writing $u^{\,\theta}_{\,x\,x\,x}\ =\ \nu\,u^{\,\theta}_{\,x\,x\,x}\ +\ (1\ -\ \nu)\,u^{\,\theta}_{\,x\,x\,x}$ and $u^{\,\theta}_{\,x\,x\,t}\ =\ \mu\,u^{\,\theta}_{\,x\,x\,t}\ +\ (1\ -\ \mu)\,u^{\,\theta}_{\,x\,x\,t}$ for $\mu,\,\nu\ \in\ \mathds{R}\,$, and using \eqref{bbm1a} and \eqref{bbm2a} into the last terms of \eqref{masst} and \eqref{momentumt} respectively we get the system:
\begin{align}
  & \eta_{\,t}\ +\ \frac{1}{2}\;(r_{\,0}\ +\ \varepsilon\,\eta)\,u^{\,\theta}_{\,x}\ +\ ({r_{\,0}}_{\,x}\ +\ \varepsilon\,\eta_{\,x})\,u^{\,\theta}\ -\ \delta^{\,2}\,\mathcal{A}\,(x)\,u^{\,\theta}_{\,x\,x}\ +\ \delta^{\,2}\,\mathcal{B}\,(x)\,u^{\,\theta}_{\,x\,x\,x}\label{massg} \\
  &\quad -\ \delta^{\,2}\,\mathcal{C}\,(x)\,(5\,{r_{\,0}}_{\,x\,x}\,u^{\,\theta}_{\,x}\ +\ 4\,{r_{\,0}}_{\,x}\,u^{\,\theta}_{\,x\,x}\ +\ 2\,{r_{\,0}}_{\,x\,x\,x}\,u^{\,\theta}\ +\ 2\,\eta_{\,x\,x\,t})\ =\ \O\,(\varepsilon\,\delta^{\,2},\,\delta^{\,4})\,, \nonumber \\
  & (1\ -\ \alpha\,\delta^{\,2}\,{r_{\,0}}_{\,x\,x})\,u^{\,\theta}_{\,t}\ +\ [\,\beta\,\eta\,]_{\,x}\ +\ \varepsilon\,u^{\,\theta}\,u^{\,\theta}_{\,x}\ -\ \delta^{\,2}\,\mathcal{D}\,(x)\,u^{\,\theta}_{\,x\,t}\ + \nonumber\\ 
  & \qquad \delta^{\,2}\,\mathcal{E}\,(x)\,[\,\beta\,\eta\,]_{\,x\,x\,x}\ -\ \delta^{\,2}\,\mathcal{F}\,(x)\,u^{\,\theta}_{\,x\,x\,t}\ =\ \O\,(\varepsilon\,\delta^{\,2},\,\delta^{\,4})\,, \label{momentumg}
\end{align}
where
\begin{align}
  & \mathcal{A}\,(x)\ =\ \frac{{r_{\,0}}^{\,2}\,{r_{\,0}}_{\,x}\,(1\ -\ 2\,\theta^{\,2})}{4}\,,\quad \mathcal{B}\,(x)\ =\ \frac{{r_{\,0}}^{\,3}\,(2\,\theta^{\,2}\ -\ 1)\,\nu}{16}\,, \label{pars0} \\ 
  & \mathcal{C}\,(x)\ =\ \frac{{r_{\,0}}^{\,2}\,(2\,\theta^{\,2}\ -\ 1)\,(1\ -\ \nu)}{16}\,,\quad \mathcal{D}\,(x)\ =\ \frac{{r_{\,0}}_{\,x}\,(3\,\alpha\ +\ r_{\,0})}{2}\,, \label{pars1} \\ 
  & \mathcal{E}\,(x)\ =\ \frac{[\,2\,\alpha\ +\ (1\ -\ \theta^{\,2})\,r_{\,0}\,]\,r_{\,0}\,\mu}{4}\,, \quad \mathcal{F}\,(x)\ =\ \frac{[\,2\,\alpha\ +\ (1\ -\ \theta^{\,2})\,r_{\,0}\,]\,r_{\,0}\,(1\ -\ \mu)}{4}\,. \label{pars2}
\end{align}
In dimensional variables the Equations \eqref{massg} -- \eqref{momentumg} take the form:
\begin{align}
  & \eta_{\,t}\ +\ \frac{1}{2}\;(r_{\,0}\ +\ \eta)\,u^{\,\theta}_{\,x}\ +\ ({r_{\,0}}_{\,x}\ +\ \eta_{\,x})\,u^{\,\theta}\ -\ \bar{\mathcal{A}}\,(x)\,u^{\,\theta}_{\,x\,x}\ +\ \bar{\mathcal{B}}\,(x)\,u^{\,\theta}_{\,x\,x\,x}\ -\ \nonumber \\
  &\qquad -\ \bar{\mathcal{C}}\,(x)\,(5\,{r_{\,0}}_{\,x\,x}\,u^{\,\theta}_{\,x}\ +\ 4\,{r_{\,0}}_{\,x}\,u^{\,\theta}_{\,x\,x}\ +\ 2\,{r_{\,0}}_{\,x\,x\,x}\,u^{\,\theta}\ +\ 2\,\eta_{\,x\,x\,t})\ =\ 0\,, \label{massgd} \\
  & (1\ -\ \bar{\alpha}\,{r_{\,0}}_{\,x\,x})\,u^{\,\theta}_{\,t}\ +\ [\,\bar{\beta}\,(x)\,\eta\,]_{\,x}\ +\ u^{\,\theta}\,u^{\,\theta}_{\,x}\ -\ \bar{\mathcal{D}}\,(x)\,u^{\,\theta}_{\,x\,t}\ +\nonumber \\
  & \qquad \bar{\mathcal{E}}\,(x)\,[\,\bar{\beta}\,(x)\,\eta\,]_{\,x\,x\,x}\ -\ \bar{\mathcal{F}}\,(x)\,u^{\,\theta}_{\,x\,x\,t}\ =\ 0\,, \label{momentumgd}
\end{align}
where
\begin{align}
  & \bar{\mathcal{A}}\,(x)\ =\ \frac{{r_{\,0}}^{\,2}\,{r_{\,0}}_{\,x}\,(1\ -\ 2\,\theta^{\,2})}{4}\,,\quad \bar{\mathcal{B}}\,(x)\ =\ \frac{{r_{\,0}}^{\,3}\,(2\,\theta^{\,2}\ -\ 1)\,\nu}{16}\,, \label{pars0d}\\
  & \bar{\mathcal{C}}\,(x)\ =\ \frac{{r_{\,0}}^{\,2}\,(2\,\theta^{\,2}\ -\ 1)\,(1\ -\ \nu)}{16}\,, \quad \bar{\mathcal{D}}\,(x)\ =\ \frac{{r_{\,0}}_{\,x}\,(3\,\bar{\alpha}\ +\ r_{\,0})}{2}\,,\label{pars1d} \\ 
  & \bar{\mathcal{E}}\,(x)\ =\ \frac{[\,2\,\bar{\alpha}\ +\ (1\ -\ \theta^{\,2})r_{\,0}\,]\,r_{\,0}\,\mu}{4}\,, \quad \bar{\mathcal{F}}\,(x)\ =\ \frac{[\,2\,\bar{\alpha}\ +\ (1\ -\ \theta^{\,2})\,r_{\,0}\,]\,r_{\,0}\,(1\ -\ \mu)}{4}\,. \label{pars2d}
\end{align}

\begin{remark}
The simplest of all these systems can be obtained by choosing $\theta^{\,2}\ =\ 1/2$ and $\mu\ =\ \nu\ =\ 0\,$, which is:
\begin{align}
  & \eta_{\,t}\ +\ \frac{1}{2}\;(r_{\,0}\ +\ \varepsilon\,\eta)\,u^{\,\theta}_{\,x}\ +\ ({r_{\,0}}_{\,x}\ +\ \varepsilon\,\eta_{\,x})\,u^{\,\theta}\ =\ \O\,(\varepsilon\,\delta^{\,2},\,\delta^{\,4})\,, \label{massp} \\
  & (1\ -\ \alpha\,\delta^{\,2}\,{r_{\,0}}_{\,x\,x})\,u^{\,\theta}_{\,t}\ +\ [\,\beta\,\eta\,]_{\,x}\ +\ \varepsilon\,u^{\,\theta}\,u^{\,\theta}_{\,x}\nonumber \\
  & \qquad -\ \delta^{\,2}\;\frac{(3\,\alpha\ +\ r_{\,0})\,{r_{\,0}}_{\,x}}{2}\;u^{\,\theta}_{\,x\,t}\ -\ \delta^{\,2}\;\frac{(4\,\alpha\ +\ r_{\,0})\,r_{\,0}}{8}\;u^{\,\theta}_{\,x\,x\,t}\ =\ \O(\varepsilon\,\delta^{\,2},\,\delta^{\,4})\,. \label{momentump}
\end{align}
This system has many similarities with \textsc{Peregrine}'s system of water wave theory, \cite{Peregrine1967}. The rest of the systems \eqref{massg} -- \eqref{momentumg} are relevant to the \textsc{Boussinesq} systems derived in the context of water waves in \cite{BCS, Mitsotakis2007}. In dimensional and unscaled variables, and discarding the high-order terms the system can be written as:
\begin{align}
  & \eta_{\,t}\ +\ \frac{1}{2}\;(r_{\,0}\ +\ \eta)\,u^{\,\theta}_{\,x}\ +\ ({r_{\,0}}_{\,x}\ +\ \eta_{\,x})\,u^{\,\theta}\ =\ 0\,, \label{masspd} \\
  & (1\ -\ \bar{\alpha}\,{r_{\,0}}_{\,x\,x})\,u^{\,\theta}_{\,t}\ +\ [\,\bar{\beta}\,(x)\,\eta\,]_{\,x}\ +\ u^{\,\theta}\,u^{\,\theta}_{\,x} \nonumber \\ 
  &\qquad -\ \frac{(3\,\bar{\alpha}\ +\ r_{\,0})\,{r_{\,0}}_{\,x}}{2}\;u^{\,\theta}_{\,x\,t}\ -\ \frac{(4\,\bar{\alpha}\ +\ r_{\,0})\,r_{\,0}}{8}\;u^{\,\theta}_{\,x\,x\,t}\ =\ 0\,, \label{momentumpd}
\end{align}
where $\alpha\,$, $\bar{\alpha}\,$, $\beta$ and $\bar{\beta}$ are defined in \eqref{eq:parsv1} and \eqref{eq:parsv2}.
\end{remark}

\begin{remark}
Further alternatives can be derived by replacing the terms (or some of the terms) $\delta^{\,2}\,u^{\,\theta}_{\,t}$ by $-\,\delta^{\,2}\,[\,\beta\,(x)\,\eta\,]_{\,x}$ due to the low order approximation $u^{\,\theta}_{\,t}\ =\ -\,[\,\beta\,(x)\,\eta\,]_{\,x}\ +\ \O(\varepsilon,\,\delta^{\,2})\,$. For example, system \eqref{masspd} -- \eqref{momentumpd} can be written in the alternative form:
\begin{align}
  & \eta_{\,t}\ +\ \frac{1}{2}\;(r_{\,0}\ +\ \eta)\,u^{\,\theta}_{\,x}\ +\ ({r_{\,0}}_{\,x}\ +\ \eta_{\,x})\,u^{\,\theta}\ =\ 0\,, \label{amasspd} \\
  & (1\ -\ \bar{\alpha}\,{r_{\,0}}_{\,x\,x})\,u^{\,\theta}_{\,t}\ +\ [\,\bar{\beta}\,(x)\,\eta\,]_{\,x}\ +\ u^{\,\theta}\,u^{\,\theta}_{\,x}\ + \nonumber \\
  & \qquad \frac{(3\,\bar{\alpha}\ +\ r_{\,0})\,{r_{\,0}}_{\,x}}{2}\;[\,\bar{\beta}\,(x)\,\eta\,]_{\,x\,x}\ -\ \frac{(4\,\bar{\alpha}\ +\ r_{\,0})\,r_{\,0}}{8}\;u^{\,\theta}_{\,x\,x\,t}\ =\ 0\,. \label{amomentumpd}
\end{align}
The last system appears to be more convenient for practical use for reasons explained in the next section. We will also refer to this system as the {\em classical} \textsc{Boussinesq} system due to its structure.
\end{remark}

\begin{remark}
Dropping the high-order terms of $\O\,(\delta^{\,2})\,$, the new asymptotic models are reduced to the commonly used hyperbolic one-dimensional model equations, \cite{Sherwin2003, Formaggia2003}, but now we can see that dissipative effects due to vessel's variations can be lost.
\end{remark}

%%% ------------------------------------------------------------------------ %%%

\section{Dissipative Boussinesq systems}
\label{sec:dissipative}

The motivation for the derivation of the previously mentioned reduced model equations is the accurate description of the blood flow in arteries. Although blood in large arteries can be considered as a \textsc{Newtonian} fluid, in order to simulate the blood flow more accurately one should take into consideration the viscous nature of the flow (it is known that blood can become almost two times more viscous than water). The governing equations for viscous flow are known to as the \textsc{Navier--Stokes} equations. Assuming once again cylindrical symmetry of the vessel and using the same change of variables as in \eqref{eq:ndv} we can write the \textsc{Navier--Stokes} equations in the following non-dimensional and scaled form:
\begin{align}
  & u_{\,t}\ +\ \varepsilon\,u\,u_{\,x}\ +\ \varepsilon\,v\,u_{\,r}\ +\ p_{\,x}\ =\ \frac{1}{\delta^{\,2}}\;\frac{1}{Re}\;\biggl[\,\frac{1}{r}\;(r\,u_{\,r})_{\,r}\ +\ \delta^{\,2}\,u_{\,x\,x}\,\biggr]\,, \label{eq:vmom1} \\
  & \delta^{\,2}\,(v_{\,t}\ +\ \varepsilon\,u\,v_{\,x}\ +\ \varepsilon\,v\,v_{\,r})\ +\ p_{\,r}\ =\ \frac{1}{Re}\;\biggl[\,\frac{1}{r}\;(r\,v_{\,r})_{\,r}\ -\ \frac{v}{r^{\,2}}\ +\ \delta^{\,2}\,v_{\,x\,x}\,\biggr]\,, \ \mbox{ for } 0\ \leq\ r\ \leq\ r^{\,w}\,, \label{eq:vmom2}\\
  & r\,u_{\,x}\ +\ (r\,v)_{\,r}\ =\ 0\,, \label{eq:vmas}
\end{align}
where if $\kappa$ denotes the kinematic viscosity coefficient then $Re$ is the \textsc{Reynolds} number defined as
\begin{equation*}
  Re\ =\ \frac{\lambda\,\tilde{c}}{\kappa}\,.
\end{equation*}
Deriving asymptotic approximations to the \textsc{Navier--Stokes} equations is in general very difficult and usually results in very complicated systems that are difficult to use in practice, \cite{LeMeur2015, Dutykh2014f}. In order to derive dissipative \textsc{Boussinesq} systems from the \textsc{Navier--Stokes} equations without using the boundary layer theory and simplifying the analysis we assume that the viscosity of the fluid is not significant, \ie the kinematic viscosity of the fluid is very small and that $1/Re\ =\ \O\,(\varepsilon\,\delta^{\,2})\,$. This assumption is realistic for blood flow problems since it is known that $Re\ \approx\ 10^{\,3}$ for the blood in aorta for example. Then following \cite{Smith2002, Dutykh2007} we discard terms of order $\O\,(\varepsilon\,\delta^{\,2})$ and consider the reduced equations:
\begin{align}
  & u_{\,t}\ +\ \varepsilon\,u\,u_{\,x}\ +\ \varepsilon\,v\,u_{\,r}\ +\ p_{\,x}\ =\ \frac{1}{\delta^{\,2}}\;\frac{1}{Re}\;\biggl[\,\frac{1}{r}\;(r\,u_{\,r})_{\,r}\,\biggr]\,, \label{eq:rvmom1} \\
  & \delta^{\,2}\,(v_{\,t}\ +\ \varepsilon\,u\,v_{\,x}\ +\ \varepsilon\,v\,v_{\,r})\ +\ p_{\,r}\ =\ 0\,, \ \mbox{ for } 0\ \leq\ r\ \leq\ r^{\,w}\,, \label{eq:rvmom2} \\
  & r\,u_{\,x}\ +\ (r\,v)_{\,r}\ =\ 0\,, \label{eq:rvmas}
\end{align}
To further simplify these equations we employ some classical heuristic arguments that have been proposed in \cite{VandeVosse2011, Smith2002, Formaggia2003, Fung1997a} to derive analogous dissipative models for the study of blood flow problems. First, we assume that the flow is smooth (laminar) and that the velocity profile is given by the formula
\begin{equation*}
  u\,(x,\,r,\,t)\ =\ u^{\,w}\,(x,\,t)\,\phi\,(x,\,r,\,t)\ +\ \O\,(\delta^{\,2})\,,
\end{equation*}
with $\phi\,(x,\,r,\,t)\ =\ 2\,\frac{{r^{\,w}}^{\,2}\ -\ r^{\,2}}{{r_{\,0}}^{\,2}}\,$. Then we observe that the viscous term is reduced to the simple
\begin{equation*}
  \frac{1}{r}\;(r\,u_{\,r})_{\,r}\ =\ -\,8\;\frac{u^{\,w}}{{r_{\,0}}^{\,2}}\,,
\end{equation*}
and thus we consider the reduced dissipative system:
\begin{align}
  & u_{\,t}\ +\ \varepsilon\,u\,u_{\,x}\ +\ \varepsilon\,v\,u_{\,r}\ +\ p_{\,x}\ =\ -\,\frac{1}{\delta^{\,2}}\;\frac{8}{Re}\;\frac{u^{\,w}}{{r_{\,0}}^{\,2}}\,, \label{eq:nrvmom1} \\
  & \delta^{\,2}\,(v_{\,t}\ +\ \varepsilon\,u\,v_{\,x}\ +\ \varepsilon\,v\,v_{\,r})\ +\ p_{\,r}\ =\ 0\,, \ \mbox{ for } 0\ \leq\ r\ \leq\ r^{\,w}\,, \label{eq:nrvmom2}\\
  & r\,u_{\,x}\ +\ (r\,v)_{\,r}\ =\ 0\,, \label{eq:nrvmas}
\end{align}
where again we consider $u\ =\ u\,(x,\,r,\,t)\,$. These equations differ from the \textsc{Euler} equations \eqref{eq:mom1} -- \eqref{eq:mas} only to the dissipative term of \eqref{eq:nrvmom1} and most of the analysis carried in Sections~\ref{sec:deriv1} and \ref{sec:deriv2} can be repeated for the new set of equations without imposing new boundary conditions or assuming rotational flow. Specifically, the only difference occurs in Equation~\eqref{eq:mom1c} which is now written in the form:
\begin{equation}\label{eq:vmom1c}
  u^{\,w}_{\,t}\ +\ \varepsilon\,u^{\,w}\,u^{\,w}_{\,x}\ +\ [\,\beta\,\eta\,]_{\,x}\ -\ \delta^{\,2}\;\frac{r^{\,w}\,r^{\,w}_{\,x}}{2}\;u^{\,w}_{\,x\,t}\ +\ \alpha\,\delta^{\,2}\,\eta_{\,x\,t\,t}\ =\ -\,\frac{1}{\delta^{\,2}}\;\frac{8}{Re}\;\frac{u^{\,w}}{{r_{\,0}}^{\,2}}\ +\ \O\,(\varepsilon\,\delta^{\,2})\,,
\end{equation}
and therefore we obtain the following dissipative \textsc{Boussinesq} system in lieu of \eqref{mass} -- \eqref{momentum}
\begin{align}
  & \eta_{\,t}\ +\ \frac{1}{2}\;(r_{\,0}\ +\ \varepsilon\,\eta)\,u^{\,w}_{\,x}\ +\ ({r_{\,0}}_{\,x}\ +\ \varepsilon\,\eta_{\,x})\,u^{\,w}\ +\nonumber \\ 
  & \qquad \frac{\delta^{\,2}}{4}\;{r_{\,0}}^{\,2}\,{r_{\,0}}_{\,x}\,u^{\,w}_{\,x\,x}\ +\ \frac{\delta^{\,2}}{16}\;{r_{\,0}}^{\,3}\,u^{\,w}_{\,x\,x\,x}\ =\ \O\,(\varepsilon\,\delta^{\,2},\,\delta^{\,4})\,, \label{vmass} \\
  & u^{\,w}_{\,t}\ +\ [\,\beta\,\eta\,]_{\,x}\ +\ \varepsilon\,u^{\,w}\,u^{\,w}_{\,x}\ -\ \alpha\,\delta^{\,2}\,({r_{\,0}}_{\,x}\,u^{\,w}_{\,t})_{\,x}\nonumber \\
  & \qquad -\ \frac{\delta^{\,2}\,r_{\,0}\,{r_{\,0}}_{\,x}}{2}\;u^{\,w}_{\,x\,t}\ -\ \frac{\alpha\,\delta^{\,2}}{2}\;\left(r_{\,0}\,u^{\,w}_{\,x\,t}\right)_{\,x}\ =\ -\,\frac{1}{\delta^{\,2}}\;\frac{8}{Re}\;\frac{u^{\,w}}{r_{\,0}^{\,2}}\ +\ \O\,(\varepsilon\,\delta^{\,2},\,\delta^{\,4})\,. \label{vmomentum}
\end{align}
System \eqref{vmass} -- \eqref{vmomentum} can be written in dimensional form and discarding the high-order terms as:
\begin{align}
  & \eta_{\,t}\ +\ \frac{1}{2}\;(r_{\,0}\ +\ \eta)\,u^{\,w}_{\,x}\ +\ ({r_{\,0}}_{\,x}\ +\ \eta_{\,x})\,u^{\,w}\ +\ \frac{{r_{\,0}}^{\,2}\,{r_{\,0}}_{\,x}}{4}\;u^{\,w}_{\,x\,x}\ +\ \frac{{r_{\,0}}^{\,3}}{16}\;u^{\,w}_{\,x\,x\,x}\ =\ 0\,, \label{dvmass} \\
  & u^{\,w}_{\,t}\ +\ \left[\,\frac{E\,h}{r_{\,0}^{\,2}\,\rho}\;\eta\,\right]_{\,x}\ +\ u^{\,w}\,u^{\,w}_{\,x}\ -\ \frac{\rho^{\,w}\,h}{\rho}\;({r_{\,0}}_{\,x}\,u^{\,w}_{\,t})_{\,x} \nonumber \\
  & \qquad -\ \frac{r_{\,0}\,{r_{\,0}}_{\,x}}{2}\;u^{\,w}_{\,x\,t}\ -\ \frac{\rho^{\,w}\,h}{2\,\rho}\;\left(r_{\,0}\,u^{\,w}_{\,x\,t}\right)_{\,x}\ =\ -8\,\kappa\;\frac{u^{\,w}}{r_{\,0}^{\,2}}\,. \label{dvmomentum}
\end{align}
Furthermore, since $u^{\,w}\ =\ u^{\,\theta}\ +\ \O\,(\delta^{\,2})\,$, and since we have assumed that $1/Re\ =\ \O\,(\varepsilon\,\delta^{\,2})$ we can extend all the \textsc{Boussinesq} models derived in Sections~\ref{sec:deriv1} and \ref{sec:deriv2} to the dissipative systems by just adding the term $-8\,\kappa\,u/r_{\,0}^{\,2}$ on the right-hand side of momentum equations. Since the classical \textsc{Boussinesq}-type System~\eqref{amasspd} -- \eqref{amomentumpd} is the simplest system and for other reasons that we explain later we rewrite it here in its dissipative form
\begin{align}
  & \eta_{\,t}\ +\ \frac{1}{2}\;(r_{\,0}\ +\ \eta)\,u^{\,\theta}_{\,x}\ +\ ({r_{\,0}}_{\,x}\ +\ \eta_{\,x})\,u^{\,\theta}\ =\ 0\,, \label{vamasspd} \\
  & (1\ -\ \bar{\alpha}\,{r_{\,0}}_{\,x\,x})\,u^{\,\theta}_{\,t}\ +\ [\,\bar{\beta}\,\eta\,]_{\,x}\ +\ u^{\,\theta}\,u^{\,\theta}_{\,x}\ \nonumber \\ 
  & \qquad +\ \frac{(3\,\bar{\alpha}\ +\ r_{\,0})\,{r_{\,0}}_{\,x}}{2}\;[\,\bar{\beta}\,\eta\,]_{\,x\,x}\ -\ \frac{(4\,\bar{\alpha}\ +\ r_{\,0})\,r_{\,0}}{8}\;u^{\,\theta}_{\,x\,x\,t}\ +\ 8\,\kappa\,\frac{u^{\,\theta}}{r_{\,0}^{\,2}}\ =\ 0\,. \label{vamomentumpd}
\end{align}
This system will be used later in some practical applications to verify the applicability of the analysis of this paper.

\begin{remark}
The dynamic viscosity for the blood is usually around $0.4\times 10^{-2}~ Kg/(m\cdot sec)$ on $37~^{\circ}C$ and so the kinematic viscosity is then $\kappa\ =\ 0.4\times 10^{-5}~m^2/sec$. We will see bellow that for practical use in blood flow problems the dissipative \textsc{Boussinesq} systems are very useful and the results are very close to {\em in vivo} observations of the blood pressure and flow for a common carotid artery, \cite{Alastruey2016, Nichols2011}.
\end{remark}

%%% ------------------------------------------------------------------------ %%%

\section{Unidirectional models and other conservation laws}
\label{sec:properties}

In this section we study some properties of significant importance of the systems \eqref{massg} -- \eqref{momentumg}, \eqref{pars1} -- \eqref{pars2} in the case of a vessel with constant radius $r_{\,0}\ =\ R\,$. For scaled radius $r_{\,0}\ =\ 1$ the Equations \eqref{massg} -- \eqref{momentumg} become:
\begin{align}
  & \eta_{\,t}\ +\ \frac{1}{2}\;(1\ +\ \varepsilon\,\eta)\,u^{\,\theta}_{\,x}\ +\ \varepsilon\,\eta_{\,x}\,u^{\,\theta}\ +\ \delta^{\,2}\,a\,u^{\,\theta}_{\,x\,x\,x}\ -\ \delta^{\,2}\,b\,\eta_{\,x\,x\,t}\ =\ \O\,(\varepsilon\,\delta^{\,2},\,\delta^{\,4})\,,  \label{massc}  \\
  & u^{\,\theta}_{\,t}\ +\ \beta\,\eta_{\,x}\ +\ \varepsilon\,u^{\,\theta}\,u^{\,\theta}_{\,x}\ +\ \delta^{\,2}\,\beta\,c\,\eta_{\,x\,x\,x}\ -\ \delta^{\,2}\,d\,u^{\,\theta}_{\,x\,x\,t}\ =\ \O\,(\varepsilon\,\delta^{\,2},\,\delta^{\,4})\,, \label{momentumc}
\end{align}
and in dimensional variables for general constant radius $r_{\,0}\ >\ 0\,$:
\begin{align}
  & \eta_{\,t}\ +\ \frac{1}{2}\;(r_{\,0}\ +\ \eta)\,u^{\,\theta}_{\,x}\ +\ \eta_{\,x}\,u^{\,\theta}\ +\ r_{\,0}^{\,3}\,a\,u^{\,\theta}_{\,x\,x\,x}\ -\ r_{\,0}^{\,2}\,b\,\eta_{\,x\,x\,t}\ =\ 0\,,  \label{massgc}  \\
  & u^{\,\theta}_{\,t}\ +\ \bar{\beta}\,\eta_{\,x}\ +\ u^{\,\theta}\,u^{\,\theta}_{\,x}\ +\ \bar{\beta}\,r_{\,0}^{\,2}\,c\,\eta_{\,x\,x\,x}\ -\ r_{\,0}^{\,2}\,d\,u^{\,\theta}_{\,x\,x\,t}\ =\ 0\,, \label{momentumgc}
\end{align}
where
\begin{multline*}
  a\ =\ \frac{(2\,\theta^{\,2}\ -\ 1)\,\nu}{16}\,, \qquad b\ =\ \frac{(2\,\theta^{\,2}\ -\ 1)\,(1\ -\ \nu)}{8}\,, \\ 
  c\ =\ \frac{(2\,\alpha\ +\ 1\ -\ \theta^{\,2})\,\mu}{4}\,, \qquad d\ =\ \frac{(2\,\alpha\ +\ 1\ -\ \theta^{\,2})\,(1\ -\ \mu)}{4}\,.
\end{multline*}
We mention that because the radius is constant $R\ =\ r_{\,0}$ the parameter $\alpha$ contains the known constant radius $r_{\,0}\,$.

It is noted that the well-posedness theory of \cite{BCS, Bona2004} can be generalised for the new set of \textsc{Boussinesq} systems in a straightforward way. We only mention here that following \cite{BCS} it turns out that a  \textsc{Boussinesq} system of the form \eqref{massc} -- \eqref{momentumc} is linearly well-posed if and only if one of the following conditions holds:
\begin{enumerate}
  \item $b\ \geq\ 0\,$, $d\ \geq\ 0\,$, $a\ \leq\ 0\,$, $c\ \leq\ 0\,$,
  \item $b\ \geq\ 0\,$, $d\ \geq\ 0\,$, $2\,a\ =\ \bar{\beta}\,c\ >\ 0\,$,
  \item $b\ =\ d\ <\ 0\,$,  $2\,a\ =\ \bar{\beta}\,c\ >\ 0\,$.
\end{enumerate}
We can then distinguish the following linearly well-posed systems:
\begin{itemize}
  \item Classical \textsc{Boussinesq}-type: $\mu\ =\ \nu\ =\ 0\,$, $\theta^{\,2}\ =\ \frac{1}{2}$
  \item \textsc{Bona--Smith}-type: $\nu\ =\ 0\,$, $\mu\ \in\ \mathds{R}\,$, $(2\,\alpha\ +\ 1\ -\ \theta^{\,2})\,\mu\ \leq\ 0\,$, $(2\,\alpha\ +\ 1\ -\ \theta^{\,2})\,(1\ -\ \mu)\ \geq\ 0$
  \item BBM--BBM-type: $\nu\ =\ 0\,$, $\mu\ =\ 0\,$, $\theta^{\,2}\ \in\ (1/2,\,2\,\alpha\ +\ 1)$ 
  \item KdV--BBM-type: $\nu\ =\ 1\,$, $\mu\ =\ 0\,$, $\theta^{\,2}\ \leq\ \min(1/2,\,2\,\alpha\ +\ 1)$
  \item KdV--KdV-type: $\nu\ =\ 1\,$, $\mu\ =\ 1\,$, $\theta^{\,2}\ \in\ (1/2,\,2\,\alpha\ +\ 1)$
\end{itemize}
where because of the linear well-posedness conditions more restrictions apply on $\alpha\,$. We first study one-way propagation models that can be derived from these systems.

%%% ------------------------------------------------------------------------ %%%

\subsection{Unidirectional models with constant radius}

Considering the scaling parameters \eqref{eq:ndv1} -- \eqref{eq:ndv2} then the simplest of the Systems \eqref{massgc} -- \eqref{momentumgc} with $\theta^{\,2}\ =\ 1/2\,$, $\mu\ =\ 0\,$, (\ie when $a\ =\ b\ =\ c\ =\ 0$ and $d\ =\ (4\,\alpha\ +\ 1)\,/\,8$) can be written in the following non-dimensional and scaled form:
\begin{align}
  \eta^{\,\ast}_{\,t^{\,\ast}}\ +\ 2\,\varepsilon\,\eta^{\,\ast}_{\,x^{\,\ast}}\,u^{\,\ast}\ +\ \varepsilon\,\eta^{\,\ast}\,u^{\,\ast}_{\,x^{\,\ast}}\ +\ u^{\,\ast}_{\,x^{\,\ast}}\ =\ 0\,,\label{neweq1b} \\
  u^{\,\ast}_{\,t^{\,\ast}}\ +\ \eta^{\,\ast}_{\,x^{\,\ast}}\ +\ 2\,\varepsilon\,u^{\,\ast}\,u^{\,\ast}_{\,x^{\,\ast}}\ -\ \delta^{\,2}\,d\,u^{\,\ast}_{\,x^{\,\ast}\,x^{\,\ast}\,t^{\,\ast}}\ =\ 0\,,\label{neweq2b}
\end{align}
which has the same form with system \eqref{casc1} -- \eqref{casc2} but with coefficient $d$ instead of $\alpha\,$. Using the same arguments as in Section~\ref{sec:review}, we derive the KdV-type equation:
\begin{equation}\label{kdvnew}
  \eta^{\,\ast}_{\,t^{\,\ast}}\ +\ \frac{5}{2}\;\varepsilon\,\eta^{\,\ast}\,\eta^{\,\ast}_{\,x^{\,\ast}}\ +\ \eta^{\,\ast}_{\,x^{\,\ast}}\ +\ \delta^{\,2}\;\frac{d}{2}\;\eta^{\,\ast}_{\,x^{\,\ast}\,x^{\,\ast}\,x^{\,\ast}}\ =\ 0\,,
\end{equation}
and the BBM-type equation
\begin{equation}\label{bbmnew}
  \eta^{\,\ast}_{\,t^{\,\ast}}\ +\ \frac{5}{2}\;\varepsilon\,\eta^{\,\ast}\,\eta^{\,\ast}_{\,x^{\,\ast}}\ +\ \eta^{\,\ast}_{\,x^{\,\ast}}\ -\ \delta^{\,2}\;\frac{d}{2}\;\eta^{\,\ast}_{\,x^{\,\ast}\,x^{\,\ast}\,t^{\,\ast}}\ =\ 0\,,
\end{equation}
which in dimensional variables are written in the form
\begin{equation*}
  \eta_{\,t}\ +\ \tilde{c}\,\eta_{\,x}\ +\ \frac{5}{2\,r_{\,0}}\;\tilde{c}\,\eta\,\eta_{\,x}\ +\ \frac{d}{2}\;r_{\,0}^{\,2}\;\tilde{c}\,\eta_{\,x\,x\,x}\ =\ 0\,,
\end{equation*}
and
\begin{equation*}
  \eta_{\,t}\ +\ \tilde{c}\,\eta_{\,x}\ +\ \frac{5}{2\,r_{\,0}}\;\tilde{c}\,\eta\,\eta_{\,x}\ -\ \frac{d}{2}\;r_{\,0}^{\,2}\,\eta_{\,x\,x\,t}\ =\ 0\,,
\end{equation*}
where $\tilde{c}\ =\ \sqrt{E\,h\,/\,2\,\rho\,r_{\,0}}$ is the \textsc{Moens--Korteweg} characteristic speed.

It is well-known, the KdV equation is integrable and, thus, it admits infinite number of (independent) conservation laws. Here we provide only a few of them:
\begin{equation*}
  \eta_{\,t}\ +\ \left[\,\tilde{c}\,\left(\eta\ +\ \frac{1}{2}\;\tilde{\alpha}\,\eta^{\,2}\ +\ \tilde{\beta}\,\eta_{\,x\,x}\right)\right]_{\,x}\ =\ 0\,,
\end{equation*}
\begin{equation*}
  \left(\eta^{\,2}\right)_{\,t}\ +\ \left[\,\tilde{c}\,\left(\frac{1}{2}\;\eta^{\,2}\ +\ \frac{1}{3}\;\tilde{\alpha}\,\eta^{\,3}\ -\ \frac{1}{2}\;\tilde{\beta}\,\eta_{\,x}^{\,2}\ +\ \tilde{\beta}\,\eta\,\eta_{\,x\,x}\right)\,\right]_{\,x}\ =\ 0\,,
\end{equation*}
\begin{multline*}
  \left(\eta\,(\tilde{\alpha}\,\eta^{\,2}\ +\ 3\,\tilde{\beta}\,\eta_{\,x\,x})\right)_{\,t}\ +\ \left[\,\tilde{c}\,\left(\tilde{\alpha}\,\eta^{\,3}\ +\ \frac{3}{4}\;\tilde{\alpha}^{\,2}\,\eta^{\,4}\ +\ 3\,\tilde{\alpha}\,\tilde{\beta}\,\eta^{\,2}\,\eta_{\,x\,x}\ +\ \right. \right.\\
  \left.\left. 3\,\tilde{\beta}^{\,2}\,\eta_{\,x\,x}^{\,2}\ +\ 3\,\tilde{\beta}\,\eta_{\,x}^{\,2}\ -\ 3\,\tilde{\beta}\,\eta\,\eta_{\,x\,t}\ +\ 3\,\tilde{\beta}\,\eta_{\,t}\,\eta_{\,x}\right)\,\right]_{\,x}\ =\ 0\,,
\end{multline*}
\begin{multline*}
  \left(\frac{1}{2}\;\eta\,\bigl(\tilde{\alpha}\,t\,\eta\ +\ 2\,t\ -\ \frac{2}{c}\;x\bigr)\right)_{\,t}\ +\ \left[\,\tilde{\beta}\,\eta_{\,x}\ +\ \tilde{c}\,\tilde{\alpha}\,t\,\eta^{\,2}\ +\ \frac{1}{3}\;\tilde{c}\,\tilde{\alpha}^{\,2}\,t\,\eta^{\,3}\ -\ \frac{1}{2}\;\tilde{\alpha}\,x\,\eta^{\,2}\ +\ \right.\\ 
  \left.\tilde{c}\,\tilde{\alpha}\,\tilde{\beta}\,t\,\eta\,\eta_{\,x\,x}\ -\ \frac{1}{2}\;\tilde{c}\,\tilde{\alpha}\,\tilde{\beta}\,t\,\eta_{\,x}^{\,2}\ -\ (x\ -\ \tilde{c}\,t)\,\eta\ -\ \tilde{\beta}\,(x\ -\ \tilde{c}\,t)\,\eta_{\,x\,x}\,\right]_{\,x}\ =\ 0\,,
\end{multline*}
where
\begin{equation*}
  \tilde{\alpha}\ \doteq\ \frac{5}{2\,r_{\,0}}\,, \qquad
  \tilde{\beta}\ \doteq\ \frac{\rho^{\,w}\,h\,r_{\,0}}{4\,\rho}\,.
\end{equation*}
Conservation laws and symmetries were computed using \texttt{GeM} package in \textsc{Maple}, \cite{Cheviakov2007}. Let us compute also the symmetry group of the KdV equation. Its infinitesimal generators are given below:
\begin{align*}
  X_{\,1}\ &=\ \partial_{\,t}\,, \\
  X_{\,2}\ &=\ \partial_{\,x}\,, \\
  X_{\,3}\ &=\ \tilde{c}\,\tilde{\alpha}\,t\,\partial_{\,x}\ +\ \partial_{\,\eta}\,, \\
  X_{\,4}\ &=\ -\,\frac{3}{2}\;t\,\partial_{\,t}\ -\ \left(\frac{1}{2}\;x\ +\ \tilde{c}\,t\right)\,\partial_{\,x}\ +\ \eta\,\partial_{\,\eta}\,.
\end{align*}
The corresponding point transformations are given here:
\begin{align*}
  & t^{\,\prime}\ =\ t\ +\ \varepsilon_{\,1}\,, \qquad x^{\,\prime}\ =\ x\,, \qquad \eta^{\,\prime}\ =\ \eta\,, \\
  & t^{\,\prime}\ =\ t\,, \qquad x^{\,\prime}\ =\ x\ +\ \varepsilon_{\,2}\,, \qquad \eta^{\,\prime}\ =\ \eta\,, \\
  & t^{\,\prime}\ =\ t\,, \qquad x^{\,\prime}\ =\ x\ +\ \varepsilon_{\,3}\,\tilde{c}\,\tilde{\alpha}\,t\,, \qquad \eta^{\,\prime}\ =\ \eta\ +\ \varepsilon_{\,3}\,,\\
  & t^{\,\prime}\ =\ \tilde{c}\,\ue^{\,-\,3/2\,\varepsilon_{\,4}}\,t\,, \qquad x^{\,\prime}\ =\ (x\ -\ \tilde{c}\,t)\,\ue^{\,-\,1/2\,\varepsilon_{\,4}}\ +\ \tilde{c}\,\ue^{\,-\,3/2\,\varepsilon_{\,4}}\,t\,, \qquad \eta^{\,\prime}\ =\ \ue^{\,\varepsilon_{\,4}}\,\eta\,.
\end{align*}
The first two transformations are time and space translations correspondingly. The third transformation is the \textsc{Galilean} boost and the last transformation is the scaling.

Nonlinear and dispersive wave equations usually admit solitary waves propagating in one direction without change in their shape and with constant phase speed $s\,$. In the case of unidirectional models it is usually easy to compute analytical formulas following \cf \eg \cite{Whitham1999}. The exact solitary wave to the particular KdV equation can be found easily to be of the form:
\begin{equation*}
  \eta\,(x,\,t)\ =\ a\,{\rm sech}^{\,2}\,\left(\frac{1}{2}\;\kappa\,\xi\,\right)\,,
\end{equation*}
where
\begin{equation*}
  \xi\ =\ x\ -\ s\,t\,, \quad \kappa\ =\ \sqrt{\frac{a\,\tilde{\alpha}}{3\,\tilde{\beta}}}\,, \quad s\ =\ \tilde{c}\,\left(1\ +\ \tilde{\beta}\,\kappa^{\,2}\right)\ \equiv\ \tilde{c}\,\left(1\ +\ \frac{a\,\tilde{\alpha}}{3}\right)\,.
\end{equation*}
Periodic cnoidal-type travelling waves can also be found in the form:
\begin{equation*}
  \eta\,(x,\,t)\ =\ a\,\mathrm{cn}^{\,2}\,\left(\frac{1}{2}\;\kappa\,\xi,\,m\right)\,, \quad m\ \in\ (0,\,1)\,,
\end{equation*}
where
\begin{equation*}
  \xi\ =\ x\ -\ s\,t\,, \quad \kappa\ =\ \sqrt{\frac{a\,\tilde{\alpha}}{3\,\tilde{\beta}\,m^{\,2}}}\,, \quad s\ =\ \tilde{c}\,\left(1\ +\ a\,\tilde{\alpha}\;\frac{2\,m^{\,2}\ -\ 1}{3\,m^{\,2}}\right)\,.
\end{equation*}
From these analytical results one can see that when the solution amplitude $a$ tends to $0\,$, the travelling wave speed tends to the speed of linear waves, \ie $s\ \to\ \tilde{c}\,$, where $\tilde{c}$ is the \textsc{Moens--Korteweg} velocity.

Contrary to the KdV equation, the BBM equation is not integrable. Thus, it can have only a finite number of conservation laws. Writing the BBM equation into the following conservative form:
\begin{equation*}
  \left(\eta\ -\ \tilde{\beta}\,\eta_{\,x\,x}\right)_{\,t}\ +\ \left[\,\tilde{c}\,\eta\ +\ \frac{1}{2}\;\tilde{c}\,\tilde{\alpha}\,\eta^{\,2}\,\right]_{\,x}\ =\ 0\,,
\end{equation*}
we were able to find an additional conservation law:
\begin{equation*}
  \left(\frac{1}{2}\;\eta^{\,2}\ +\ \frac{1}{6}\;\tilde{\beta}\,\eta_{\,x}^{\,2}\ -\ \frac{1}{3}\;\tilde{\beta}\,\eta\,\eta_{\,x\,x}\right)_{\,t}\ +\ \left[\,\frac{1}{2}\;\tilde{c}\,\eta^{\,2}\ +\ \frac{1}{3}\;\tilde{\alpha}\,\tilde{c}\,\eta^{\,3}\ -\ \frac{2}{3}\;\tilde{\beta}\,\eta\,\eta_{\,x\,t}\ +\ \frac{1}{3}\;\tilde{\beta}\,\eta_{\,x}\,\eta_{\,t}\right]_{\,x}\ =\ 0\,.
\end{equation*}
The infinitesimal generators of the BBM equation symmetries are given here:
\begin{align*}
  X_{\,1}\ &=\ \partial_{\,t}\,, \\
  X_{\,2}\ &=\ \partial_{\,x}\,, \\
  X_{\,3}\ &=\ -\,\tilde{\alpha}\,t\,\partial_{\,t}\ +\ (1\ +\ \tilde{\alpha}\,\eta)\,\partial_{\,\eta}\,.
\end{align*}
The corresponding point transformations, which keep invariant the set of the solutions to the BBM equation are given below:
\begin{align*}
  & t^{\,\prime}\ =\ t\ +\ \varepsilon_{\,1}\,, \qquad x^{\,\prime}\ =\ x\,, \qquad
  \eta^{\,\prime}\ =\ \eta\,, \\
  & t^{\,\prime}\ =\ t\,, \qquad x^{\,\prime}\ =\ x\ +\ \varepsilon_{\,2}\,, \qquad \eta^{\,\prime}\ =\ \eta\,, \\
  & t^{\,\prime}\ =\ \ue^{\,-\,\tilde{\alpha}\,\varepsilon_{\,3}}\,t\,, \qquad x^{\,\prime}\ =\ x\,, \qquad \eta^{\,\prime}\ =\ \frac{1}{\tilde{\alpha}}\;\left(\ue^{\,\tilde{\alpha}\,\varepsilon_{\,3}}\,(1\ +\ \tilde{\alpha}\,\eta)\ -\ 1\right)\,.
\end{align*}
The first two symmetries are time and space translations respectively. The third transformation is the scaling transformation.

The exact solitary wave solution to this equation can be easily derived:
\begin{equation*}
  \eta\,(x,\,t)\ =\ a\,{\rm sech}^{\,2}\,\left(\frac{1}{2}\;\kappa\,\xi\right)  \,,
\end{equation*}
where
\begin{equation*}
 \xi\ =\ x\ -\ s\,t\,, \quad \kappa\ =\ \sqrt{\frac{a\,\tilde{\alpha}}{(3\ +\ a\,\tilde{\alpha})\,\tilde{\beta}}}\,, \quad s\ =\ \tilde{c}\,\left(1\ +\ \frac{a\,\tilde{\alpha}}{3}\right)\,.
\end{equation*}
Periodic cnoidal-type travelling waves to the BBM equation can also be found in the form:
\begin{equation*}
  \eta\,(x,\,t)\ =\ a\,\mathrm{cn}^{\,2}\,\left(\frac{1}{2}\;\kappa\,\xi,\,m\right)\,, \qquad x_{\,0}\ \in\ \mathds{R}\,, \qquad m\ \in\ (0,\,1)\,,
\end{equation*}
where
\begin{equation*}
 \xi\ =\ x\ -\ s\,t\,, \quad \kappa\ =\ \sqrt{\frac{a\,\tilde{\alpha}}{\left(3\,m^{\,2}\ +\ a\,\tilde{\alpha}\,\left(2\,m^{\,2}\ -\ 1\right)\right)\,\tilde{\beta}}}\,, \quad s\ =\ \tilde{c}\,\left(1\ +\ a\,\tilde{\alpha}\,\frac{2\,m^{\,2}\ -\ 1}{3\,m^{\,2}}\right)\,.
\end{equation*}
From the expression for $\kappa$ it follows that cnoidal-type solutions for the BBM equation exist provided that
\begin{equation*}
  3\,m^{\,2}\ +\ a\,\tilde{\alpha}\,\left(2\,m^{\,2}\ -\ 1\right)\ >\ 0\,.
\end{equation*}

\begin{figure}
  \centering
  \bigskip
  \includegraphics[width=0.8\textwidth]{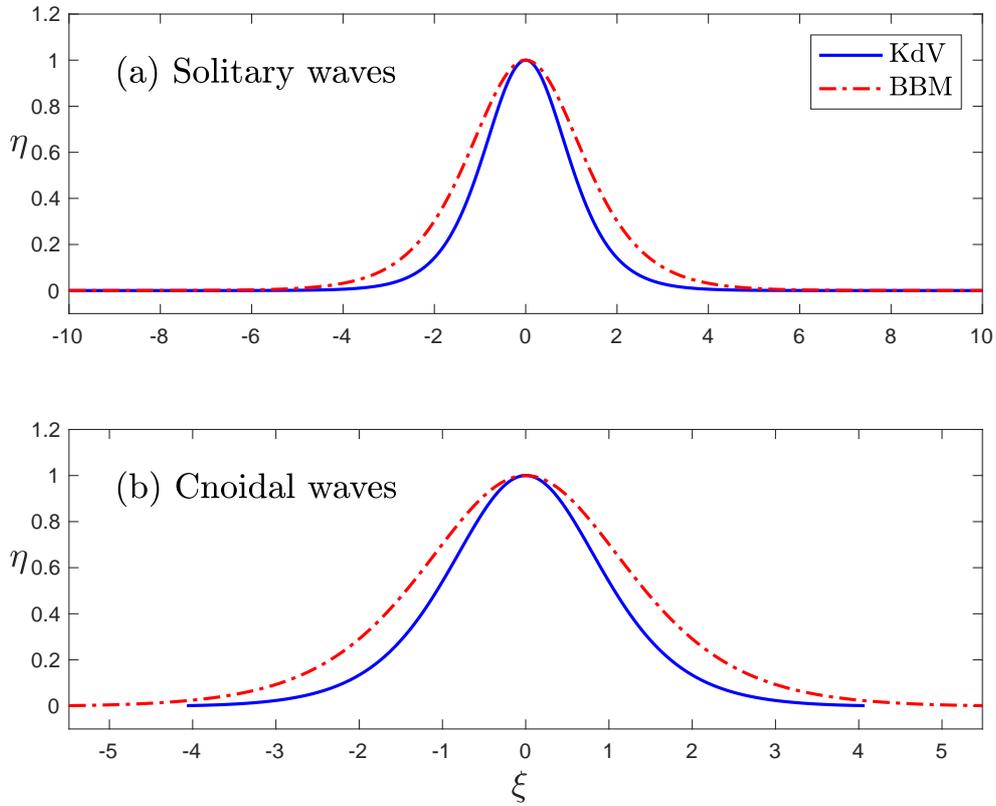}
  \caption{\small\em Comparison between traveling waves solution of the KdV and BBM equations. The cnoidal waves have been computed for $m\ =\ 0.99\,$.}
  \label{fig:KdVBBMsols}
\end{figure}

Figure~\ref{fig:KdVBBMsols} presents solitary and cnoidals waves (for $m\ =\ 0.99$) of the KdV and BBM equations of amplitude $a\ =\ 1$ in the case of $E\ =\ h\ =\ r_{\,0}\ =\ \rho\ =\ \rho^{\,w}\ =\ 1\,$. We observe that traveling waves of the BBM equation are wider (have larger support) compared to the analogous waves of the KdV equation. We will discuss traveling wave solutions again later.

It is interesting to notice that the speed/amplitude relations $s\ =\ s\,(a)$ are identical for the KdV and BBM equations. Moreover, the shapes of travelling waves are slightly different. However, these differences vanish again in the limit of linear waves, since $\lim_{a\ \to\ +0} \kappa_{\,\mathrm{KdV}}/\kappa_{\,\mathrm{BBM}}\ =\ 1\,$.

\begin{remark}
All the previously derived model equations are usually accompanied by appropriate initial and boundary conditions to form a well-posed problem in the \textsc{Hadamard} sense. These conditions can vary among the systems and the exact number of boundary conditions required by the system is known for some of the Systems including \eqref{amasspd} -- \eqref{amomentumpd} in case of vessel with constant initial radius, \cite{FP, AD3, ADM1}. In applications it is usually required to provide the velocity and the deviation of the vessel radius at an inlet position while similar conditions should be specified at the end of the vessel. These are enough to form well-posed problems for the commonly used systems such as the BBM--BBM system, or the classical \textsc{Boussinesq} and these are the boundary conditions we use in Section~\ref{sec:application} to demonstrate the capacity to use the classical \textsc{Boussinesq} model in practical problems. 
\end{remark}

%%% ------------------------------------------------------------------------ %%%

\subsection{Symmetries and conservation laws for the Boussinesq systems of constant radius}

Now we consider the general \textsc{Boussinesq} systems \eqref{massgc} -- \eqref{momentumgc} with constant radius $r_{\,0}$ and we explore some symmetries and conservation laws of these systems.

Equation \eqref{momentumgc} can be written in the conservative form:
\begin{equation*}
  \left(u\ -\ d\,r_{\,0}^{\,2}\,u_{\,x\,x}\right)_{\,t}\ +\ \left[\,\bar{\beta}\,\eta\ +\ \frac{1}{2}\;u^{\,2}\ +\ \bar{\beta}\,\tilde{c}\,r_{\,0}^{\,2}\,\eta_{\,x\,x}\,\right]_{\,x}\ =\ 0\,.
\end{equation*}
As always, $\bar{\beta}\ =\ E\,h\,/\,\rho\,r_{\,0}^{\,2}\,$. However, it is interesting to find less trivial conservation laws. Let us assume that $a\ \equiv\ 0\,$. In this case we are able to find an additional conservation law, which plays the r\^ole of the potential energy balance:
\begin{multline*}
  \left(\frac{1}{2}\;\eta^{\,2}\ +\ r_{\,0}\,\eta\ -\ b\,r_{\,0}^{\,2}\,\eta\,\eta_{\,x\,x}\ -\ \frac{1}{3}\;b\,r_{\,0}^{\,3}\,\eta_{\,x\,x}\ +\ \frac{1}{6}\;b\,r_{\,0}^{\,2}\,\eta_{\,x}^{\,2}\right)_{\,t}\ +\\
  \left[\,\frac{1}{2}\;u\,\eta^{\,2}\ +\ r_{\,0}\,u\,\eta\ +\ \frac{1}{2}\;r_{\,0}^{\,2}\,u\ +\ \frac{1}{3}\;b\,r_{\,0}^{\,2}\,\eta_{\,x}\,\eta_{\,t}\ -\ \frac{2}{3}\;b\,r_{\,0}^{\,2}\,\eta\,\eta_{\,t\,x}\,\right]_{\,x}\ =\ 0\,.
\end{multline*}
If additionally we assume that $a\ =\ c\ \equiv\ 0\,$, then we have an additional conservation law, which can be interpreted physically as the kinetic energy conservation:
\begin{multline*}
  \left(\frac{1}{2}\;u^{\,2}\ -\ 2\,\bar{\beta}\,\eta\ +\ \frac{1}{6}\;d\,r_{\,0}\,u_{\,x}^{\,2}\ +\ 2\,\bar{\beta}\,b\,r_{\,0}^{\,2}\,\eta_{\,x\,x}\right)_{\,t} +\\
  \left[\,\frac{1}{3}\;u^{\,3}\ -\ \bar{\beta}\,(\eta\ +\ r_{\,0})\,u\ +\ \frac{1}{3}\;d\,r_{\,0}^{\,2}\,u_{\,x}\,u_{\,t}\ -\ \frac{2}{3}\;d\,r_{\,0}\,u\,u_{\,t\,x}\,\right]_{\,x}\ =\ 0\,.
\end{multline*}

We also computed the symmetry point transformations of Equations~\eqref{massgc}, \eqref{momentumgc} for general values of the free parameters $\mu\,$, $\nu$ and $\theta\,$. Without any further assumptions, the solutions remain symmetric only with respect to time and space translations. However, if we assume that $\nu\ \equiv\ 0$ or $\theta^{\,2}\ \equiv\ \frac{1}{2}\,$, an additional symmetry transformation appears. The complete set of the infinitesimal generators in this particular case is given here:
\begin{align*}
  X_{\,1}\ &=\ \frac{\partial}{\partial x}\,, \\
  X_{\,2}\ &=\ \frac{\partial}{\partial t}\,, \\
  X_{\,3}\ &=\ -t\frac{\partial}{\partial t}\ +\ 2\,(\eta\ +\ r_{\,0})\;\frac{\partial}{\partial \eta}\ +\ u\,\frac{\partial}{\partial u}\,.
\end{align*}
The corresponding point transformations are given below in the same order:
\begin{align*}
  & t^{\,\prime}\ =\ t\,, \quad x^{\,\prime}\ =\ x\ +\ \varepsilon_{\,1}\,, \quad
  \eta^{\,\prime}\ =\ \eta\,, \quad u^{\,\prime}\ =\ u\,, \\
  & t^{\,\prime}\ =\ t\ +\ \varepsilon_{\,2}\,, \quad x^{\,\prime}\ =\ x\,, \quad \eta^{\,\prime}\ =\ \eta\,, \quad u^{\,\prime}\ =\ u\,, \\
  & t^{\,\prime}\ =\ \ue^{\,-\,\varepsilon_{\,3}}\,t\,, \quad x^{\,\prime}\ =\ x\,, \quad \eta^{\,\prime}\ =\ -\,r_{\,0}\ +\ \ue^{\,2\,\varepsilon_{\,3}}\,\bigl(r_{\,0}\ +\ \eta\bigr)\,, \quad u^{\,\prime}\ =\ \ue^{\,\varepsilon_{\,3}}\,u\,,
\end{align*}
where $\varepsilon_{\,1,\,2,\,3}\ \in\ \mathds{R}$ are free parameters. The last transformation is a scaling transformation. It is noted that in order to achieve further symmetries to the solutions one should include higher order terms even if they break the asymptotic order of the models, \cf \cite{Duran2013}.

%%% ------------------------------------------------------------------------ %%%

\section{Dispersion characteristics and solitary waves}
\label{sec:dispersion}

In this section we study the linear dispersion characteristics of the new \textsc{Boussinesq} systems (with constant initial radius) while we show a method to choose the parameters $\mu$ and $\nu$ in order to obtain a system with improved linear dispersion relation. Finally, we explore the solitary waves solutions of these systems.

%%% ------------------------------------------------------------------------ %%%

\subsection{Dispersion characteristics}

We start with the linear dispersion characteristics and compare with those of the linearised \textsc{Euler} equations:
\begin{align}
  & u_{\,t}\ +\ \frac{1}{\rho}\;p_{\,x}\ =\ 0\,, \label{eq:leuler1}\\
  & v_{\,t}\ +\ \frac{1}{\rho}\;p_{\,r}\ =\ 0\,, \label{eq:leuler2}\\
  & u_{\,x}\ +\ v_{\,r}\ +\ \frac{1}{r}\;v\ =\ 0\,, \label{eq:leuler3}
\end{align}
with linearised boundary conditions
\begin{align}
  & v\,(x,\,r^{\,w},\,t)\ =\ \eta_{\,t}\,(x,\,t)\,, \label{eq:ldbc1} \\
  & p^{\,w}\,(x,\,t)\ =\ \rho^{\,w}\,h\,\eta_{\,t\,t}\,(x,\,t)\ +\ \frac{E\,h}{\rho_{\,0}^{\,2}}\,\eta\,(x,\,t)\,, \label{eq:ldbc2} \\
  & v\,(x,\,0,\,t)\ =\ 0\,. \label{eq:ldbc3}
\end{align}
So considering solutions of the form $u\,(x,\,r,\,t)\ =\ u_{\,0}\,(r)\,\exp(\ui\,(k\,x\ -\ \omega\,t))\,$, $\eta\,(x,\,r,\,t)\ =\ \eta_{\,0}\,(r)\,\exp(\ui\,(k\,x\ -\ \omega\,t))\,$, $v\,(x,\,r,\,t)\ =\ v_{\,0}\,(r)\,\exp(\ui\,(k\,x\ -\ \omega\,t))\,$, $p\,(x,\,r,\,t)\ =\ p_{\,0}\,(r)\,\exp(\ui\,(k\,x\ -\ \omega\,t))$ and substituting into the linearised \textsc{Euler} equations we obtain the equation for $u_{\,0}\,(r)\,$:
\begin{equation}\label{besselm}
  r\,u_{\,0}^{\,\prime\prime}\,(r)\ +\ u_{\,0}^{\,\prime}\,(r)\ -\ r\,k^{\,2}\,u_{\,0}\,(r)\ =\ 0\,,
\end{equation}
which is the modified \textsc{Bessel} equation. Similarly, the boundary conditions take form:
\begin{align}
  & u^{\,\prime}_{\,0}\,(0)\ =\ 0\,, \label{eq:lldbc1} \\
  & \rho^{\,w}\,h\,\omega^{\,2}\,\eta_{\,0}\,(r_{\,0})\ +\ \frac{\rho\,\omega}{k}\;u_{\,0}\,(r_{\,0})\ -\ \frac{E\,h}{r_{\,0}^{\,2}}\;\eta_{\,0}\,(r_{\,0})\ =\ 0\,, \label{eq:lldbc2} \\
  & u^{\,\prime}_{\,0}\,(r_{\,0})\ =\ \omega\,k\,\eta_{\,0}\,. \label{eq:lldbc3}
\end{align}
A solution of \eqref{besselm} that satisfies also the boundary conditions \eqref{eq:lldbc1}, \eqref{eq:lldbc3} is the function $u_{\,0}\,(r)\ =\ \eta_{\,0}\,\omega\,I_{\,0}\,(r\,k)\,/\,I_{\,1}\,(r_{\,0}\,k)\,$, where $I_{\,0}\,(z)\,$, $I_{\,1}\,(z)$ are modified \textsc{Bessel} functions of first kind. Finally, substituting into the boundary condition \eqref{eq:lldbc2} we obtain the linear dispersion relation
\begin{equation}\label{eq:EulerDispRel}
  \omega_{\,\mathcal{E}}^{\,2}\,(k)\ =\ \frac{E\,h}{\rho\,r_{\,0}^{\,3}}\;\frac{r_{\,0}\,k\,I_{\,1}\,(r_{\,0}\,k)}{\frac{\rho^{\,w}\,h}{\rho\,r_{\,0}}\;r_{\,0}\,k\,I_{\,1}\,(r_{\,0}\,k)\ +\ I_{\,0}\,(r_{\,0}\,k)}\,.
\end{equation}
Similarly, taking $u^{\,\theta}\,(x,\,t)\ =\ u_{\,0}\,\exp(\ui\,(k\,x\ -\ \omega\,t))$ and $\eta\,(x,\,t)\ =\ \eta_{\,0}\,\exp(\ui\,(k\,x\ -\ \omega\,t))$ solutions to the linearised System \eqref{massgc} -- \eqref{momentumgc}
\begin{align}
  & \eta_{\,t}\ +\ \frac{1}{2}\;r_{\,0}\,u^{\,\theta}_{\,x}\ +\ r_{\,0}^{\,3}\,a\,u^{\,\theta}_{\,x\,x\,x}\ -\ r_{\,0}^{\,2}\,b\,\eta_{\,x\,x\,t}\ =\ 0\,, \label{lmassgc} \\
  & u^{\,\theta}_{\,t}\ +\ \bar{\beta}\,\eta_{\,x}\ +\ \bar{\beta}\,r_{\,0}^{\,2}\,c\,\eta_{\,x\,x\,x}\ -\ r_{\,0}^{\,2}\,d\,u^{\,\theta}_{\,x\,x\,t}\ =\ 0\,, \label{lmomentumgc}
\end{align}
with $\bar{\beta}\ =\ E\,h\,/\,\rho\,r_{\,0}^{\,2}\,$, we obtain the dispersion relation
\begin{equation}\label{eq:BousDispRel}
  \omega_{\,\mathcal{B}}^{\,2}\,(k)\ =\ \frac{E\,h}{\rho\,r_{\,0}^{\,3}}\;\frac{(1\ -\ c\,(r_{\,0}\,k)^{\,2})\cdot(1/2\ -\ a\,(r_{\,0}\,k)^{\,2})}{(1\ +\ b\,(r_{\,0}\,k)^{\,2})\cdot(1\ +\ d\,(r_{\,0}\,k)^{\,2})}\cdot(r_{\,0}\,k)^{\,2}\,.
\end{equation}

Instead of using the frequency function $\omega$ we consider the phase velocity, which describes the propagation speed of various spectral components $c\ =\ \omega\,(k)/k\,$. The \textsc{Taylor} expansions of the \textsc{Euler} equations dispersion relation \eqref{eq:EulerDispRel} in the long wave limit $k\ \rightarrow\ 0$ provide us with the reference approximation for the phase velocity:
\begin{multline}\label{eq:phaseveleul}
  \frac{c_{\,\mathcal{E}}\,(k)}{\tilde{c}}\ =\ 1\ -\ \frac{4\,\alpha\,+\,1}{2^{\,4}}\;(r_{\,0}\,k)^{\,2}\ +\ \frac{144\,\alpha^{\,2}\,+\,72\,\alpha\,+\,13}{3\,\times\,2^{\,9}}\;(r_{\,0}\,k)^{\,4} \\
  -\,\frac{960\,\alpha^{\,3}\,+\,720\,\alpha^{\,2}\,+\,228\,\alpha\,+\,31}{3\,\times\,2^{\,13}}\;(r_{\,0}\,k)^{\,6} \\
  +\ \frac{403200\,\alpha^{\,4}\,+\,403200\,\alpha^{\,3}\,+\,180000\,\alpha^{\,2}\,+\,42480\,\alpha\,+\,4591}{5\,\times\,3\,\times\,2^{\,19}}\;(r_{\,0}\,k)^{\,8}\ +\ \O\,(k^{\,10})\,,
\end{multline}
where $\tilde{c}$ is the \textsc{Moens--Korteweg} characteristic speed defined as 
\begin{equation}\label{MKvelocity}
  \tilde{c}\ =\ \sqrt{\frac{E\,h}{2\,\rho\,r_{\,0}}}\,.
\end{equation}

The same \textsc{Taylor}'s expansion can be obtained for the \textsc{Boussinesq} system dispersion relation as well:
\begin{multline}\label{eq:phasevelbouss}
  \frac{c_{\,\mathcal{B}}\,(k)}{\tilde{c}}\ =\ 1\ -\ \frac{4\,\alpha\,+\,1}{2^{\,4}}\;(r_{\,0}\,k)^{\,2}\ -\ \frac{1}{2^{\,9}}\;\Bigl[\bigl(64\,\mu\,-\,48\,\bigr)\alpha^{\,2}\ +\\
  \bigl(64\,\mu\,(1\,-\,\theta^{\,2})\,+\,32\,\theta^{\,2}\,-\,40\bigr)\,\alpha\ +\\
  16\,\bigl(\mu\,+\,\nu\,-\,1\bigr)\,\theta^{\,4}\,-\,8\bigl(4\,\mu\,+\,2\,\nu\,-\,3\bigr)\,\theta^{\,2}\,+\,16\,\mu\,+\,4\,\nu\,-\,11\Bigr]\;(r_{\,0}\,k)^{\,4}\\
  -\ \frac{\mathcal{P}_{\,3}\,(\alpha;\,\mu,\,\nu,\,\theta^{\,2})}{2^{\,13}}\;(r_{\,0}\,k)^{\,6}\ +\ \frac{\mathcal{P}_{\,4}\,(\alpha;\,\mu,\,\nu,\,\theta^{\,2})}{2^{\,19}}\;(r_{\,0}\,k)^{\,8}\ +\ \O\,(k^{\,10})\,,
\end{multline}
where $\mathcal{P}_{\,3,\,4}\,(\alpha;\,\mu,\,\nu,\,\theta^{\,2})$ are some well-defined cubic and quartic polynomials in the variable $\alpha\,$.

From the expansions \eqref{eq:phaseveleul} and \eqref{eq:phasevelbouss} we observe that the \textsc{Boussinesq} equations approximate the full \textsc{Euler} dispersion relation to the second order uniformly for all values of the parameters $(\mu,\,\nu,\,\theta^{\,2})\,$. However by choosing these parameters appropriately, we can further improve its dispersion characteristics.

\begin{figure}
  \bigskip
  \centering
  \includegraphics[width=0.99\textwidth]{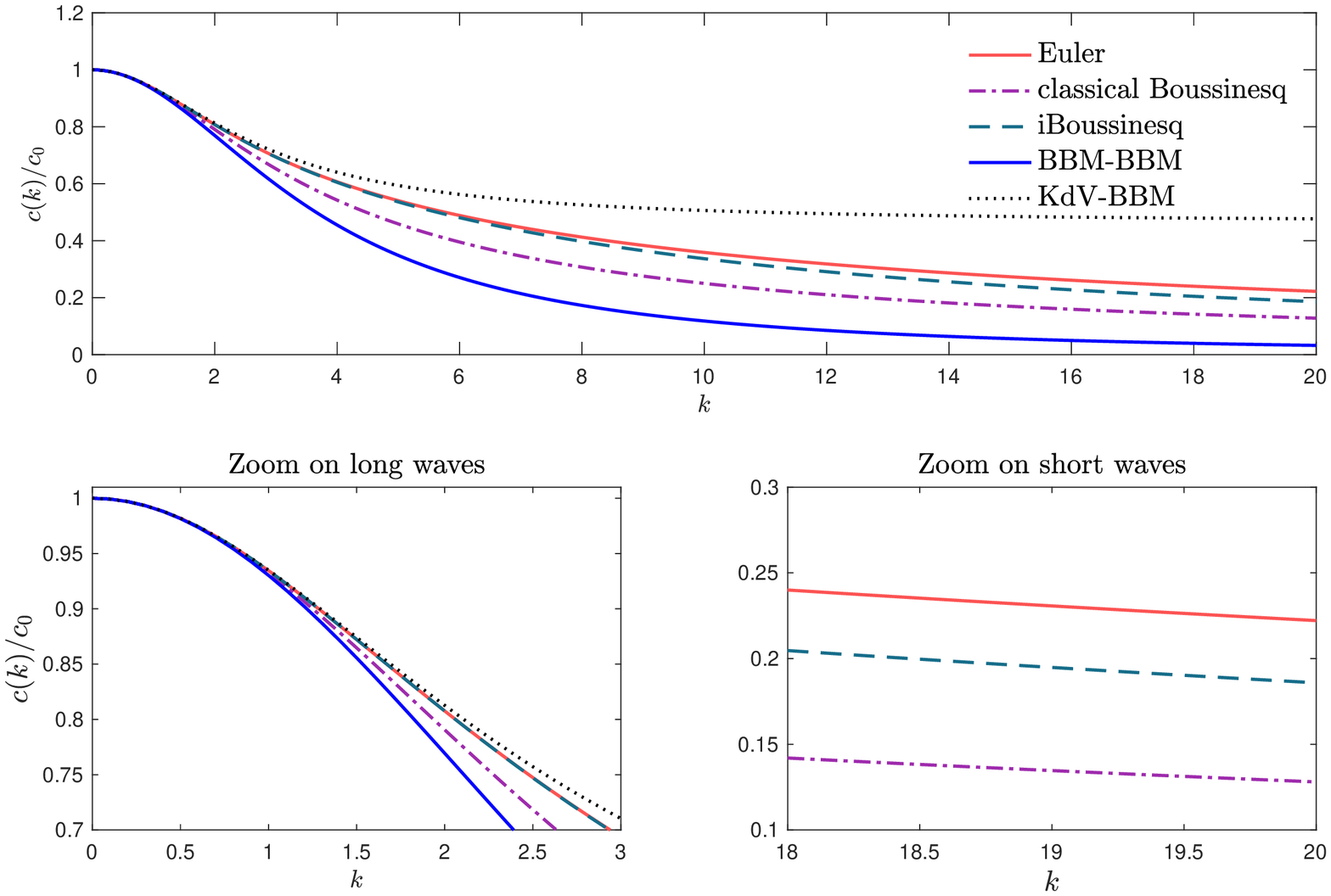}
  \caption{\small\em Comparison of dispersion relation of various \textsc{Boussinesq} systems against the reference solution provided by the linearised \textsc{Euler} model.}
  \label{fig:disprels}
\end{figure}

\begin{remark}
Even if the \textsc{Boussinesq} models were initially derived to represent long waves ($k\ \rightarrow\ 0$), it is not forbidden to look at its phase speed behaviour in the short waves limit $k\ \rightarrow\ +\infty\,$. First, we compute the reference asymptotic behaviour from the full \textsc{Euler} model:
\begin{equation*}
  c_{\,\mathcal{E}}\,(k)\ =\ \frac{1}{r_{\,0}}\;\sqrt{\frac{E}{\rho^{\,w}}}\cdot\biggl[\,\frac{1}{k}\ -\ \frac{\rho}{2\,\rho^{\,w}\,h}\;\frac{1}{k^{\,2}}\,\biggr]\ +\ \O\,\biggl(\frac{1}{k^{\,3}}\biggr)\,.
\end{equation*}
The quantity $\sqrt{E\,/\,\rho^{\,w}}$ has the dimension of the velocity and is called the first velocity, while the coefficient $1/r_{\,0}\,\sqrt{E\,/\,\rho^{\,w}}$ is the first frequency. The dispersion relation of the \textsc{Boussinesq} systems has the following asymptotic behaviour (provided that $\theta^{\,2}\ \neq\ 1/2\,$, $\mu\ \neq\ 1$ and $\nu\ \neq\ 1$):
\begin{equation*}
  c_{\,\mathcal{B}}\,(k)\ =\ c_{\,0}\,\sqrt{\frac{\mu\,\nu}{(\mu\ -\ 1)\,(\nu\ -\ 1)}}\ +\ \O\,\biggl(\frac{1}{k}\biggr)\,.
\end{equation*}
If $\theta^{\,2}\ =\ 1/2\,$, then the value of $\nu$ is immaterial and we obtain a similar asymptotic behaviour:
\begin{equation*}
  c_{\,\mathcal{B}}\,(k)\ =\ c_{\,0}\,\sqrt{\frac{\mu}{\mu\ -\ 1}}\ +\ \O\,\biggl(\frac{1}{k}\biggr)\,.
\end{equation*}
Thus, the phase speed of short waves in the \textsc{Boussinesq} model tends to a finite value, while \textsc{Euler}'s equations phase speed decays monotonically to zero. This drawback can be corrected by taking $\nu\ =\ 0$ or $\mu\ =\ 0\,$. The value $\mu\ =\ 0$ seems to be preferable since it works for all values of the parameter $\theta\,$. By making a different choice of other parameters, the decay rate of the phase velocity $c_{\,\mathcal{B}}\,(k)$ as $k\ \to\ +\infty$ may be corrected as well. However, the precise rate might be of secondary importance provided that the trend goes into the right direction, \ie $\lim_{\,k\ \to\ +\infty}\,c_{\,\mathcal{B}}\,(k)\ =\ 0\,$. Additionally, as we observe in Figure~\ref{fig:disprels} among other classical candidates the classical \textsc{Boussinesq} system with $\mu\ =\ 0\,$, $\nu\ =\ 1$ and $\theta^{\,2}\ =\ 1/2$ is one of the best choices as far as it concerns the dispersion characteristics. In what follows we take $\mu\ \equiv\ 0\,$.
\end{remark}

By equating the coefficients of $\O\,(k^{\,4})$ terms of the Expansions~\eqref{eq:phaseveleul} and \eqref{eq:phasevelbouss} we obtain the equation:
\begin{equation}\label{eq:equat1}
  \mathcal{E}_{\,1}\,(\theta,\,\nu;\,\alpha)\ \doteq\ (2\,\theta^{\,2}\ -\ 1)\,\alpha\ +\ (\nu\ -\ 1)\,\theta^{\,4}\ -\ \left(\nu\ -\ \frac{3}{2}\right)\,\theta^{\,2}\ +\ \frac{3\,\nu\ -\ 5}{12}\ =\ 0\,.
\end{equation}
It can be easily verified that the value $\theta^{\,2}\ =\ 1/2$ does not satisfy the Equation~\eqref{eq:equat1}. Specifically, we have $\mathcal{E}_{\,1}\,(\sqrt{1/2},\,\nu;\,\alpha)\ =\ 4$ for all $\nu$ and $\alpha\,$. Therefore, although the choice $\theta^{\,2}\ =\ 1/2$ leads to the simplest System~\eqref{amasspd} -- \eqref{amomentumpd} it cannot be accurate beyond the standard second order term.

\begin{figure}
  \centering
  \includegraphics[width=0.61\textwidth]{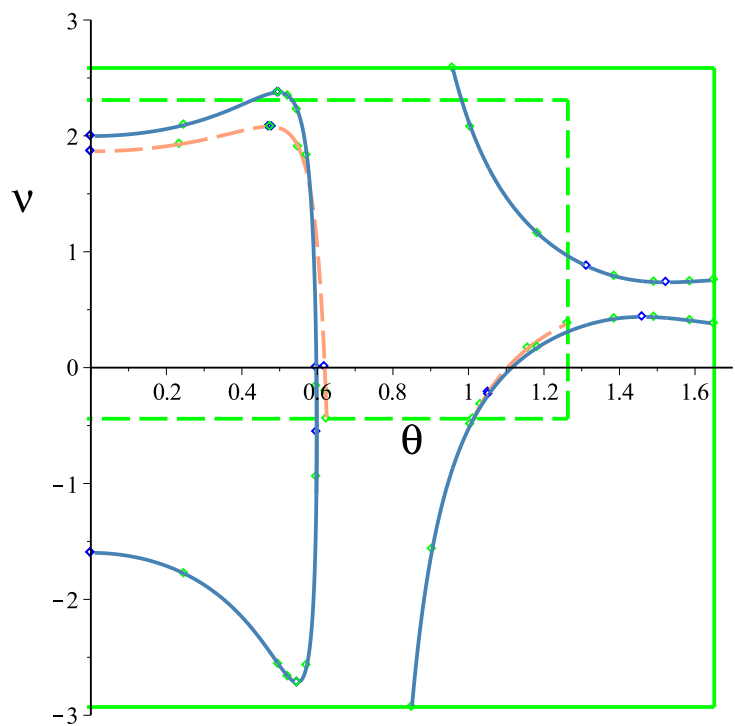}
  \caption{\small\em Loci of algebraic curves $\mathcal{E}_{\,1,\,2}\,(\theta,v\,\nu)\ =\ 0$ in the admissible strip for the physical parameter $\alpha\ =\ 0.05\,$. The dashed line ($- -$) represents the curve $\mathcal{E}_{\,1}\,(\theta,\,\nu)\ =\ 0$ while the solid line (---) shows $\mathcal{E}_{\,2}\,(\theta,\,\nu)\ =\ 0\,$.}
  \label{fig:zoom}
\end{figure}

Similarly, equating the terms of order $\O\,(k^{\,6})$ we obtain the equation:
\begin{multline}\label{eq:equat2}
  \mathcal{E}_{\,2}(\theta,\,\nu;\,\alpha)\ \doteq\ (2\,\theta^{\,2}\ -\ 1)\,\alpha^{\,2}\ +\ \left[\,\left(\frac{1}{3}\;\nu\ -\ 1\right)\,\theta^{\,4}\ -\ \left(\frac{1}{3}\;\nu\ -\ 2\right)\,\theta^{\,2}\ +\ \frac{\nu\ -\ 8}{12}\,\right]\,\alpha\\
  -\ \frac{1}{3}\;\nu\,\left(1\ -\ \nu\right)\,\theta^{\,6}\ +\ \left(\frac{1}{2}\;\nu^{\,2}\ -\ \frac{5}{12}\;\nu\ -\ \frac{1}{4}\right)\,\theta^{\,4}\
  -\ \left(\frac{1}{4}\;\nu^{\,2}\ -\ \frac{1}{6}\;\nu\ -\ \frac{1}{8}\right)\,\theta^{\,2}\ +\ \frac{1}{24}\;\nu^{\,2}\ -\ \frac{1}{48}\;\nu\ -\ \frac{7}{72}\ =\ 0\,.
\end{multline}
In order for the dispersion relations \eqref{eq:phaseveleul} and \eqref{eq:phasevelbouss} to match up to the sixth order of accuracy the Equations~\eqref{eq:equat1}, \eqref{eq:equat2} have to be solved simultaneously for a fixed value of the parameter $\alpha\,$:
\begin{align}
  \mathcal{E}_{\,1}\,(\theta,\,\nu;\,\alpha)\ &=\ 0\,,  \label{eq:sys1} \\
  \mathcal{E}_{\,2}\,(\theta,\,\nu;\,\alpha)\ &=\ 0\,.\label{eq:sys2}
\end{align}
Only solutions that belong to the infinite strip $\bigl(\theta,\,\nu\bigr)\ \in\ [\,0,\,1\,]\,\times\,\mathds{R}$ are admissible. Geometrically these solutions correspond to the intersection of two real algebraic curves defined by Equations~\eqref{eq:equat1}, \eqref{eq:equat2} in the $(\theta,\,\nu)$-plane. Let us consider a biologically realistic value of the parameter $\alpha\ =\ 0.05\,$. In Figure~\ref{fig:zoom} we show the loci of two algebraic curves in the admissible strip. We can see that there is only one intersection in the admissible strip. The coordinates of this intersection point $(\theta^{\,\star},\,\nu^{\,\star})$ for $\alpha\ =\ 0.05$ is given here (correct to $30$ digits):
\begin{align*}
  \nu^{\,\star}\    &=\ 1.53773145502645468091997801621\ldots\,, \\
  \theta^{\,\star}\ &=\ 0.58088109654602731473797307004\ldots\,.
\end{align*}
The topology of algebraic curves defined in \eqref{eq:sys1}, \eqref{eq:sys2} does not depend on the parameter $\alpha\,$, \ie for any positive value of $\alpha$ we have a unique intersection point in the admissible strip. If we increase the parameter $\alpha\ \to\ +\infty\,$, the solution $\nu^{\,\star}\ \to\ +\infty$ as well, while the other parameter tends to a finite value $\theta^{\,\star}\ \to\ 0.70\cdots$. The \textsc{Boussinesq} model with parameters $\bigl(\theta^{\,\star}\,(\alpha),\,\nu^{\,\star}\,(\alpha),\,\mu^{\,\star}\ \equiv\ 0\bigr)$ will be called the \emph{improved} \textsc{Boussinesq} system.

The phase speeds for the full \textsc{Euler}, classical and improved \textsc{Boussinesq} equations are depicted in Figure~\ref{fig:disprels}. We can see that the judicious choice of free parameters results in the substantial improvement of the dispersion relation properties for long waves (infrared region, which is achieved naturally by construction), but also for short waves (`ultraviolet' region, which is rather surprising). We computed the \textsc{Taylor} expansion coefficients of dispersion relations in the long wave limit $k\ \to\ 0$ for $\alpha\ =\ 0.05$ for the classical and improved \textsc{Boussinesq} models. Their numerical values are reported in Table~\ref{tab:coefs}. One can see that the coefficients of the improved model coincide with the reference values up to the sixth order. It is not a surprise, but rather a verification. However, we can see that beyond sixth order the coefficients remain close to the exact values. The classical \textsc{Boussinesq} system starts to differ from the fourth order.

\begin{table}
\begin{tabular}{c|c|c|c}
\hline
\textit{Degree of the term} & \textit{Euler} & \textit{iBoussinesq} & \textit{classical Boussinesq} \\
\hline
$k^{0}$ &  $1$                         & $1$                         & $ 1$ \\
$k^{2}$ &  $-0.075$                    & $-0.075$                    & $-0.075$ \\
$k^{4}$ &  $0.011041(6)$               & $0.011041(6)$               & $ 0.00843750$ \\
$k^{6}$ &  $-0.00180338$               & $-0.00180338$               & $-0.00105469$ \\
$k^{8}$ &  $ 0.00030594$               & $ 0.00030526$               & $ 0.00013843$ \\
$k^{10}$ & $-0.00005304$               & $-0.00005279$               & $-0.00001868$ \\
$k^{12}$ & $ 0.00000934$               & $ 0.00000927$               & $ 0.00000257$ \\
$k^{14}$ & $-0.00000166$               & $-0.00000165$               & $-3.57903671\times 10^{-7}$ \\
$k^{16}$ & $ 2.98218393\times 10^{-7}$ & $ 2.94810072\times 10^{-7}$ & $ 5.03302038\times 10^{-8}$ \\
$k^{18}$ & $-5.38927682\times 10^{-8}$ & $-5.31665759\times 10^{-8}$ & $-7.13011220\times 10^{-9}$ \\
$k^{20}$ & $ 9.79265927\times 10^{-9}$ & $ 9.64105154\times 10^{-9}$ & $ 1.01604099\times 10^{-10}$ \\
\hline
\end{tabular}
\bigskip
\caption{\small\em Comparison of coefficients in \textsc{Taylor} expansion of dispersion relations at the long wave limit $k\ \to\ 0\,$.}
\label{tab:coefs}
\end{table}

Let us check now how the improved system behaves asymptotically for short waves $k\ \to\ +\infty\,$. First of all, the reference solution is
\begin{equation*}
  \frac{c_{\,\mathcal{E}}\,(k)}{c_{\,0}}\ =\ \frac{\mathbf{1}}{k}\ -\ \frac{10}{k^{\,2}}\ +\ \frac{145}{k^{\,3}}\ +\ \O\,\Bigl(\frac{1}{k^{\,4}}\Bigr)\,.
\end{equation*}
The classical \textsc{Boussinesq} system gives
\begin{equation*}
  \frac{c_{\,\mathrm{c}\mathcal{B}}\,(k)}{c_{\,0}}\ =\ \frac{\mathbf{0.4082}}{k}\ -\ \frac{1.3608}{k^{\,3}}\ +\ \O\,\Bigl(\frac{1}{k^{\,4}}\Bigr)\,,
\end{equation*}
while the improved system gives
\begin{equation*}
  \frac{c_{\,\mathrm{i}\mathcal{B}}\,(k)}{c_{\,0}}\ =\ \frac{\mathbf{0.7422}}{k}\ +\ \frac{19.3358}{k^{\,3}}\ +\ \O\,\Bigl(\frac{1}{k^{\,4}}\Bigr)\,.
\end{equation*}
We can observe that the improved system outperforms the classical one in the ultraviolet region as well.

\begin{figure}
  \bigskip
  \centering
  \includegraphics[width=0.65\textwidth]{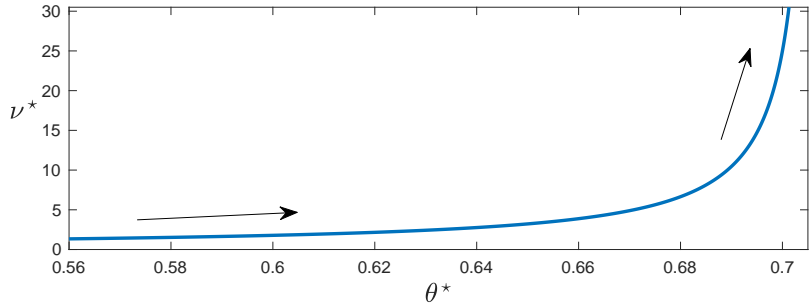}
  \caption{\small\em The optimal parameters $\bigl(\theta^{\,\star},\,\nu^{\,\star})$ as functions of $\alpha\ \in\ \mathds{R}^{\,+}\,$. The arrows indicate the direction of increasing $\alpha\,$.}
  \label{fig:homo}
\end{figure}

The question that one could ask now is how the optimal parameters $\bigl(\theta^{\,\star},\,\nu^{\,\star})$ change when we let $\alpha$ vary in $\mathds{R}^{\,+}$? We addressed this problem numerically and the continuation curve in the space $\bigl(\theta^{\,\star},\,\nu^{\,\star})$ is shown in Figure~\ref{fig:homo}. It is not difficult to estimate also the limiting points of this curve. Namely, in the limit $\alpha\ \to\ 0\,$, we depart from the point
\begin{align*}
  \theta^{\,\star}\,(0)\ &=\ \frac{\sqrt{5\ -\ \frac{1}{3}\;\sqrt{57}}}{2\,\sqrt{2}}\ \approx\ 0.5572\ldots\,, \\
  \nu^{\,\star}\,(0)\ &=\ \frac{9}{4}\ -\ \frac{3}{8}\;\bigl(5\ -\ \frac{1}{3}\;\sqrt{57}\bigr)\ \approx\ 1.3187\ldots\,.
\end{align*}
On the other hand, when the parameter $\alpha\ \to\ +\infty\,$, we tend to the point:
\begin{align*}
  \theta^{\,\star}\,(+\infty)\ &=\ \frac{\sqrt{2}}{2}\ \approx\ 0.7071\ldots\,, \\
  \nu^{\,\star}\,(+\infty)\ &=\ +\infty\,.
\end{align*}

We continue with the study of solitary wave solutions of the new \textsc{Boussinesq} systems.

%%% ------------------------------------------------------------------------ %%%

\subsection{Solitary waves}

Finding analytical formulas for the solitary wave solutions of unidirectional model equations such as the KdV and BBM equations is usually straightforward. The computation of the solitary waves to \textsc{Boussinesq} systems such as for the System~\eqref{massgc} -- \eqref{momentumgc} is very difficult and there are only a few cases were analytical formulas can be found, \cite{Chen2000a}. Using the appropriate {\em ansatz} we were able to compute exact solitary waves in a particular case when $\mu\ =\ \nu\ \equiv\ 0$ and $\theta^{\,2}\ \in\ \left[\,0,\,1\,\right]\ \setminus\ \left\{1/2\right\}\,$. For these values the coefficients become:
\begin{equation*}
  a\ =\ 0\,, \qquad
  b\ =\ \frac{2\,\theta^{\,2}\ -\ 1}{8}\,, \qquad
  c\ =\ 0\,, \qquad
  d\ =\ \frac{2\,\alpha\ +\ 1\ -\ \theta^{\,2}}{4}\,.
\end{equation*}
The exact solution reads:
\begin{align}
  & \eta\,(x,\,t)\ =\ 9\,(\tilde{a}\ +\ r_{\,0})\,{\rm sech}^{\,4}\,(\kappa\,\xi)\ -\ 6\,(\tilde{a}\ +\ r_{\,0})\,{\rm sech}^{\,2}\,(\kappa\,\xi)\ -\ 2\,\tilde{a}\ -\ 3\,r_{\,0}\,, \label{eq:swb1}\\
  & u\,(x,\,t)\ =\ 3\,s\,{\rm sech}^{\,2}(\kappa\,\xi)\,, \label{eq:swb2}
\end{align}
where $\xi\ =\ x\ -\ s\,t\ -\ x_{\,0}$ and
\begin{equation*}
  \kappa\ =\ \frac{1}{r_{\,0}\,\sqrt{2\,\theta^{\,2}\ -\ 1}}\,, \qquad
  s\ =\ \sqrt{\frac{2\,\bar{\beta}\,(\tilde{a}\ +\ r_{\,0})\,(2\,\theta^{\,2}\ -\ 1)}{2\,(1\ +\ \alpha)\ -\ 3\,\theta^{\,2}}}\,.
\end{equation*}
By $\bar{\beta}$ we denote as usual the coefficient $\bar{\beta}\ =\ E\,h\,/\,\rho\,r_{\,0}^{\,2}$ while $\tilde{a}$ is a parameter that defines the amplitude of the solitary wave. It is worth mentioning that the $\eta-$component of the solution \eqref{eq:swb1} does not tend to zero as $\abs{\xi}\ \to\ \infty$ except for the case where $2\,\tilde{a}\ +\ 3\,r_{\,0}\ =\ 0$ which defines a unique solitary wave for each value of $\theta\,$. These traveling wave solution are not classical solitary waves and can be rather unphysical and are expected to be rather unstable. We also note that the classical \textsc{Boussinesq} system with $\theta^{\,2}\ =\ 1/2$ does not admit solitary waves of the form \eqref{eq:swb1} -- \eqref{eq:swb2}. On the other hand, other traveling wave solutions can be found for the specific system by assuming that $\eta\ =\ 0\,$.

Although it is difficult to find analytical formulas for the traveling wave solutions of the \textsc{Boussinesq} systems at hand, one can see that dynamical systems techniques used in \cite{DMII} can be applied in a straightforward manner to the general system \eqref{massgc} -- \eqref{momentumgc} and conclude that all admissible systems with coefficients as shown above and in \cite{DMII} possess classical solitary waves that decay exponentially to zero as $\abs{x}\ \to\ \infty\,$. Because the classical \textsc{Boussinesq} system is of significant importance for applications due to its simplicity and because it has favourable dispersion characteristic properties we compute approximately solitary waves with high accuracy using the \textsc{Petviashvili} method, \cite{Petviashvili1976, Pelinovsky2004}. The \textsc{Petviashvili} method is a modification of the classical fixed point iteration but with several advantages. For example, it is known that the classical fixed point iteration applied to solitary wave problems usually diverges in contrast with the \textsc{Petviashvili} method which is usually convergent, \cf \eg \cite{Alvarez2014}. This method was applied successfully to similar systems of differential equations such as the analogous classical \textsc{Boussinesq} system of water wave theory, \cite{Duran2013}.

We consider the System~\eqref{massgc} -- \ref{momentumgc} with $a\ =\ b\ =\ c\ =\ 0$ and $d\ >\ 0\,$. In what follows we drop $\theta$ in our notation. Using the {\em ansatz} $\eta\,(x,\,t)\ =\ \eta\,(\xi)$ and $u\,(x,\,t)\ =\ u\,(\xi)$ where $\xi\ =\ x\ -\ s\,t$ to denote traveling wave solutions propagating with constant speed $s\,$, the classical \textsc{Boussinesq} system can be transformed into the system of ordinary differential equations
\begin{align}
  & 2\;\frac{\eta^{\,\prime}}{r_{\,0}\ +\ \eta}\ =\ -\,\frac{u^{\,\prime}}{u\ -\ s}\,, \label{eq:odecb1} \\
  & -\,s\,u^{\,\prime}\ +\ \bar{\beta}\,\eta^{\,\prime}\ +\ u\,u^{\,\prime}\ +\ r_{\,0}^{\,2}\,d\,s\,u^{\,\prime\prime\prime}\ =\ 0\,, \label{eq:odecb2}
\end{align}
where in \eqref{eq:odecb1} we have separated the variables, while ${}^\prime$ stands for the derivative $\od{}{\xi}\,$. Integrating both equations and using the assumption that the solutions decrease to zero as $\abs{\xi}\ \to\ \infty$ we have that
\begin{align}
  & \eta\ =\ r_{\,0}\left(-1\ \pm\ \sqrt{\frac{s}{\abs{u\ -\ s}}}\right)\,, \label{eq:odecb1b} \\
  &-\,s\,u\ +\ \bar{\beta}\,\eta\ +\ \frac{1}{2}\;u^{\,2}\ +\ r_{\,0}^{\,2}\,d\,s\,u^{\,\prime\prime}\ =\ 0\,. \label{eq:odecb2b}
\end{align}
Combining \eqref{eq:odecb1b} and \eqref{eq:odecb2b} we get a single equation for the unknown velocity profile $u$
\begin{equation}\label{eq:petm1}
  -\,s\,u\ +\ \bar{\beta}\,r_{\,0}\,\left(-1\ \pm\ \sqrt{\frac{s}{\abs{u\ -\ s}}}\right)\ +\ \frac{1}{2}\;u^{\,2}\ +\ r_{\,0}^{\,2}\,d\,s\,u^{\,\prime\prime}\ =\ 0\,.
\end{equation}
Due to physical restrictions we assume that $u\ \leq\ s$ for $s\ >\ 0$ otherwise the wave would break and also assume that $\eta\ >\ 0\,$. Then \eqref{eq:petm1} becomes:
\begin{equation}\label{eq:petm2}
  -\,s\,u\ +\ \bar{\beta}\,r_{\,0}\,\left(-1\ +\ \sqrt{\frac{s}{s\ -\ u}}\right)\ +\ \frac{1}{2}\;u^{\,2}\ +\ r_{\,0}^{\,2}\,d\,s\,u^{\,\prime\prime}\ =\ 0\,,
\end{equation}
while
\begin{equation}\label{eq:peteta}
  \eta\ =\ r_{\,0}\,\left(-\,1\ +\ \sqrt{\frac{s}{s\ -\ u}}\right)\,.
\end{equation}
Other assumptions on $s\,$, $u$ and $\eta$ will result to other traveling waves, such as solitary waves of depression, and the analysis and the methods are the same so we focus only on classical solitary waves of elevation. Using the classical \textsc{Fourier} transform to \eqref{eq:petm2} we get an equation of the form
\begin{equation*}
  \mathcal{L}\,U\ =\ \mathcal{N}\,(U)\,,
\end{equation*}
where $U\ =\ \mathcal{F}\,u$ denotes the \textsc{Fourier} transform of $u\,$, while $\mathcal{L}\,U\ =\ (1\ +\ r_{\,0}^{\,2}\,d\,s\,k^{\,2})\,U$ and $\mathcal{N}\,(U)\ =\ \mathcal{F}\,\left[\bar{\beta}\,r_{\,0}\,\left(-\,1\ +\ \sqrt{\frac{s}{s\ -\ u}}\right)\ +\ \frac{1}{2}\;u^{\,2}\,\right]$ are the linear and nonlinear parts of the Equation~\eqref{eq:petm1} respectively, and $u\ =\ \mathcal{F}^{\,-\,1}\,U$ is the inverse \textsc{Fourier} transform of $U\,$.  Here $k$ denotes the \textsc{Fourier} modes. The \textsc{Petviashvili} method is formulated then as follows: Given an arbitrary initial profile $U_{\,0}\,$, the $(n\,+\,1)-$th approximation $U_{\,n\,+\,1}$ of the \textsc{Fourier} transform of the solution is given by the recursive formula:
\begin{equation*}
  \mathcal{L}\,U_{\,n\,+\,1}\ =\ M_{\,n}^{\,\gamma}\,\mathcal{N}\,(U_{\,n})\,,
\end{equation*}
where
\begin{equation*}
  M_{\,n}\ =\ \frac{\langle\mathcal{L}\,U_{\,n},\,U_{\,n}\rangle}{\langle\mathcal{N}\,(U_{\,n}),\,U_{\,n}\rangle}\,.
\end{equation*}
is known as the stabilising factor. Here $\langle \cdot,\,\cdot \rangle$ denotes the usual $L^{\,2}$ inner product and $\gamma$ is a free parameter that controls the convergence of the method and is usually taken $\gamma\ =\ 2\,$. In order to apply this iterative scheme in practice we consider periodic boundary conditions in a large domain $[\,-L,\,L\,]$ while the \textsc{Fourier} transform is approximated using the discrete \textsc{Fourier} transform. As a termination criterion for the iterative proceedure we consider the error $\|U_{\,n\,+\,1}\ -\ U_{\,n}\|_{\,\infty}\ <\ tol\,$, which in our computation the tolerance has been taken $tol\ =\ 10^{\,-15}\,$. Therefore, we compute an approximation $u\ \approx\ \mathcal{F}^{\,-\,1}\,(U)$ while the approximation for $\eta$ can be computed using the exact Equation~\eqref{eq:odecb1b}. Initial approximations $U_{\,0}$ of the solitary wave can be computed using the exact solitary wave solutions computed for the KdV and BBM equations as well as for the \textsc{Boussinesq} systems with $\theta^{\,2}\ \not\ =\ 1/2$ since they are all expected to be similar in shape.

\begin{figure}
  \bigskip
  \centering
  \includegraphics[width=0.8\columnwidth]{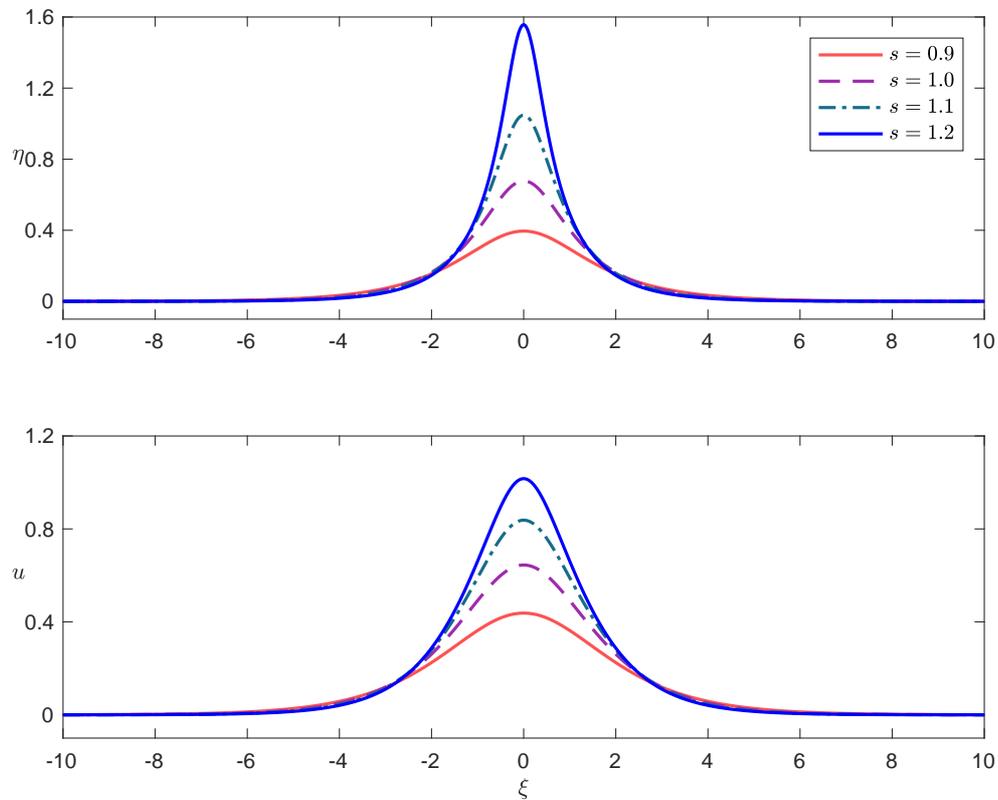}
  \caption{\small\em Solitary wave profiles for various values of speed $s\,$.}
  \label{fig:swbous}
\end{figure}

Figure~\ref{fig:swbous} presents several solitary wave profiles of the classical \textsc{Boussinesq} system in the case of $E\ =\ h\ =\ r_{\,0}\ =\ \rho\ =\ \rho^{\,w}\ =\ 1$ for various values of the speed $s$ acquired using the \textsc{Petviashvili} method. We observe that solitary waves of higher amplitude propagate faster as it was expected. Moreover, due to the unusual relationship between the wall vessel excitation $\eta$ and the velocity profile $u\,$, their profiles differ in shape by a nonlinear manner.

\begin{figure}
  \centering
  \includegraphics[width=0.8\columnwidth]{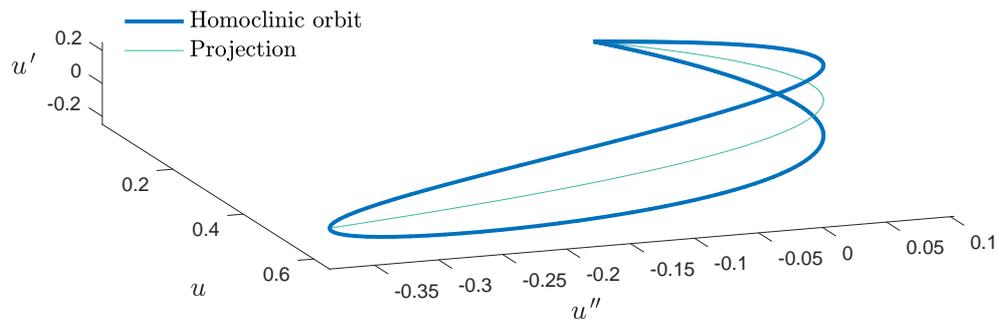}
  \caption{\small\em Phase diagram of a solitary wave with $s\ =\ 1\,$.}
  \label{fig:phase3d}
\end{figure}

Using also \eqref{eq:petm2} we are able to plot the solution curve in the phase space $(u,\,u^{\,\prime},\,u^{\,\prime\prime})$ of numerically computed solitary waves of the classical \textsc{Boussinesq} system as they can be represented by homoclinic orbits to the origin in the three dimensional space. Figure~\ref{fig:phase3d} presents the homoclinic orbit of a solitary wave with $s\ =\ 1$ in the three-dimensional space along with its projection on the $(u,\,u^{\,\prime\prime})$ plane. Various projections can be drawn from this data onto two-dimensional spaces resulting to other homoclinic orbits, \cf \eg \cite{DMII}.

\begin{figure}
  \bigskip
  \centering
  \includegraphics[width=0.8\columnwidth]{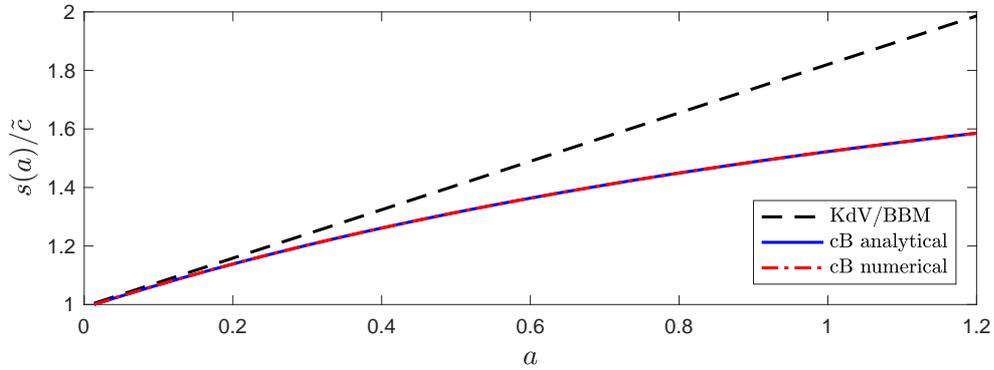}
  \caption{\small\em The speed $s$ of the solitary wave as a function of its amplitude for the unidirectional propagation models (KdV/BBM) and the classical \textsc{Boussinesq} system (cB).}
  \label{fig:spdampl}
\end{figure}

It is noted that the relationship between the amplitude $a$ of a solitary wave and its speed $s$ can be found analytically for the classical \textsc{Boussinesq} system. Specifically, multiplying \eqref{eq:petm2} with $u^{\,\prime}\,$, integrating (using the fact that solitary waves and their derivatives tend exponentially to zero), using \eqref{eq:peteta} and solving for $s$ when $u^{\,\prime}\ =\ \eta^{\,\prime}\ =\ 0$ we obtain:
\begin{equation}\label{eq:dpdampl}
  s\ =\ \sqrt{6\,\bar{\beta}\,r_{\,0}\;\frac{(2\ -\ \zeta)\ -\ 2\,\sqrt{1\ -\ \zeta}}{\zeta^{\,2}\,(3\ -\ \zeta)}}\,,
\end{equation}
where
\begin{equation*}
  \zeta\ =\ 1\ -\ \frac{r_{\,0}^{\,2}}{(a\ +\ r_{\,0})^{\,2}}\,.
\end{equation*}
We observe that the speed-amplitude relation is universal for all classical \textsc{Boussinesq} systems of the specific form since it is independent of the coefficient $d\,$. Similar behaviour has been observed also between the KdV and BBM equations which share the same speed-amplitude relation. Figure~\ref{fig:spdampl} shows the speed of the solitary wave as a function of its amplitude for the classical \textsc{Boussinesq} system and the unidirectional model equations.

In this figure also we present the computational values of the speed obtained during the computation of the solitary waves with amplitudes in the range $[\,0.01,\,1.2\,]$ with the \textsc{Petviashvili} method. We observe that the numerical values coincide with the analytical solution while the speed of the solitary wave of the unidirectional model equations diverges from the analogous speed of the solitary wave of the classical \textsc{Boussinesq} system as the amplitude of the solitary wave increases.

%%% ------------------------------------------------------------------------ %%%

\section{Numerical results and applications to blood flow}
\label{sec:application}

In this Section we consider the System~\eqref{vamasspd} -- \eqref{vamomentumpd} in a finite domain defined by a vessel of finite length $L$ (located in the interval $[\,0,\,L\,]$) with appropriate time dependent \textsc{Dirichlet} boundary conditions $u\,(0,\,t)\,$, $\eta\,(0,\,t)$ and $u\,(L,\,t)\,$, $\eta\,(L,\,t)$ and initial conditions $u\,(x,\,0)$ and $\eta\,(x,\,0)\,$. Here we drop the $\theta$ from the notation since it is fixed to $\theta\ =\ 1/2\,$. This particular system is a dissipative version of the classical \textsc{Boussinesq} system \cite{BCS, Peregrine1967}. After constructing a simple and efficient numerical method for the numerical solution of this classical \textsc{Boussinesq} system, we validate the constructions of the previous sections by studying the propagation and the reflection from a wall of a solitary wave. We close this section by demonstrating the applicability of this system in blood flow studies.

%%% ------------------------------------------------------------------------ %%%

\subsection{Numerical validation of the model}
\label{sec:validation}

We start with the presentation of the numerical method. \textsc{Bousisnesq} systems in the context of water waves have been discretised successfully using finite difference schemes, \cite{Peregrine1967} and other sophisticated numerical methods such as spectral, finite element/continuous and discontinuous \textsc{Galerkin}, and finite volume methods, \cf \eg \cite{Eskilsson2005, AD3, ADM2, Dutykh2011e, Dutykh2011a, Mitsotakis2017c}. For the purposes of this paper we consider a simple and efficient numerical method based on finite-difference discretisation. Specifically, we discretise the classical \textsc{Boussinesq} system using the method of lines comprised by a second order finite difference scheme in space and a third order adaptive \textsc{Runge--Kutta} method in time. The temporal discretisation is implemented using the \textsc{Matlab} \texttt{ode23} method. In order to discretise the spatial derivatives we consider a uniform subdivision $x_{\,i}\ =\ i\cdot\Delta x\,$, $i\ =\ 0,\,\cdots,\,N$ of the computation domain $[\,0,\,L\,]\,$, where $\Delta x\ =\ L\,/\,N$ is the constant mesh length. We will approximate the solution $u\,(x_{\,i},\,t)$ by the values $U_{\,i}\,(t)\,$, $\eta\,(x_{\,i},\,t)$ by $H_{\,i}\,(t)$ while $R_{\,i}\ =\ r_{\,0}\,(x_{\,i})\,$, $\bar{\alpha}_{\,i}\ =\ \bar{\alpha}\,(x_{\,i})$ and $\bar{\beta}_{\,i}\ =\ \bar{\beta}\,(x_{\,i})\,$. We will also consider the solution vectors $V\ =\ (H,\,U)^{\,T}\ =\ (H_{\,0},\,\cdots,\,H_{\,N},\,U_{\,0},\,\cdots,\,U_{\,N})^{\,T}\,$.

To describe the spatial discretisation we introduce the following notation for the approximation of the spatial derivatives with finite-difference operators:
\begin{align}
  & \delta V_{\,i}\ =\ \frac{V_{\,i\,+\,1}\ -\ V_{\,i\,-\,1}}{2\,\Delta x}\,, \\
  & \Delta V_{\,i}\ =\ \frac{V_{\,i\,+\,1}\ -\ 2\,V_{\,i}\ +\ V_{\,i\,-\,1}}{\Delta x^{\,2}}\,,
\end{align}
Then the semi-discrete problems can be written as:
\begin{align}
  & \partial_{\,t}\,H_{\,i}\ +\ f\,(H_{\,i},\,U_{\,i})\ =\ 0\,, \label{eq:7p3} \\
  & \left(1\ -\ \bar{\alpha}_{\,i}\,\Delta R_{\,i}\ -\ \frac{(4\,\bar{\alpha}_{\,i}\ +\ R_{\,i})\,R_{\,i}}{8}\;\Delta\right)\partial_{\,t}\,U_{\,i}\ +\ g\,(H_{\,i},\,U_{\,i})\ =\ 0\,, \label{eq:7p4}
\end{align}
where $\partial_{\,t}$ denotes the temporal derivative and
\begin{align*}
  & f\,(H_{\,i},\,U_{\,i})\ =\ \frac{1}{2}\;(R_{\,i}\ +\ H_{\,i})\,\delta U_{\,i}\ +\ \delta(R_{\,i}\ +\ H_{\,i})\,U_{\,i}\,,\\
  & g\,(H_{\,i},\,U_{\,i})\ =\ \delta\,[\,\bar{\beta}_{\,i}\,H_{\,i}\,]\ +\ U_{\,i}\,\delta U_{\,i}\ +\ \frac{(3\,\bar{\alpha}_{\,i}\ +\ R_{\,i})\,\delta R_{\,i}}{2}\;\Delta[\,\bar{\beta}_{\,i}\,H_{\,i}\,]\ +\ 8\,\kappa\;\frac{U_{\,i}}{R_{\,i}^{\,2}}\,.
\end{align*}
The System \eqref{eq:7p4} can be written as a system of ordinary differential equations of the form
\begin{equation*}
  A\,(R)\,\partial_{\,t}\,U\ =\ F\,(U)\,.
\end{equation*}
where $A$ is a tridiagonal $(N\,+\,1)\,\times\,(N\,+\,1)$ matrix that depends on $R_{\,i}\,$, $\bar{\alpha}\,$, and $\Delta x\,$. After solving for $\partial_{\,t}\,U$ using a tridiagonal algorithm, the resulting system of ordinary differential equations is being integrated in time using the adaptive \textsc{Runge--Kutta} method. It is noted that when inhomogeneous boundary conditions are provided the system can be transformed to a similar one with homogeneous \textsc{Dirichlet} boundary conditions using a linear transformation of the solution and modifying the function $F$ and the matrix $A$ appropriately. Moreover, we mention that in the case constant radius we only need to impose boundary data for the velocity $u$ in order to determine the solution of the classical \textsc{Boussinesq} system, \cite{FP}. We also mention that the derivatives on the boundary are treated appropriately in order to comply with the given boundary conditions.

\begin{figure}
  \bigskip
  \centering
  \includegraphics[width=0.99\columnwidth]{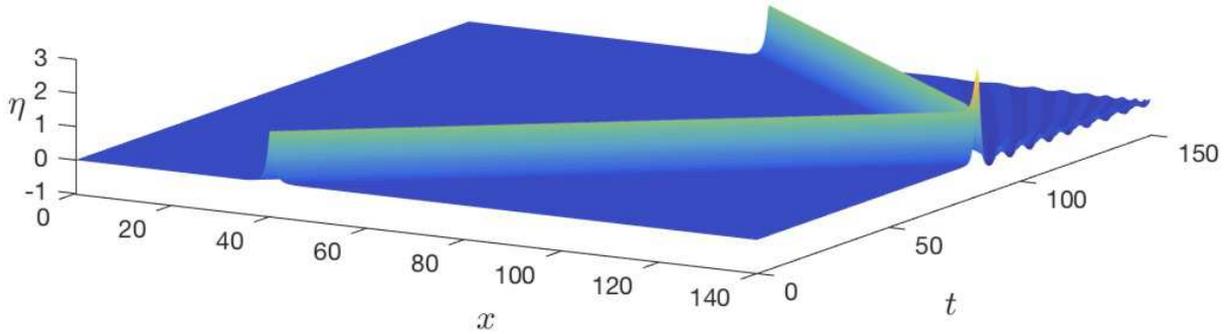}
  \caption{\small\em Propagation and reflection of a solitary wave of the classical \textsc{Boussinesq} system ($s\ =\ 1.2$).}
  \label{fig:swprop}
\end{figure}

In order to verify the efficiency of the numerical method we consider the case of the propagation of a solitary wave with $s\ =\ 1.2$ in a vessel of constant length $L\ =\ 140$ and radius $r_{\,0}\ =\ 1\,$. For simplicity we take in this experiment $E\ =\ h\ =\ \rho\ =\ \rho^{\,w}\ =\ 1\,$. We impose reflective boundary conditions $u\ =\ 0$ on both boundaries while we take the viscosity coefficient $\kappa\ =\ 0\,$. After we constructed the solitary wave numerically using \textsc{Petviashvili}'s method in the specified domain and using $\Delta x\ =\ 0.02$ we move the solitary wave to $x\ =\ 40$ and let it propagate until the final time $T\ =\ 150\,$. For the adaptive time-stepping method we consider the relative error tolerance $10^{\,-\,6}$ which appeared to be sufficient to compute the solution accurately. Figure~\ref{fig:swprop} shows the propagation of the solitary wave and its reflection on the right boundary.

We observe that the solitary wave propagates without variations in its shape. The reflection is not elastic as it was expected since for \textsc{Boussinesq} systems the result of the reflection coincides with the head-on collision of two solitary waves of the same amplitude that propagate in different directions. Similar results have been observed for the analogous classical \textsc{Boussinesq} system of water wave theory, \cite{AD3, ADM2}.

%%% ------------------------------------------------------------------------ %%%

\subsection{Application to a blood flow problem}
\label{sec:blood}

In the rest of this section we use the prescribed numerical method to study the blood flow in a vessel that represents an ideal common carotid artery. Analogous experiments have been considered in \cite{Figueroa2006, Xiao2014}. In this experiment we consider an ideal blood flow in an artery of a fixed length $L$ with a constant nominal radius $R$ where we also consider variations of this vessel with variable radius as well. The required characteristic quantities for this particular experiment are summarised in Table~\ref{tab:carotid}.

\begin{table}
\begin{tabular}{l c c}
\hline
Vessel's length & $L$ & $12.6\times 10^{-2}~m$ \\
Vessel's nominal radius & $R$ & $0.3\times 10^{-2}~m$ \\
Vessel's thickness & $h$ & $0.03\times 10^{-2}~m$\\
Wall density & $\rho^w$ & $1000~kg/m^3$\\
Fluid density & $\rho$ & $1060~kg/m^3$\\
Young's modulus & $E$ & $4.07\times 10^5~kg/m\cdot sec^2$\\
Kinematic viscosity & $\kappa$ & $0.4\times 10^{-5}~m^2/sec$\\
\hline\\
\end{tabular}
\caption{\small\em Characteristic values of an ideal blood flow in a carotid artery in S.I.}\label{tab:carotid}
\end{table}

\begin{figure}
  \bigskip
  \centering
  \includegraphics[width=0.99\columnwidth]{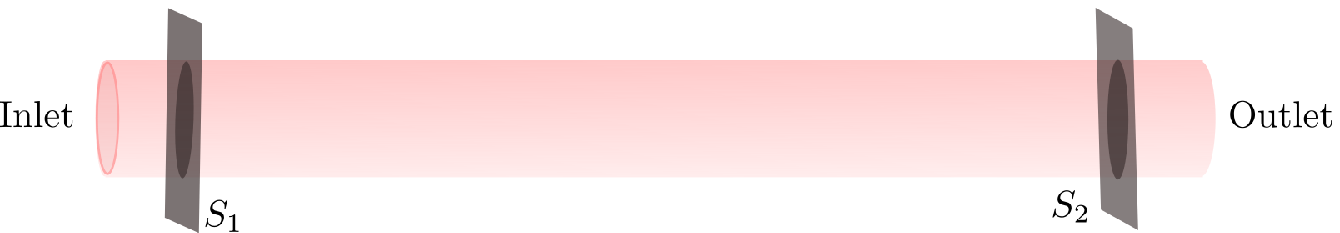}
  \includegraphics[width=0.99\columnwidth]{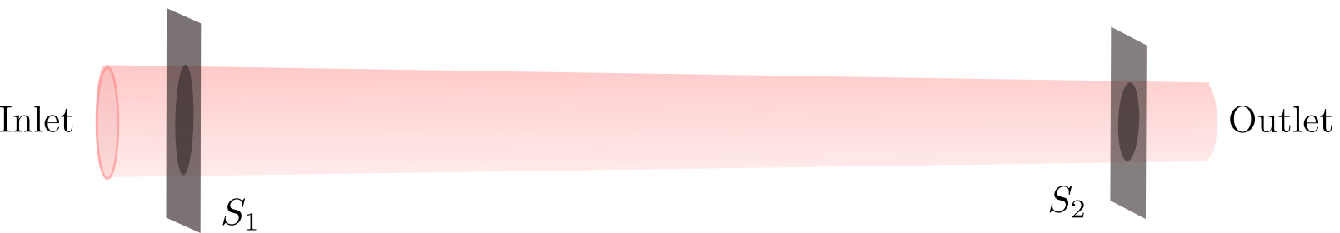}
  \caption{\small\em Sketch of ideal and tapered carotid arteries.}
  \label{fig:carotid1}
\end{figure}

We consider two cases: (1) an ideal and (2) a tapered carotid-like artery. In the case (1) the vessel has constant radius $r_{\,0}\ =\ 0.3\,\times\,10^{\,-\,2}~ m$. The tapered artery initially has a luminal radius $r_{\,0}\,(0)\ =\ 0.003~m$ while the outflow boundary has radius $r_{\,0}\,(0.126)\ =\ 0.0026~m$. A sketch of the two different vessels is presented in Figure~\ref{fig:carotid1}. In all cases the inlet is located at $x\ =\ 0$ while the outlet at $x\ =\ L$ where all the vessel parameters are described in Table~\ref{tab:carotid}.

\begin{figure}
  \bigskip
  \centering
  \includegraphics[width=0.8\columnwidth]{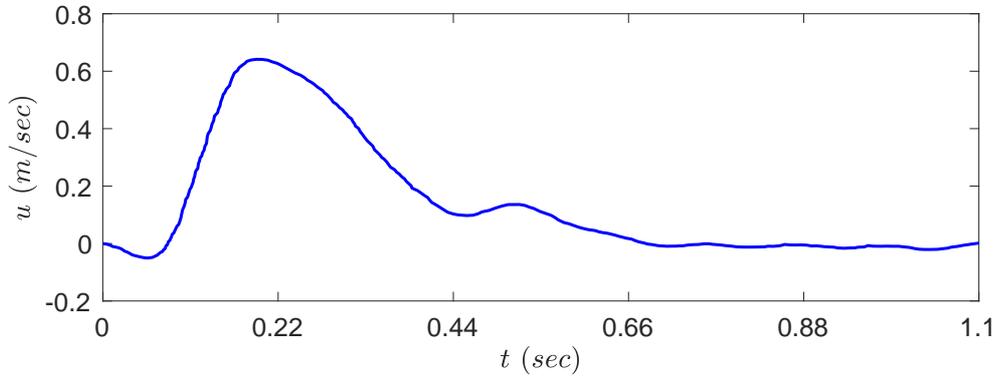}
  \caption{\small\em The velocity wave at the inlet of an ideal carotid artery model.}
  \label{fig:inlet}
\end{figure}

In each case the blood flow is determined at the inlet by a standard velocity waveform during a cardiac cycle for the left common carotid (\cf \eg \cite{Alastruey2016}) which is presented in Figure~\ref{fig:inlet} and it is used as a left boundary condition of the horizontal velocity for the numerical simulation. While boundary conditions based on lumped models \cite{Alastruey2008, Milisic2004} can be employed to describe the radiation of the flow to other vessels in a more realistic setting, we choose to use a standard absorbing layer by extending the computational domain sufficiently and imposing standard wall boundary conditions on the right boundary of the extended domain. We also assume that initially there is no flow in the vessel while the vessel wall is at rest. The initial pressure can also be prescribed using realistic data and here we consider initially that the pressure in the vessel is $10.26~kPa$ (or $76~mmHg$). Due to the boundary requirements of the classical \textsc{Boussinesq} system (and since the vessel initially has constant radius) at the inlet it is not necessary to prescribe boundary conditions for the motion of the vessel wall, \ie for the $\eta$ component of the solution but we compute it using the mass conservation equation \eqref{vamasspd}, \cite{FP}. In the case of variable radius at the inlet the additional boundary conditions on $\eta$ are prescribed using low-order approximations such as \eqref{eq:p1}.

\begin{figure}
  \bigskip
  \centering
  \includegraphics[width=0.8\columnwidth]{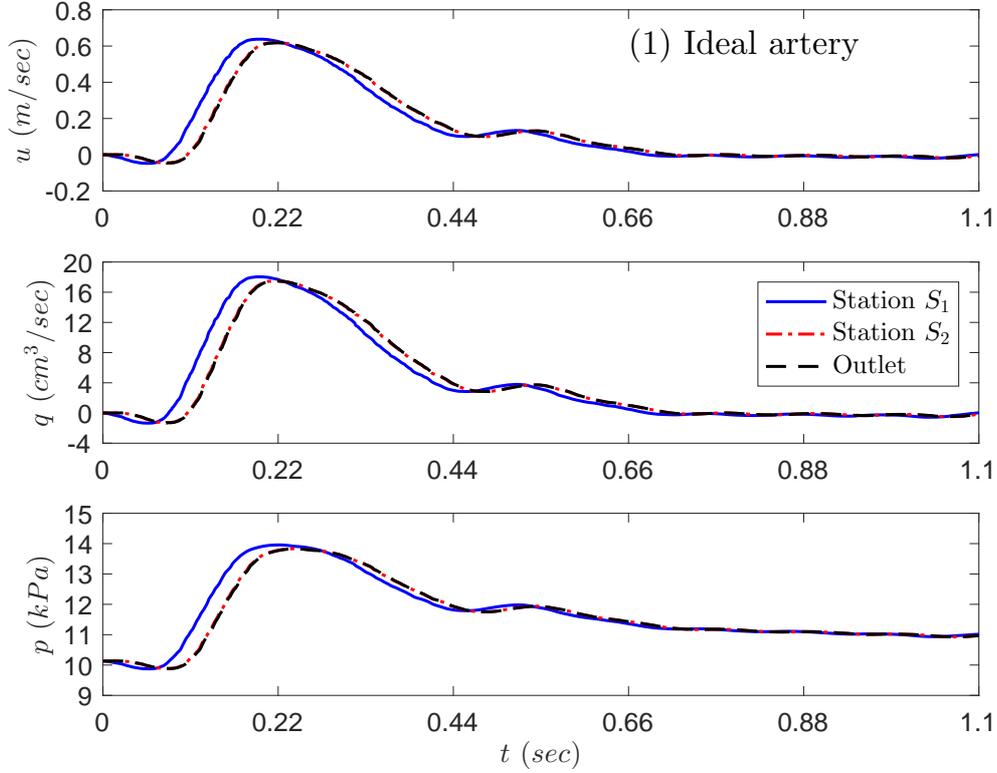}
  \caption{\small\em The velocity, flow rate and pressure at the inlet, outlet and stations $S_{\,1}$ and $S_{\,2}$ of an ideal carotid artery model.}
  \label{fig:results1}
\end{figure}

\begin{figure}
  \bigskip
  \centering
  \includegraphics[width=0.8\columnwidth]{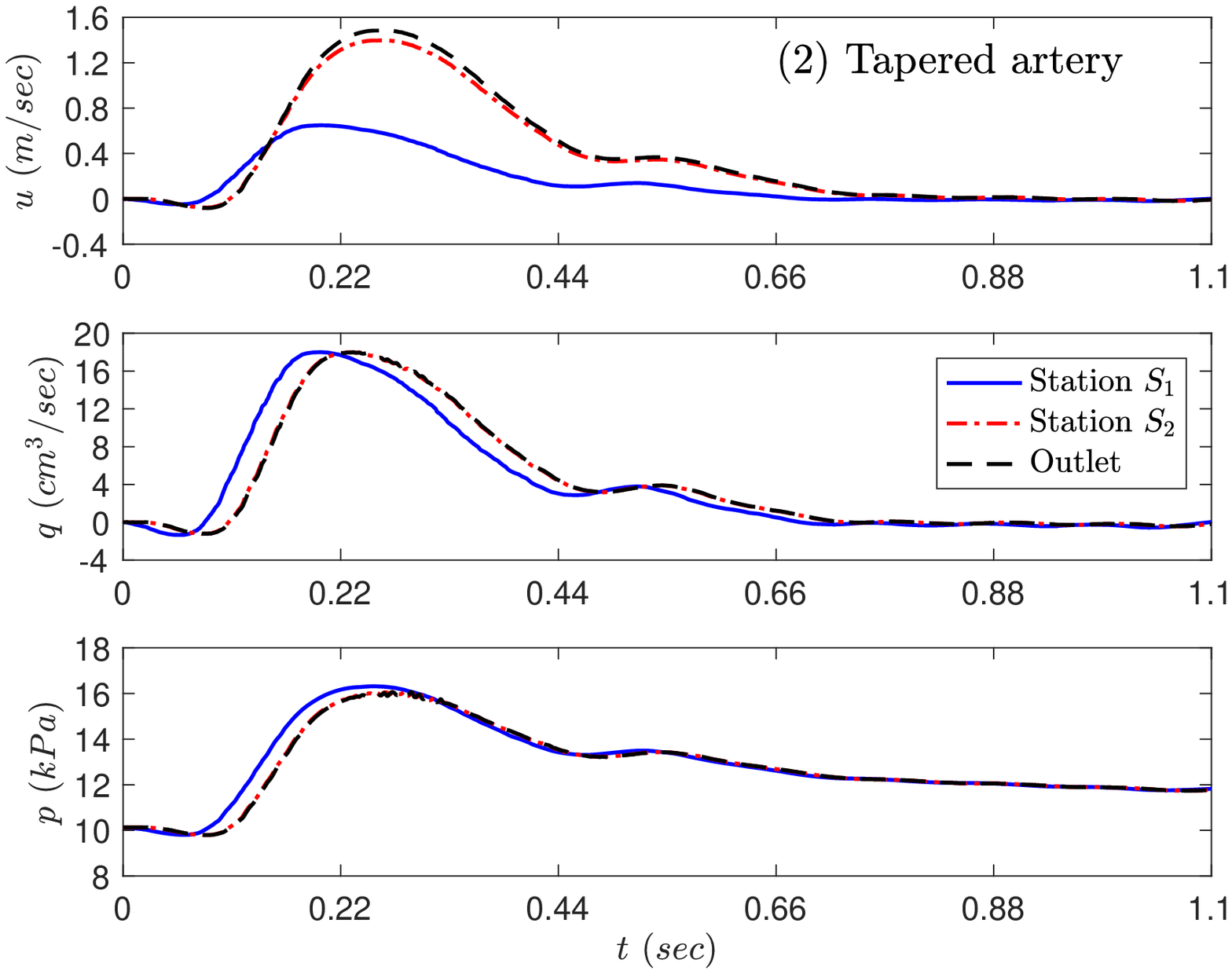}
  \caption{\small\em The velocity, flow rate and pressure at the inlet, outlet and stations $S_{\,1}$ and $S_{\,2}$ of a tapered carotid artery model.}
  \label{fig:results3}
\end{figure}

We compute numerically the velocity and the pressure at two different locations denoted by $S_{\,1}$ and $S_{\,2}$ in Figure~\ref{fig:carotid1} and at the inlet and outlet of the vessel. We take the stations $S_{\,1}$ and $S_{\,2}$ at $x\ =\ 0.006~m$ and $x\ =\ 0.12~m$ respectively. We also compute an approximation of the flow rate 
\begin{equation*}
  q\ =\ 2\,\pi\,\int_{\,0}^{\,r^{\,w}}\,s\,u\;\ud r\ =\ \pi\,r_{\,0}^{\,2}\,u^{\,\theta}\ +\ \O\,(\varepsilon\,\delta^{\,2})\,,
\end{equation*} 
at all stations. The results for the case (1) are presented in Figure~\ref{fig:results1} and for the case (2) in Figure~\ref{fig:results3}. It is noted that pressure has been computed using Equation~\eqref{eq:dbc2} which is an $\O\,(\delta^{\,2})$ approximation of the pressure $p\,(x,\,r,\,t)$ due to the asymptotic relation \eqref{E11}.

It is noted that we studied also the case of a stenosed vessel with a symmetric stenosis. We observed that when the radius of the vessel at the centre of the stenosed vessel was decreased by a small amount (for example by $0.00005~m$) then the effects of the stenosis were insignificant. The flow and pressure weren't affected significantly by the geometry of the vessel but only insignificant reduction to their absolute value  of $\O\,(10^{\,-\,2})$ from the case (1) was observed. This reduction to the flow and pressure were mainly observed because of the partial reflection of some portion of the transmitted pulse. It is also noted that the location of the stenosis wasn't important and its effects were always similar. For smaller values of the diameter of the stenosed vessel results to an unstable flow due to turbulence and therefore cannot be studied with the current model. A remarkable change of the velocity and the pressure can be observed in the case of the tapered vessel. Although the flow rate is decreasing slightly, we observe a noticeable increase in the velocity of the fluid while we observe lower pressure values due to the decreasing values of the luminal radius of the vessel. Due to the high speed values the differences between the station $S_{\,2}$ and the outlet cannot be observed with eye accuracy except for the cases of the tapered vessel where the tapering effects are more effective compared to a local stenosis. The changes observed in the pressure can also be justified by considering Bernouli's principle of fluid flow in pipes. In all the experiments we used $\Delta x\ =\ 5\times 10^{\,-\,4}~m$ in the spatial discretisation while for the adaptive time-stepping method we considered the relevant error tolerance $10^{\,-\,6}\,$. We also implemented the implicit numerical method of \cite{Peregrine1967} and we verified the validity of our numerical experiments. Finally, we mention that in all the previously mentioned experiments a typical deformation of the vessel was of $\O\,(10^{\,-\,4})$ and taking into account the size of the prescribed vessel we verified that the $\varepsilon\ \approx\ 0.07\,$, $\delta^{\,2}\ \approx\ 0.03$ are of the same order, while $Re\ =\ \O\,(10^{\,-\,6})\,$, and therefore the examples fall into the regime modelled by the model equations. For more realistic results and comparisons with {\em in vitro} and {\em in vivo} measurements the inclusion of the viscoelastic properties of the vessels along with branching effects \cite{Bona2008} should be considered but we leave this for future studies. Turbulent phenomena can also be handled either by discarding the dispersive terms when the flow becomes critical and by using other numerical techniques that can handle the propagation of discontinuities, \cite{Sherwin2003}, but this is beyond the purposes of this paper.

%%% ------------------------------------------------------------------------ %%%

\section{Conclusions}

In this paper we derived a new set of weakly nonlinear and weakly dispersive asymptotic model equations that describe the irrotational flow of an ideal fluid in an elastic vessel. Extensions including dissipative effects due to viscosity have also been considered. The new systems verify other commonly used hyperbolic equations by asymptotic means and improve previously derived \textsc{Boussinesq} equations in the same context. Fundamental properties of the new models such as dispersive and nonlinear characteristics were studied in depth. A particular model similar to the classical \textsc{Boussinesq} system of water wave theory appeared to have favourable properties and attracted some special attention. This model was tested against the propagation of solitary waves using a simple finite-difference algorithm. Finally, we used this model to study the propagation of a single pulse in an ideal blood vessel modelling a carotid artery and study the effects of tapering and stenosis. The results were qualitatively in the range of observed {\em in vivo} measurements (\cite{VandeVosse2011}) confirming the applicability of the new model equations to regular blood flow problems. More realistic situations require the inclusion of viscoelastic effects of the vessel wall along with shock capturing numerical techniques so as to simulate unstable flows. The new models can also be used to provide initial conditions to three-dimensional models that have been developed to study blood flow in complicated geometrical settings.

%%% ------------------------------------------------------------------------ %%%

\subsection*{Acknowledgments}
\addcontentsline{toc}{subsection}{Acknowledgments}

The work of D.~\textsc{Mitsotakis} was supported by the \textsc{Marsden} Fund administered by the Royal Society of \textsc{New Zealand}.

%%% ------------------------------------------------------------------------ %%%

%%% Bibliography
\bigskip
\addcontentsline{toc}{section}{References}
\bibliographystyle{abbrv}
\bibliography{biblio}

\begin{thebibliography}{10}

\bibitem{Alastruey2011}
J.~Alastruey, A.~W. Khir, K.~S. Matthys, P.~Segers, S.~J. Sherwin, P.~R.
  Verdonck, K.~H. Parker, and J.~Peir{\'{o}}.
\newblock {Pulse wave propagation in a model human arterial network: Assessment
  of 1-D visco-elastic simulations against in vitro measurements}.
\newblock {\em Journal of Biomechanics}, 44(12):2250--2258, aug 2011.

\bibitem{Alastruey2008}
J.~Alastruey, K.~H. Parker, and S.~J. Sherwin.
\newblock {Lumped parameter outflow models for 1-D blood flow simulations:
  effect on pulse waves and parameter estimation}.
\newblock {\em Comm. Comp. Phys.}, 4(2):317--336, 2008.

\bibitem{Alastruey2016}
J.~Alastruey, N.~Xiao, H.~Fok, T.~Schaeffter, and C.~A. Figueroa.
\newblock {On the impact of modelling assumptions in multi-scale,
  subject-specific models of aortic haemodynamics}.
\newblock {\em Journal of The Royal Society Interface}, 13(119):20160073, jun
  2016.

\bibitem{Alvarez2014}
J.~{\'{A}}lvarez and A.~Dur{\'{a}}n.
\newblock {Petviashvili type methods for traveling wave computations: I.
  Analysis of convergence}.
\newblock {\em J. Comp. Appl. Math.}, 266:39--51, aug 2014.

\bibitem{ADM1}
D.~C. Antonopoulos, V.~A. Dougalis, and D.~E. Mitsotakis.
\newblock {Initial-boundary-value problems for the Bona-Smith family of
  Boussinesq systems}.
\newblock {\em Advances in Differential Equations}, 14:27--53, 2009.

\bibitem{ADM2}
D.~C. Antonopoulos, V.~A. Dougalis, and D.~E. Mitsotakis.
\newblock {Numerical solution of Boussinesq systems of the Bona-Smith family}.
\newblock {\em Appl. Numer. Math.}, 30:314--336, 2010.

\bibitem{AD3}
D.~C. Antonopoulos and V.~D. Dougalis.
\newblock {Numerical solution of the `classical' Boussinesq system}.
\newblock {\em Math. Comp. Simul.}, 82:984--1007, 2012.

\bibitem{Berntsson2016}
F.~Berntsson, M.~Karlsson, V.~Kozlov, and S.~A. Nazarov.
\newblock {A one-dimensional model of viscous blood flow in an elastic vessel}.
\newblock {\em Applied Mathematics and Computation}, 274:125--132, feb 2016.

\bibitem{Bessems2007}
D.~Bessems, M.~Rutten, and F.~{Van de Vosse}.
\newblock {A wave propagation model of blood flow in large vessels using an
  approximate velocity profile function}.
\newblock {\em J. Fluid Mech.}, 580:145, jun 2007.

\bibitem{Bona2008}
J.~L. Bona and R.~Cascaval.
\newblock {Nonlinear dispersive waves on trees}.
\newblock {\em Canadian Applied Mathematics Quarterly}, 16(1):1--18, 2008.

\bibitem{BCS}
J.~L. Bona, M.~Chen, and J.-C. Saut.
\newblock {Boussinesq equations and other systems for small-amplitude long
  waves in nonlinear dispersive media. I: Derivation and linear theory}.
\newblock {\em J. Nonlinear Sci.}, 12:283--318, 2002.

\bibitem{Bona2004}
J.~L. Bona, M.~Chen, and J.-C. Saut.
\newblock {Boussinesq equations and other systems for small-amplitude long
  waves in nonlinear dispersive media: II. The nonlinear theory}.
\newblock {\em Nonlinearity}, 17:925--952, 2004.

\bibitem{BS}
J.~L. Bona and R.~Smith.
\newblock {A model for the two-way propagation of water waves in a channel}.
\newblock {\em Math. Proc. Camb. Phil. Soc.}, 79:167--182, 1976.

\bibitem{Cascaval2003}
R.~Cascaval.
\newblock {Variable coefficient KdV equations and waves in elastic tubes}.
\newblock In G.~R. Goldstein, R.~Nagel, and S.~Romanelli, editors, {\em
  Evolution Equations}, pages 57--70. Marcel Dekker, Inc., New York, Basel,
  2003.

\bibitem{Cascaval2012}
R.~C. Cascaval.
\newblock {A Boussinesq model for pressure and flow velocity waves in arterial
  segments}.
\newblock {\em Math. Comp. Simul.}, 82(6):1047--1055, feb 2012.

\bibitem{Chandran2012}
K.~B. Chandran, S.~E. Rittgers, and A.~P. Yoganathan.
\newblock {\em {Biofluid Mechanics: The Human Circulation}}.
\newblock CRC Press, Boca Raton, FL, 2 edition, 2012.

\bibitem{Chen2000a}
M.~Chen.
\newblock {Solitary-wave and multi-pulsed traveling-wave solutions of
  boussinesq systems}.
\newblock {\em Applicable Analysis}, 75(1-2):213--240, jun 2000.

\bibitem{Cheviakov2007}
A.~F. Cheviakov.
\newblock {GeM software package for computation of symmetries and conservation
  laws of differential equations}.
\newblock {\em Comp. Phys. Comm.}, 176(1):48--61, jan 2007.

\bibitem{Demiray2007}
H.~Demiray.
\newblock {Waves in fluid-filled elastic tubes with a stenosis: Variable
  coefficients KdV equations}.
\newblock {\em J. Comp. Appl. Math.}, 202(2):328--338, may 2007.

\bibitem{DMII}
V.~A. Dougalis and D.~E. Mitsotakis.
\newblock {Theory and numerical analysis of Boussinesq systems: A review}.
\newblock In N.~A. Kampanis, V.~A. Dougalis, and J.~A. Ekaterinaris, editors,
  {\em Effective Computational Methods in Wave Propagation}, pages 63--110. CRC
  Press, 2008.

\bibitem{Duran2013}
A.~Duran, D.~Dutykh, and D.~Mitsotakis.
\newblock {On the Galilean Invariance of Some Nonlinear Dispersive Wave
  Equations}.
\newblock {\em Stud. Appl. Math.}, 131(4):359--388, nov 2013.

\bibitem{Dutykh2011a}
D.~Dutykh, D.~Clamond, P.~Milewski, and D.~Mitsotakis.
\newblock {Finite volume and pseudo-spectral schemes for the fully nonlinear 1D
  Serre equations}.
\newblock {\em Eur. J. Appl. Math.}, 24(05):761--787, 2013.

\bibitem{Dutykh2007}
D.~Dutykh and F.~Dias.
\newblock {Dissipative Boussinesq equations}.
\newblock {\em C. R. Mecanique}, 335:559--583, 2007.

\bibitem{Dutykh2014f}
D.~Dutykh and O.~Goubet.
\newblock {Derivation of dissipative Boussinesq equations using the
  Dirichlet-to-Neumann operator approach}.
\newblock {\em Math. Comp. Simul.}, 127:80--93, sep 2016.

\bibitem{Dutykh2011e}
D.~Dutykh, T.~Katsaounis, and D.~Mitsotakis.
\newblock {Finite volume schemes for dispersive wave propagation and runup}.
\newblock {\em J. Comput. Phys.}, 230(8):3035--3061, apr 2011.

\bibitem{Erbay1992}
H.~A. Erbay, S.~Erbay, and S.~Dost.
\newblock {Wave propagation in fluid filled nonlinear viscoelastic tubes}.
\newblock {\em Acta Mechanica}, 95(1-4):87--102, mar 1992.

\bibitem{Eskilsson2005}
C.~Eskilsson and S.~J. Sherwin.
\newblock {Discontinuous Galerkin Spectral/hp Element Modelling of Dispersive
  Shallow Water Systems}.
\newblock {\em J. Sci. Comput.}, 22:269--288, 2005.

\bibitem{Figueroa2006}
C.~A. Figueroa, I.~E. Vignon-Clementel, K.~E. Jansen, T.~J.~R. Hughes, and
  C.~A. Taylor.
\newblock {A coupled momentum method for modeling blood flow in
  three-dimensional deformable arteries}.
\newblock {\em Computer Methods in Applied Mechanics and Engineering},
  195(41-43):5685--5706, aug 2006.

\bibitem{FP}
A.~S. Fokas and B.~Pelloni.
\newblock {Boundary value problems for Boussinesq type systems}.
\newblock {\em Math. Phys. Anal. Geom.}, 8:59--96, 2005.

\bibitem{Formaggia2003}
L.~Formaggia, D.~Lamponi, and A.~Quarteroni.
\newblock {One-dimensional models for blood flow in arteries}.
\newblock {\em Journal of Engineering Mathematics}, 47(3/4):251--276, dec 2003.

\bibitem{Fung1993}
Y.-C. Fung.
\newblock {\em {Biomechanics: Mechanical Properties of Living Tissues}}.
\newblock Springer New York, New York, NY, 1993.

\bibitem{Fung1997a}
Y.~C. Fung.
\newblock {\em {Biomechanics: circulation}}.
\newblock Springer New York, New York, NY, 1997.

\bibitem{Hughes1973}
T.~J.~R. Hughes and J.~Lubliner.
\newblock {On the one-dimensional theory of blood flow in the larger vessels}.
\newblock {\em Mathematical Biosciences}, 18(1-2):161--170, oct 1973.

\bibitem{Korteweg1878}
D.~J. Korteweg.
\newblock {Ueber die Fortpflanzungsgeschwindigkeit des Schalles in elastischen
  R{\"{o}}hren}.
\newblock {\em Annalen der Physik und Chemie}, 241(12):525--542, 1878.

\bibitem{Landau1987}
L.~D. Landau and E.~M. Lifshitz.
\newblock {\em {Fluid Mechanics}}.
\newblock Pergamon Press, Oxford, 2nd edition, 1987.

\bibitem{LeMeur2015}
H.~V.~J. {Le Meur}.
\newblock {Derivation of a viscous Boussinesq system for surface water waves}.
\newblock {\em Asymptotic Analysis}, 94(3-4):309--345, sep 2015.

\bibitem{Milisic2004}
V.~Mili{\v{s}}i{\'{c}} and A.~Quarteroni.
\newblock {Analysis of lumped parameter models for blood flow simulations and
  their relation with 1D models}.
\newblock {\em ESAIM: M2AN}, 38(4):613--632, jul 2004.

\bibitem{Mitsotakis2017c}
D.~Mitsotakis, C.~Synolakis, and M.~McGuinness.
\newblock {A modified Galerkin/finite element method for the numerical solution
  of the Serre-Green-Naghdi system}.
\newblock {\em Int. J. Num. Meth. Fluids}, 83(10):755--778, apr 2017.

\bibitem{Mitsotakis2007}
D.~E. Mitsotakis.
\newblock {Boussinesq systems in two space dimensions over a variable bottom
  for the generation and propagation of tsunami waves}.
\newblock {\em Math. Comp. Simul.}, 80:860--873, 2009.

\bibitem{Moens1877}
A.~I. Moens.
\newblock {\em {Over de voortplantingssnelheid van den pols}}.
\newblock SC Van Doesburgh, Leiden, 1877.

\bibitem{Muller2013}
L.~O. M{\"{u}}ller and E.~F. Toro.
\newblock {Well-balanced high-order solver for blood flow in networks of
  vessels with variable properties}.
\newblock {\em Int. J. Numer. Methods Biomed. Eng.}, 29(12):1388--1411, dec
  2013.

\bibitem{Mynard2008}
J.~P. Mynard and P.~Nithiarasu.
\newblock {A 1D arterial blood flow model incorporating ventricular pressure,
  aortic valve and regional coronary flow using the locally conservative
  Galerkin (LCG) method}.
\newblock {\em Commun. Numer. Meth. Engng.}, 24(5):367--417, mar 2008.

\bibitem{Nichols2011}
W.~Nichols, M.~O'Rourke, and C.~Vlachopoulos.
\newblock {\em {McDonald's Blood Flow in Arteries}}.
\newblock CRC Press, London, sixth edition, 2011.

\bibitem{Nwogu1993}
O.~Nwogu.
\newblock {Alternative form of Boussinesq equations for nearshore wave
  propagation}.
\newblock {\em J. Waterway, Port, Coastal and Ocean Engineering}, 119:618--638,
  1993.

\bibitem{Olufsen2000}
M.~S. Olufsen, C.~S. Peskin, W.~Y. Kim, E.~M. Pedersen, A.~Nadim, and
  J.~Larsen.
\newblock {Numerical Simulation and Experimental Validation of Blood Flow in
  Arteries with Structured-Tree Outflow Conditions}.
\newblock {\em Annals of Biomedical Engineering}, 28(11):1281--1299, nov 2000.

\bibitem{Pelinovsky2004}
D.~E. Pelinovsky and Y.~A. Stepanyants.
\newblock {Convergence of Petviashvili's iteration method for numerical
  approximation of stationary solutions of nonlinear wave equations}.
\newblock {\em SIAM J. Num. Anal.}, 42:1110--1127, 2004.

\bibitem{Peregrine1967}
D.~H. Peregrine.
\newblock {Long waves on a beach}.
\newblock {\em J. Fluid Mech.}, 27:815--827, 1967.

\bibitem{Perktold1995}
K.~Perktold and G.~Rappitsch.
\newblock {Computer simulation of local blood flow and vessel mechanics in a
  compliant carotid artery bifurcation model}.
\newblock {\em Journal of Biomechanics}, 28(7):845--856, jul 1995.

\bibitem{Petviashvili1976}
V.~I. Petviashvili.
\newblock {Equation of an extraordinary soliton}.
\newblock {\em Sov. J. Plasma Phys.}, 2(3):469--472, 1976.

\bibitem{Quarteroni2004}
A.~Quarteroni and L.~Formaggia.
\newblock {Mathematical Modelling and Numerical Simulation of the
  Cardiovascular System}.
\newblock In {\em Handbook of Numerical Analysis}, pages 3--127. Elsevier B.V.,
  2004.

\bibitem{Ravindran1979}
R.~Ravindran and P.~Prasad.
\newblock {A mathematical analysis of nonlinear waves in a fluid filled
  visco-elastic tube}.
\newblock {\em Acta Mechanica}, 31(3-4):253--280, sep 1979.

\bibitem{Reymond2011}
P.~Reymond, Y.~Bohraus, F.~Perren, F.~Lazeyras, and N.~Stergiopulos.
\newblock {Validation of a patient-specific one-dimensional model of the
  systemic arterial tree}.
\newblock {\em American Journal of Physiology-Heart and Circulatory
  Physiology}, 301(3):H1173--H1182, sep 2011.

\bibitem{Sherwin2003}
S.~J. Sherwin, V.~Franke, J.~Peir{\'{o}}, and K.~Parker.
\newblock {One-dimensional modelling of a vascular network in space-time
  variables}.
\newblock {\em Journal of Engineering Mathematics}, 47(3/4):217--250, dec 2003.

\bibitem{Smith2002}
N.~P. Smith, A.~J. Pullan, and P.~J. Hunter.
\newblock {An Anatomically Based Model of Transient Coronary Blood Flow in the
  Heart}.
\newblock {\em SIAM J. Appl. Math.}, 62(3):990--1018, jan 2002.

\bibitem{Stergiopulos1992}
N.~Stergiopulos, D.~F. Young, and T.~R. Rogge.
\newblock {Computer simulation of arterial flow with applications to arterial
  and aortic stenoses}.
\newblock {\em Journal of Biomechanics}, 25(12):1477--1488, dec 1992.

\bibitem{Tait1981}
R.~J. Tait, T.~B. Moodie, and J.~B. Haddow.
\newblock {Wave propagation in a fluid-filled elastic tube}.
\newblock {\em Acta Mechanica}, 38(1-2):71--83, mar 1981.

\bibitem{Taylor1998}
C.~A. Taylor, T.~J.~R. Hughes, and C.~K. Zarins.
\newblock {Finite element modeling of blood flow in arteries}.
\newblock {\em Computer Methods in Applied Mechanics and Engineering},
  158(1-2):155--196, may 1998.

\bibitem{VandeVosse2011}
F.~N. van~de Vosse and N.~Stergiopulos.
\newblock {Pulse Wave Propagation in the Arterial Tree}.
\newblock {\em Ann. Rev. Fluid Mech.}, 43(1):467--499, jan 2011.

\bibitem{Wang2015}
X.~Wang, J.-M. Fullana, and P.-Y. Lagr{\'{e}}e.
\newblock {Verification and comparison of four numerical schemes for a 1D
  viscoelastic blood flow model}.
\newblock {\em Computer Methods in Biomechanics and Biomedical Engineering},
  18(15):1704--1725, nov 2015.

\bibitem{Whitham1999}
G.~B. Whitham.
\newblock {\em {Linear and nonlinear waves}}.
\newblock John Wiley {\&} Sons Inc., New York, 1999.

\bibitem{Xiao2014}
N.~Xiao, J.~Alastruey, and C.~{Alberto Figueroa}.
\newblock {A systematic comparison between 1-D and 3-D hemodynamics in
  compliant arterial models}.
\newblock {\em Int. J. Numer. Methods Biomed. Eng.}, 30(2):204--231, feb 2014.

\bibitem{Yomosa1987}
S.~Yomosa.
\newblock {Solitary Waves in Large Blood Vessels}.
\newblock {\em Journal of the Physical Society of Japan}, 56(2):506--520, feb
  1987.

\bibitem{Zagzoule1986}
M.~Zagzoule and J.-P. Marc-Vergnes.
\newblock {A global mathematical model of the cerebral circulation in man}.
\newblock {\em Journal of Biomechanics}, 19(12):1015--1022, jan 1986.

\bibitem{Zamir2000}
M.~Zamir.
\newblock {\em {The Physics of Pulsatile Flow}}.
\newblock Springer New York, New York, NY, 2000.

\end{thebibliography}
\bigskip

\end{document}